\documentclass[
nofootinbib,
amsmath,amssymb,
aps,
]{revtex4-2}
\usepackage{amsmath,amsfonts,amsthm,amssymb}
\usepackage[dvips]{graphics,graphicx}
\usepackage[usenames,dvipsnames]{color}
\definecolor{darkblue}{RGB}{0,0,196}
\definecolor{darkgreen}{RGB}{0,120,0}
\usepackage[colorlinks=true,linktocpage=true,linkcolor=darkblue,citecolor=red,urlcolor=darkblue]{hyperref}
\usepackage{cancel}
\usepackage{bbold}
\usepackage{multirow}
\usepackage{longtable}
\usepackage{upgreek}
\usepackage{color}
\usepackage{booktabs}
\usepackage[normalem]{ulem}
\usepackage{comment}
\usepackage{hyperref}
\usepackage{bigints}
\usepackage{xparse}
\usepackage{physics}
\usepackage{verbatim}
\usepackage{minibox}
\usepackage{comment}
\usepackage{appendix}
\usepackage{slashed}
\usepackage{marginnote}
\usepackage{graphicx}
\usepackage[nice]{nicefrac}
 
\usepackage{amsmath}
\usepackage[colorinlistoftodos]{todonotes}
\usepackage{hepunits}
\usepackage{stackengine,scalerel}
\usepackage{physics}
\newcommand\hstar[1]{\ThisStyle{\ensurestackMath{%
\setbox0=\hbox{$\SavedStyle#1$}%
\stackengine{0pt}{\copy0}{\kern.2\ht0\smash{\SavedStyle\star}}{O}{c}{F}{T}{S}}}}
\definecolor {darkgreen}{rgb}{0.2,0.7,0.2}

\newcommand{\di}{\dd}

\newcommand{\be}{\begin{equation}}
\newcommand{\ee}{\end{equation}}                   
\def\bea{\begin{eqnarray}}
\def\eea{\end{eqnarray}}
\newcommand{\msc}[1]{{\color{teal}{[MS: #1]}}}
\newcommand{\ms}[1]{{\color{teal}#1}}
\def\a{\alpha}
\def\b{\beta}

\newcommand{\at}[2][]{#1|_{#2}}

\newcommand{\mcZ}{{\mathcal{Z}}}

\begin{document}
%
%\title{\textcolor{blue}{Relativistic quantum kinetic theory at $\mathcal{O}(\hbar^{2})$: Thermodynamic Consistency and the Resolution of the Pseudo-Gauge Ambiguity}}
%
\title{
%\textcolor{blue}{Relativistic Quantum Kinetic Theory at $O(\hbar^2)$:
%Mass-Shell Modification, Pseudo-Gauge Structure, and Thermodynamics}
%\textcolor{blue}{Relativistic quantum kinetic theory at $\mathcal{O}(\hbar^{2})$: Thermodynamic Consistency and the Resolution of the Pseudo-Gauge Ambiguity}\\
Global equilibrium at order $\mathcal{O}(\hbar^2)$: pseudo-gauge ambiguity, Maxwell relations, and thermodynamic consistency}

\begin{abstract}
We study massive spin-$1/2$ Boltzmann particles in global thermodynamic equilibrium with rotation and acceleration, consistently including quantum corrections up to order $\mathcal{O}(\hbar^2)$. 
For such a system, the fluid velocity becomes an independent thermodynamic variable and fundamental thermodynamic relations have to be extended.
Employing the Wigner-function formalism, we show that the standard mass-shell condition $k^2 = m^2$ is modified at order $\mathcal{O}(\hbar^2)$ by contributions quadratic in the thermal vorticity. 
Incorporating this modified mass-shell constraint, we derive the (net) particle-number current, energy-momentum tensor, spin-current tensor, and associated thermodynamic quantities in the kinetic, canonical, and de Groot--van Leeuwen--van Weert (GLW) pseudo-gauges. 
We explicitly demonstrate that, while global quantities, like the total particle number or total energy, are independent of the choice of pseudo-gauge, local quantities, like the particle-number density or the energy density, are pseudo-gauge dependent.
We then examine the thermodynamic relations in each pseudo-gauge and demonstrate that thermodynamic consistency requires that the thermodynamic Maxwell relations are satisfied. 
We find that these relations hold in the kinetic-theory pseudo-gauge, whereas they are in general violated in the canonical and GLW pseudo-gauges. 
These findings suggest a preference for using the kinetic-theory pseudo-gauge in applications such as spin hydrodynamics.
\end{abstract}

\author{Asaad Daher}
\email{daher@itp.uni-frankfurt.de}
\affiliation{Institut f\"ur Theoretische Physik, 
	Johann Wolfgang Goethe--Universit\"at,
	Max-von-Laue-Str.\ 1, D--60438 Frankfurt am Main, Germany}
\affiliation{Institute  of  Nuclear  Physics  Polish  Academy  of  Sciences,  PL-31-342  Krak\'ow, Poland}
\author{David Wagner}
\email{dwagner.research@gmail.com}
\affiliation{Institut f\"ur Theoretische Physik, 
	Johann Wolfgang Goethe--Universit\"at,
	Max-von-Laue-Str.\ 1, D--60438 Frankfurt am Main, Germany}
\author{Masoud Shokri}
\email{shokri@itp.uni-frankfurt.de}
\affiliation{Institut f\"ur Theoretische Physik, 
	Johann Wolfgang Goethe--Universit\"at,
	Max-von-Laue-Str.\ 1, D--60438 Frankfurt am Main, Germany}
\author{Dirk H.~Rischke}
\email{drischke@itp.uni-frankfurt.de}
\affiliation{Institut f\"ur Theoretische Physik, 
	Johann Wolfgang Goethe--Universit\"at,
	Max-von-Laue-Str.\ 1, D--60438 Frankfurt am Main, Germany}
\affiliation{Helmholtz Research Academy Hesse for FAIR, Campus Riedberg,\\
Max-von-Laue-Str.~12, D-60438 Frankfurt am Main, Germany}
\maketitle

%\newpage
%\textcolor{blue}{\tableofcontents}
%\newpage
%*******************************************************************************************************************************************************************************************************************************************************************************************************************************************************************************************************************************************************************************
%
\section{Introduction}
\label{sec:Introduction}
%
% STEP 1 — Motivation of the theory

The spin polarization of hadrons produced in noncentral relativistic heavy-ion collisions~\cite{STAR:2017ckg,STAR:2018gyt,Niida:2018hfw,STAR:2019erd,ALICE:2019onw,ALICE:2019aid,STAR:2020xbm,Kornas:2020qzi,STAR:2021beb,ALICE:2021pzu,HADES:2014ttv} provides a unique probe of strong-interaction matter subject to acceleration and rotation~\cite{Liang:2004ph,Becattini:2013fla,lisa2021}. 
Relativistic hydrodynamics is an effective theory for the low-frequency, long-wavelength dynamics of matter and has turned out to be enormously successful in describing bulk observables in heavy-ion collisions~\cite{Bass:2000ib,Heinz:2013th,Romatschke:2017ejr,Florkowski:2017olj}.
In order to also describe polarization phenomena in accelerating and rotating matter in a hydrodynamical framework, one needs to extend conventional hydrodynamics by spin degrees of freedom and by the conservation law for total angular momentum.
The emerging theory, relativistic spin hydrodynamics, has attracted significant attention in recent years, both from the phenomenological~\cite{Hattori:2019lfp,Fukushima:2020ucl,Li:2020eon,She:2021lhe,Cao:2022aku,Hu:2022azy,Abbasi:2022rum,Ren:2024pur,Yang:2024duc,Drogosz:2024gzv,Gallegos:2020otk,Hongo:2021ona,Floerchinger:2021uyo,Gallegos:2021bzp,Biswas:2023qsw,Daher:2024ixz,Yang:2024duc,Daher:2024bah,Abboud:2025shb,Zakharov:2025stp,Lapygin:2025ewl,Fang:2025aig,Hontarenko:2025itd,Zhang:2026zee,Singh:2026ytd,Prokhorov:2026swu,Prokhorov:2026ebn,Khakimov:2026sct,Li:2026vld,Montenegro:2026phf,Lobos:2026ctf,Drogosz:2026qbo,Wu:2026zne,Montenegro:2026phf,Lu:2026ceo,Dey:2026epy,AitTamlihat:2026crj,Lu:2026vft,Drogosz:2026egr,Zhou:2026jbh,Fukushima:2026lzd} as well as the microscopic perspective~\cite{Florkowski:2017ruc,Florkowski:2018ahw,Florkowski:2018myy,Hidaka:2017auj,Hattori:2019ahi,Gao:2019znl,Weickgenannt:2019dks,Wang:2019moi,Li:2019qkf,Kapusta:2019sad,Ayala:2019iin,Kapusta:2020npk,Liu:2019krs,Yang:2020hri,Ayala:2023vgv,Liu:2020flb,Shi:2020htn,Bhadury:2020cop,Speranza:2020ilk,Weickgenannt:2020aaf,Peng:2021ago,Cartwright:2021qpp,Sheng:2021kfc,Weickgenannt:2021cuo,Weickgenannt:2022zxs,Weickgenannt:2022qvh,Hongo:2022izs,Hidaka:2022dmn,Weickgenannt:2022jes,Weickgenannt:2023bss,Becattini:2019poj,Becattini:2023ouz,Amano:2023bhg,Wang:2022yli,Wagner:2024fhf,Palermo:2024tza,Chiarini:2024cuv,Weickgenannt:2024esg,Weickgenannt:2024ibf, Palermo:2025imv,Becattini:2018duy,Hu:2021lnx,Tiwari:2024trl,Daher:2025pfq,Giacalone:2025bgm,Buzzegoli:2025zud,Singha:2025bda,Carrington:2025xws,Sung:2025gke,Arslan:2025tan,Chen:2026yej,Florkowski:2026ofs,Singh:2026iew,She:2025qri,Oliva:2026wbo,Kiefer:2025xdp}.

Nevertheless, three important issues remain unresolved. 
\textbf{(i)} The first concerns the microscopic foundations of the thermodynamic relations employed in the phenomenological formulation of spin hydrodynamics~\cite{Hattori:2019lfp,Fukushima:2020ucl,Biswas:2023qsw}.
This aspect has recently attracted considerable attention, in particular in view of a possible violation of thermodynamic relations~\cite{Florkowski:2024bfw,Becattini:2025oyi,Ambrus:2025dca,Armas:2026bmw,Matthaiakakis:2026qpz,Palermo:2026mwu}. 
These thermodynamic relations are relevant to a broad range of physical systems, such as rotating hadronic matter and quark-gluon plasma~\cite{Pradhan:2023rvf,Sahoo:2023xnu,Mukherjee:2023ijv,Padhan:2024edf,Chernodub:2017ref,Chernodub:2020qah,Braguta:2023yjn,Chernodub:2022veq}, rotating black holes~\cite{Gibbons:2004ai,Altamirano:2014tva,Garbiso:2020puw,Gao:2021xtt,Gong:2023ywu,Zhen:2025mqz,Ladino:2026dru}, and magnetized relativistic plasmas~\cite{Bhadury:2022ulr,Biswas:2020rps,Dommes:2020ktk,Kushwah:2024zgd,deBrito:2025jaz,Satapathy:2025jjx}.
\textbf{(ii)} The second issue is the freedom of choosing different pseudo-gauges in the definition of the hydrodynamic currents and thermodynamic quantities~\cite{Hehl:1976vr,Becattini:2011ev,Florkowski:2018fap,Weickgenannt:2022jes,Drogosz:2024rbd,Buzzegoli:2024mra,Becattini:2025twu}. 
In particular, different pseudo-gauge choices lead to different forms of the (net) particle-number current, energy-momentum tensor, and spin-current tensor, which are nevertheless conserved for all choices and are eventually supposed to describe the same physical system. 
This raises the question which pseudo-gauge should be regarded as physically appropriate in a hydrodynamic description~\cite{Armas:2026bmw}.
\textbf{(iii)} The third open topic is the numerical implementation of relativistic spin hydrodynamics. 
Recent results~\cite{Singh:2024cub,Sapna:2025yss,Matsuda:2026hwn} have highlighted the need for further theoretical investigations. 
Addressing these three issues is essential for establishing a consistent theoretical framework of relativistic spin hydrodynamics.
%

% STEP 3 — Our approach addressing the gap 
In this work, we address the first and second questions described above using the concrete example of a system of massive spin-1/2 Boltzmann particles in global thermodynamic equilibrium with nonvanishing acceleration and rotation. 
In such a system, the fluid velocity becomes an independent thermodynamic state variable and fundamental thermodynamic relations receive additional terms.
Acceleration and rotation influence the equation of state of such a system at order $\order{\hbar^{2}}$, therefore we consistently include such corrections~\cite{Buzzegoli:2017cqy,Florkowski:2024bfw}.
Quantum kinetic theory~\cite{Vasak:1987um} shows that at this order in $\hbar$, one also has to account for modifications of the standard mass-shell condition $k^2 = m^2$.
In global thermodynamic equilibrium, this generates contributions which are quadratic in thermal vorticity and which were not included in previous work~\cite{Florkowski:2024bfw}.
We then derive the (net) particle-number current, energy-momentum tensor, spin-current tensor, and associated thermodynamic quantities at order $\mathcal{O}(\hbar^2)$ for various choices of pseudo-gauges: besides the canonical and de Groot--van Leeuwen--van Weert (GLW) pseudo-gauges, we also study a modification of the latter~\cite{Wagner:2024fry}, which we term ``kinetic-theory pseudo-gauge'', because it results in the same expressions for the energy-momentum and spin-current tensors as in kinetic theory~\cite{israel1973electrodynamics}.While the grand partition and global quantities, like the total particle number, total four-momentum, total angular momentum, or total entropy, are independent of the choice of pseudo-gauge in global thermodynamic equilibrium, our calculations explicitly demonstrate that local thermodynamic quantities like the particle density, energy density, pressure, entropy density, or spin density do depend on the choice of pseudo-gauge.

For each pseudo-gauge, we check whether the thermodynamic Maxwell relations are fulfilled. 
These relations are derived from the commutativity of second derivatives of the thermodynamic pressure with respect to the independent thermodynamic variables.
In the presence of spin degrees of freedom, besides the standard relation involving the derivative of the particle-number density with respect to temperature and that of the entropy density with respect to chemical potential, we also obtain two additional relations involving the derivatives of the spin density.
The validity of the Maxwell relations ensures that the thermodynamic pressure is a thermodynamic potential in the sense that when computing it via an integral representation, the result does not depend on the path chosen in the space of independent thermodynamic variables.
This also ensures thermodynamic consistency, i.e., that thermodynamic quantities can be computed as partial derivatives of the pressure with respect to the thermodynamic variables.
We find that the kinetic-theory pseudo-gauge is thermodynamically consistent, i.e., all Maxwell relations are satisfied, different ways to compute the thermodynamic pressure give identical results, and all thermodynamic relations are fulfilled. 
By contrast, the canonical and GLW pseudo-gauges in general violate the Maxwell relations, leading to  inconsistencies of some thermodynamic relations.
At least in the canonical pseudo-gauge, thermodynamic consistency can be re-established by an \textit{ad hoc} (and thus somewhat unsatisfactory) implementation of the conditions for global thermodynamic equilibrium.
This suggests a preference for employing the kinetic-theory pseudo-gauge in applications such as spin hydrodynamics.

%

% STEP 6 — Paper roadmap
This paper is organized as follows. 
Section~\ref{sec:thermodynamics} discusses the thermodynamics of systems in global thermodynamic equilibrium with acceleration and rotation. 
In Sec.~\ref{sec:Wigner} we introduce the Wigner-function formalism for spin-$\tfrac{1}{2}$ particles and derive the $\mathcal{O}(\hbar^{2})$ modification of the mass-shell condition.
Section~\ref{sec:pseudogauge} summarizes the expressions for the (net) particle-number current, the energy-momentum tensor, and the spin-current tensors in terms of momentum-space integrals of the Clifford components of the Wigner function in different pseudo-gauges. 
Adopting Boltzmann statistics, Secs.~\ref{sec:em_s_kin_theory},~\ref{Sec-can-psuedo}, and~\ref{Sec:GLWPG} present the explicit computation of the hydrodynamic currents and thermodynamic quantities in the kinetic-theory, canonical, and GLW pseudo-gauges, respectively. 
Section~\ref{conclusion} concludes our work with a summary and an outlook.
Details of the calculations are relegated to several appendices.
%

%STEP 7 - notation 
We work in natural units $c=k_{B}=1$, keeping the reduced Planck constant $\hbar$ explicit to count the order of quantum effects. 
The metric tensor is defined as $g_{\mu\nu}=\mathrm{diag}(+,-,-,-)$. The scalar product between two four-vectors $x^{\mu}$ and $y^{\mu}$ is denoted by $x \cdot y := x^{\mu}y_{\mu}$. 
%The symmetrization of a rank-two tensor $X^{\mu\nu}$ is defined as $X^{(\mu\nu)}:= X^{\mu\nu}+X^{\nu\mu}$, while its anti-symmetrization reads $X^{[\mu\nu]} := X^{\mu\nu}-X^{\nu\mu}$. 
The fluid four-velocity is normalized $u \cdot u = 1$. 
We denote the comoving derivative of a quantity $X$ as $\dot{X}:=u^{\mu}\partial_{\mu}X$. 
The projector onto the three-space orthogonal to $u^{\mu}$ is $\Delta^{\mu\nu}:=g^{\mu\nu}-u^{\mu}u^{\nu}$.
%, and the projected four-vector is denoted $x^{\langle \mu \rangle} := \Delta^{\mu\nu}x_{\nu}$. 
%The projector onto the symmetric traceless three-subspace orthogonal to $u^{\mu}$ of a rank-2 tensor is $\Delta^{\mu\nu}_{\alpha\beta}:=\Delta^{(\mu}_{\alpha}\Delta^{\nu)}_{\beta}/2-\Delta^{\mu\nu}\Delta_{\alpha\beta}/3$, and the corresponding projected tensor is denoted $X^{\langle\mu\nu\rangle}:=\Delta^{\mu\nu}_{\alpha\beta}X^{\alpha\beta}$.}
%
%*************************************************************************************************************************************************************************************************************************************************************************************************************************************************************************************************************************************************************************************************************************************************************************************************
%
\section{Thermodynamics of accelerating and rotating systems in global thermodynamic equilibrium}
\label{sec:thermodynamics}

In this section, we review the conservation laws and thermodynamic relations for a system with nonvanishing acceleration and rotation in global thermodynamic equilibrium~(for reviews, see Refs.~\cite{Florkowski:2018fap,Becattini:2020sww}).
We first discuss the conservation laws of (net) particle number, energy-momentum, and total angular momentum and the pseudo-gauge dependence of these quantities.
Then we define the grand partition function of such a system and introduce a local thermodynamic-potential four-current $\phi^\mu(x)$, the integral of which over a Cauchy hypersurface in spacetime is related to the logarithm of the grand partition function. 
We discuss the conditions of global thermodynamic equilibrium and then derive thermodynamic identities for the local particle-number current, energy-momentum tensor, and spin tensor in terms of the thermodynamic-potential current. 
The thermodynamic pressure is proportional to the component of this current pointing in the direction of the fluid velocity.
The thermodynamic-potential current can be computed from an integral over the temperature variable, where the integrand involves the particle-number current, the energy-momentum tensor, and the spin tensor.
%\textcolor{blue}{The relation between the thermodynamic-potential current and the thermodynamic pressure then gives rise to an integral identity for the latter quantity.}
This integral identity is not unique: as long as the Maxwell relations are fulfilled, an alternative integral identity to compute the pressure involving only the energy density gives an identical result and allows one to recover well-known thermodynamic identities.

\subsection{Conservation laws}
We assume that (net) particle number, energy, momentum, and total angular momentum are conserved.
The operator-valued (net) particle four-current $\hat{J}^\mu$, energy-momentum tensor $\hat{T}^{\mu \nu}$, and total angular-momentum tensor $\hat{J}^{\mu, \nu \lambda}$ thus fulfill the conservation laws
\begin{subequations} \label{eq:conservation_laws}
\begin{align}
   \partial_\mu \hat{J}^\mu & = 0 \;, \\
   \partial_\mu \hat{T}^{\mu \nu} & = 0\;, \\
   \partial_\mu \hat{J}^{\mu, \nu \lambda} & = 0\;.
   \label{eq:conservation_laws_tot_ang_mom}
\end{align}
\end{subequations}
The total angular-momentum tensor is antisymmetric in the last two indices, $\hat{J}^{\mu, \nu \lambda} = - \hat{J}^{\mu, \lambda \nu}$.
Using the decomposition of this tensor into orbital and spin parts, 
\begin{equation} \label{eq:tot_ang_mom}
\hat{J}^{\mu, \nu \lambda} \equiv x^\nu \hat{T}^{\mu \lambda} - x^\lambda \hat{T}^{\mu \nu} + \hbar\, \hat{S}^{\mu, \nu \lambda}\;,
\end{equation}  Eq.~\eqref{eq:conservation_laws_tot_ang_mom} reads
\begin{equation}
    \hbar\, \partial_\mu \hat{S}^{\mu, \nu \lambda} = \hat{T}^{\lambda \nu} - \hat{T}^{\nu \lambda}\;.
\end{equation}
The operator-valued quantities $\hat{J}^\mu$, $\hat{T}^{\mu\nu}$, and $\hat{S}^{\mu,\nu\lambda}$ are not unique and can be redefined via so-called pseudo-gauge transformations~\cite{Hehl:1976vr,Becattini:2011ev,Florkowski:2018fap,Weickgenannt:2022jes,Drogosz:2024rbd,Buzzegoli:2024mra},
\begin{subequations} \label{eq:pseudogauge_ops}
\begin{align}
\hat{J}^{\prime\mu} &= \hat{J}^{\mu} + \hbar\, \partial_{\lambda} \hat{B}^{\lambda\mu}\;,
 \label{eq:op_current_psgt} \\
\hat{T}'^{\mu\nu}& =\hat{T}^{\mu\nu}+\frac{\hbar}{2}\partial_{\lambda}\left({\hat{\Phi}}^{\lambda,\mu\nu}-{\hat{\Phi}}^{\mu,\lambda\nu}-{\hat{\Phi}}^{\nu,\lambda\mu}\right)\;,\label{eq:op_emt_psgt}\\
\hat{S}'^{\mu, \nu\lambda}& =\hat{S}^{\mu, \nu\lambda}-\hat{\Phi}^{\mu, \nu\lambda} + \hbar \,\partial_\rho \hat{Z}^{\nu \lambda, \mu \rho} \;,
\label{eq:op_spin_psgt}
\end{align}
\end{subequations}
where 
$\hat{B}^{\lambda \mu}$ is an arbitrary differentiable operator-valued antisymmetric tensor, $\hat{B}^{\lambda \mu} = -\hat{B}^{\mu\lambda}$, $\hat{\Phi}^{\lambda,\mu\nu}$ denotes an arbitrary differentiable operator-valued tensor which is antisymmetric in the last two indices, \(\hat{\Phi}^{\lambda,\mu\nu}= -\hat{\Phi}^{\lambda,\nu\mu}\), and $\hat{Z}^{\mu \nu, \lambda \rho}$ is an arbitrary differentiable operator-valued tensor which is antisymmetric in the first two and in the last two indices, $\hat{Z}^{\mu \nu, \lambda \rho} = -\hat{Z}^{\nu \mu, \lambda \rho} = -\hat{Z}^{\mu \nu, \rho \lambda}$. 
The conservation laws \eqref{eq:conservation_laws} do not change under the transformation~\eqref{eq:pseudogauge_ops}.
Furthermore, the conserved charges are pseudo-gauge invariant~\cite{Chiarini:2024cuv}.

\subsection{Grand partition function}
In a system with nonvanishing acceleration and rotation, the grand-canonical partition function is given by~\cite{Becattini:2012tc}
\begin{align}
\mcZ & = \Tr \, \exp \left[ -\int_\Sigma \dd\Sigma_\mu  \left( \hat{T}^{\mu \nu} \mathfrak{b}_\nu - \alpha \,\hat{J}^\mu - \frac{1}{2} \,\Omega_{\nu \lambda} \,\hat{J}^{\mu, \nu \lambda} \right) \right]  \;, \label{eq:grand_partition_function}
\end{align}
where the \textit{reduced chemical potential\/} $\alpha$ is a Lagrange multiplier that controls the amount of (net) particle number in the system, the four-vector $\mathfrak{b}_\nu$ is a Lagrange multiplier that controls the energy-momentum content of the system, and the \textit{reduced spin potential\/} $\Omega_{\nu \lambda}$ is a Lagrange multiplier which controls the amount of total angular momentum in the system.
Since in this work we exclusively consider a system in global thermodynamic equilibrium, $\alpha$, $\mathfrak{b}_\nu$, and $\Omega_{\nu \lambda}$ are constant (and not functions of space-time, as in local thermodynamic equilibrium).
The integration in the exponent of Eq.~\eqref{eq:grand_partition_function} is over an arbitrary Cauchy hypersurface $\Sigma$ with future-directed timelike normal vector $\dd\Sigma_\mu$.
Note that, since $\hat{J}^{\mu, \nu \lambda}$ is antisymmetric in the last two indices, $\Omega_{\nu \lambda}$ can also be taken to be antisymmetric,  $\Omega_{\nu \lambda} = - \Omega_{\lambda \nu}$.
Consequently, the factor $1/2$ in Eq.~\eqref{eq:grand_partition_function} accounts for the fact that the sum over Lorentz indices $\nu, \lambda$ runs over independent as well as dependent degrees of freedom.
For the following, we also quote the grand-canonical density matrix
\begin{align}
\hat{\rho} & = \frac{1}{\mcZ} \, \exp \left[ -\int_\Sigma \dd\Sigma_\mu  \left( \hat{T}^{\mu \nu} \mathfrak{b}_\nu - \alpha \,\hat{J}^\mu - \frac{1}{2} \,\Omega_{\nu \lambda} \,\hat{J}^{\mu, \nu \lambda} \right) \right] \;.
\label{eq:statistical_operator}
\end{align}
Inserting Eq.~\eqref{eq:tot_ang_mom}, one often writes the grand partition function and the density matrix in the alternative form~\cite{Becattini:2018duy}
\begin{subequations}
    \begin{align}
        \mcZ & = \Tr \, \exp \left[ -\int_\Sigma \dd\Sigma_\mu  \left( \hat{T}^{\mu \nu} \beta_\nu - \alpha \,\hat{J}^\mu - \frac{\hbar}{2} \,\Omega_{\nu \lambda} \,\hat{S}^{\mu, \nu \lambda} \right) \right]  \;, \label{eq:grand_partition_function_2} \\
       \hat{\rho} & = \frac{1}{\mcZ} \, \exp \left[ -\int_\Sigma \dd\Sigma_\mu  \left( \hat{T}^{\mu \nu} \beta_\nu - \alpha \,\hat{J}^\mu - \frac{\hbar}{2} \,\Omega_{\nu \lambda} \,\hat{S}^{\mu, \nu \lambda} \right) \right] \;,
\label{eq:statistical_operator_2}
    \end{align}
\end{subequations}
with the \textit{thermal four-velocity}
\begin{equation} \label{eq:thermal_vel}
    \beta_\nu := \mathfrak{b}_\nu + \Omega_{\nu \lambda} x^\lambda\;.
\end{equation}
We note that the global-equilibrium density operator and grand partition function are invariant under the pseudo-gauge transformations \eqref{eq:pseudogauge_ops}, cf.~App.~\ref{sec:pseudogauge_invariance}.

For the following, it turns out to be useful to introduce a spacetime dependent \textit{thermodynamic-potential four-current\/} $\phi^\mu(x)$, whose integral over a Cauchy hypersurface is related to the logarithm of the grand partition function as~\cite{Becattini:2019poj,Becattini:2023ouz},
\begin{equation} \label{eq:th_dyn_potential_vector_0}
    \ln \mcZ \equiv \int_\Sigma \dd \Sigma_\mu\, \left[ \phi^\mu - \left\langle 0 \left| \hat{T}^{\mu \nu} \beta_\nu - \alpha \,\hat{J}^\mu - \frac{\hbar}{2} \,\Omega_{\nu \lambda} \,\hat{S}^{\mu, \nu \lambda} \right| 0 \right\rangle \right]\;,
\end{equation}
where the vacuum $|0\rangle$ is the lowest lying state of the operator $\hat{T}^{\mu \nu} \beta_\nu - \alpha \,\hat{J}^\mu - \frac{\hbar}{2} \,\Omega_{\nu \lambda} \,\hat{S}^{\mu, \nu \lambda}$.
This state is not necessarily the standard Minkowski vacuum.
In this paper, we focus on the thermodynamics of an ideal gas of Boltzmann particles, for which the vacuum expectation value (v.e.v.) on the right-hand side vanishes.
We will therefore omit this v.e.v.~in the following, such that
\begin{equation} \label{eq:th_dyn_potential_vector}
    \ln \mcZ \equiv \int_\Sigma \dd \Sigma_\mu\,  \phi^\mu \;.
\end{equation}
To guarantee positivity of $\ln \mcZ$ for arbitrary Cauchy hypersurfaces $\Sigma$, $\phi^\mu$ is required to be future-directed timelike.
Furthermore, in global thermodynamic equilibrium, the hypersurface integral of the thermodynamic-potential four-current must be independent of the choice of the hypersurface, which in turn requires
\begin{equation}\label{eq:phi_hypsur_ind}
    \partial_\mu \phi^\mu = 0\;.
\end{equation}

\subsection{Global thermodynamic equilibrium}
In global thermodynamic equilibrium with nonvanishing acceleration and rotation, the reduced chemical potential, the four-vector $\mathfrak{b}_\nu$, and the reduced spin potential $\Omega_{\nu \lambda}$ are constant everywhere in spacetime, 
\begin{subequations} \label{eq:gteq_conditions}
\begin{equation}
    \alpha = \mathrm{const}.\;, \quad \mathfrak{b}_\nu = \mathrm{const}.\;, \quad \Omega_{\nu \lambda} =\mathrm{const}.\;. 
\end{equation}
With Eq.~\eqref{eq:thermal_vel} and the antisymmetry of the reduced spin potential, we then immediately conclude that the thermal four-velocity is a Killing vector, i.e., the \textit{thermal shear tensor} vanishes,
\begin{align}\label{eq:killing_condition} 
 \xi_{\nu \lambda} := \frac1{2}\left(\partial_\nu \beta_\lambda +\partial_\lambda \beta_\nu\right) & = \frac{1}{2}\left(
 \Omega_{\lambda \nu} + \Omega_{\nu \lambda} \right)= 0\;.
\end{align}
Moreover, the thermal vorticity
\begin{equation}\label{eq:th_vort_def}
    \varpi_{\nu \lambda} := - \frac{1}{2} \left(\partial_\nu \beta_\lambda - \partial_\lambda \beta_\nu \right) = \Omega_{\nu \lambda} = \mathrm{const}.\;.
\end{equation}
\end{subequations}
We define the (local) \textit{fluid four-velocity} $u_\nu$ via
\begin{equation} \label{eq:fluid_vel}
    \beta_\nu  \equiv \beta\, u_\nu\;,
\end{equation}
with $\beta := \sqrt{\beta \cdot \beta} \equiv 1/T$ being the (local) inverse temperature.
Thus, $u_\nu$ is a timelike unit vector, $u \cdot u = 1$.
From the definition of $\beta$ and the condition that $\beta_\nu$ is a Killing vector, one derives that the comoving derivative of the inverse temperature vanishes in global thermodynamic equilibrium, 
\begin{subequations}
\begin{equation}\label{eq:geq_bdot}
\dot{\beta} := u^\mu \partial_\mu \beta = 0\;,
\end{equation}
i.e., the local temperature does not change with time in the rest frame of a fluid element.
In global thermodynamic equilibrium, the same holds trivially for the reduced chemical potential and the reduced spin potential,
\begin{equation}\label{eq:geq_alpha_omega_dot}
    \dot{\alpha} = \dot{\Omega}_{\nu \lambda} =0\;.
\end{equation}
\end{subequations}

In the following calculations, it is advantageous to decompose the reduced spin potential as follows:
\begin{subequations}
\begin{align}
&\Omega^{\mu\nu} \equiv \beta \omega^{\mu \nu} = \beta \left(u^{\mu}\varkappa^{\nu}-u^{\nu}\varkappa^{\mu}+\epsilon^{\mu\nu\alpha\beta}u_{\alpha}\omega_\beta \right)\;,\label{eq:decomp_red_spin_potential}\\
&\varkappa^{\mu} = -T \Omega^{\mu\nu} u_{\nu} = - \omega^{\mu \nu} u_\nu\;,\quad\omega^\mu = \frac{T}{2} \, \epsilon^{\mu\nu\alpha\beta} u_{\nu} \Omega_{\alpha\beta} =  \frac{1}{2} \, \epsilon^{\mu\nu\alpha\beta} u_{\nu} \omega_{\alpha\beta}\;,
\label{eq:kappa_omega_def}
\end{align}
\end{subequations}
where $\omega^{\mu \nu}$ is the \textit{spin potential}, and $\varkappa^\mu$ and $\omega^\mu$ are spacelike vectors orthogonal to $u^\mu$, $\varkappa \cdot u = \omega \cdot u =0$.
Since they are spacelike, we define their scalar products as $\varkappa\cdot \varkappa := - \varkappa^2$, and $\omega \cdot \omega := - \omega^2$, with positive definite $\varkappa^2$ and $\omega^2$. 
As we will see, in the following calculations also a vector orthogonal to both $\varkappa^\mu$ and $\omega^\mu$ appears,
\begin{align} \label{eq:def_cap_l_vec}
L^\mu := \epsilon^{\mu \nu \alpha \beta} u_\nu \varkappa_\alpha \omega_\beta\;.
\end{align}
Note that a nontrivial $L^\mu$ is a spacelike vector, $L^2 :=- L \cdot L =  \varkappa^2 \omega^2 - (\varkappa \cdot \omega)^2 > 0$ on account of the Schwarz inequality. 
In the following, we will also use $\ell^\mu := L^\mu/\sqrt{-L \cdot L}$, which is normalized, $\ell^\mu \ell_\mu =-1$.

Similar to the reduced spin potential, we decompose the thermal vorticity $\varpi^{\mu \nu}$ as
\begin{subequations}
\begin{align}
&\varpi^{\mu\nu}=\beta \left(u^{\mu}a^{\nu}-u^{\nu}a^{\mu}+\epsilon^{\mu\nu\alpha\beta}u_{\alpha}b_{\beta}\right)\;,\label{eq:decomp_thermal_vort}\\
& a^{\mu} = - T\varpi^{\mu\nu} u_{\nu}\;,\quad b^\mu = \frac{T}{2} \,\epsilon^{\mu\nu\alpha\beta} u_{\nu} \varpi_{\alpha\beta}\;,
\label{eq:a_b}
\end{align}
\end{subequations}
where the vectors $a^\mu$ and $b^\mu$ are spacelike vectors orthogonal to $u^\mu$, $a \cdot u= b \cdot u =0$. 
As spacelike vectors, their scalar products are $a\cdot a := - a^2$, and $b\cdot b := - b^2$, with $a^2>0$ and $b^2>0$. 
Substituting the expression for the thermal vorticity in Eq.~\eqref{eq:a_b}, we obtain
\begin{align}
a^{\mu}=-u_{\nu}\partial^{\nu}u^{\mu} \equiv - \dot{u}^\mu~,~~b^{\mu}=-\frac{1}{2}\, \epsilon^{\mu\nu\alpha\beta}u_{\nu}\partial_{\alpha}u_{\beta}\;.
\end{align}
Note that $a^{\mu}$ and $b^{\mu}$ are, up to a difference in sign, the fluid acceleration and vorticity, respectively.
By analogy, we can then identify $\varkappa^{\mu}$ as an acceleration-like vector and $\omega^{\mu}$ as a vorticity-like vector.
In global thermodynamic equilibrium, $\varpi^{\mu \nu} \equiv \Omega^{\mu \nu} = \beta \omega^{\mu \nu}$, and thus $\varkappa^\mu \equiv a^\mu$ and $\omega^\mu \equiv b^\mu$.
Similar to the vector $L^\mu$, Eq.~\eqref{eq:def_cap_l_vec}, we define
\begin{subequations}
\begin{align} \label{eq:def_cap_k_vector}
K^\mu := \epsilon^{\mu \nu \alpha \beta} u_\nu a_\alpha b_\beta\;, \\
\label{eq:def_Mmu}
M^\mu := \epsilon^{\mu \nu \alpha \beta} u_\nu a_\alpha \omega_\beta\;, \\
\label{eq:def_cap_n_vector}
N^\mu := \epsilon^{\mu \nu \alpha \beta} u_\nu \varkappa_\alpha b_\beta\;.
\end{align}
\end{subequations}
All of these vectors are spacelike on account of the Schwarz inequality.
In global equilibrium, $K^\mu \equiv M^\mu \equiv N^\mu \equiv L^\mu$.

Imposing the conditions of global thermodynamic equilibrium,  one can decompose the three-space orthogonal to $u^\mu$ into an orthogonal basis. 
We choose the $\omega^\mu$ direction, the $L^\mu$ direction (which is already orthogonal to $\omega^\mu$) and the direction of the component of $\varkappa^\mu$ orthogonal to both $\omega^\mu$ and $L^\mu$, which we denote as
\begin{equation}\label{eq:def_xi}
    \xi^\mu := \varkappa^\mu + \frac{\varkappa \cdot \omega}{\omega^2}\,\omega^\mu\;.
\end{equation}
One readily verifies that $\xi \cdot u = \xi \cdot \omega = \xi \cdot L = 0$, while $\xi^2 := - \xi \cdot \xi = -\xi \cdot \varkappa =  \varkappa^2 - (\varkappa \cdot \omega)^2/\omega^2$.

In global thermodynamic equilibrium, where all Lagrange multipliers are constants, the partition function and the density matrix assume the form
\begin{subequations}
    \begin{align}
        \mcZ & = \Tr \, \exp \left( - \mathfrak{b}_\nu \hat{\mathbb{P}}^{\nu}  + \alpha \,\hat{\mathbb{N}} + \frac{1}{2} \,\Omega_{\nu \lambda} \,\hat{\mathbb{J}}^{\nu \lambda} \right)  \;, \label{eq:grand_partition_function_gteq} \\
       \hat{\rho} & = \frac{1}{\mcZ} \, \exp \left( - \mathfrak{b}_\nu \hat{\mathbb{P}}^{\nu}  + \alpha \,\hat{\mathbb{N}} + \frac{1}{2} \,\Omega_{\nu \lambda} \,\hat{\mathbb{J}}^{\nu \lambda} \right) \;,
\label{eq:statistical_operator_gteq}
    \end{align}
\end{subequations}
where the operators of the (net) particle number, the four-momentum, and the total angular momentum are defined as
\begin{subequations} \label{eq:total_conserved_operators}
    \begin{align}
        \hat{\mathbb{N}} := \int_\Sigma \dd \Sigma_\mu\, \hat{J}^\mu\;, \\
        \hat{\mathbb{P}}^\nu := \int_\Sigma \dd \Sigma_\mu\, \hat{T}^{\mu \nu}\;, \\
        \hat{\mathbb{J}}^{\nu \lambda} := \int_\Sigma \dd \Sigma_\mu\, \hat{J}^{\mu, \nu \lambda}\;.
    \end{align}
\end{subequations}

\subsection{Thermodynamic identities}
The expectation values of the (net) particle number, total four-momentum, and total angular momentum immediately follow from their definitions as
\begin{subequations} \label{eq:evals_tot_con_chg_op}
    \begin{align}
        \mathbb{N} & := \Tr \left( \hat{\rho}\,  \hat{\mathbb{N}} \right) \equiv \left. \pdv{\ln \mcZ}{\alpha} \right|_{\mathfrak{b}_\rho, \Omega_{\sigma \tau}} = \int_\Sigma \dd\Sigma_\mu \left. \pdv{\phi^\mu}{\alpha} \right|_{\mathfrak{b}_\rho, \Omega_{\sigma\tau}}\;, 
        \label{eq:eval_q}\\
        \mathbb{P}^\nu & := \Tr \left( \hat{\rho} \, \hat{\mathbb{P}}^\nu \right) \equiv - \left. \pdv{\ln \mcZ}{\mathfrak{b}_\nu} \right|_{\alpha, \Omega_{\sigma \tau}} = - \int_\Sigma \dd\Sigma_\mu \left. \pdv{\phi^\mu}{\mathfrak{b}_\nu} \right|_{\alpha, \Omega_{\sigma\tau}}\;, \label{eq:exp_val_Pnu} \\
        \mathbb{J}^{\nu \lambda} & := \Tr \left( \hat{\rho} \, \hat{\mathbb{J}}^{\nu \lambda} \right) \equiv \left. \pdv{\ln \mcZ}{\Omega_{\nu \lambda}} \right|_{\alpha, \mathfrak{b}_\rho} = \int_\Sigma \dd\Sigma_\mu \left. \pdv{\phi^\mu}{\Omega_{\nu \lambda} }\right|_{\alpha,\mathfrak{b}_\rho}\;, \label{eq:eval_rank3_j}
    \end{align}
\end{subequations}
where we have used Eq.~\eqref{eq:th_dyn_potential_vector} in the last equality on the right-hand sides. 
We note that, due to the relativistic Stokes theorem, these expectation values, which correspond to the conserved charges in the system, are invariant under a pseudo-gauge transformation~\eqref{eq:pseudogauge_ops}, provided $B^{\lambda \mu}$ , $\Phi^{\lambda,\mu \nu}$, and $Z^{\nu \lambda, \mu \rho}$ satisfy appropriate boundary conditions.

From these global quantities, one can also define the expectation values of the local (net) particle-number current, the energy-momentum tensor, and the spin-current tensor at a spacetime point $x$ on the hypersurface,
\begin{subequations} \label{eq:srf_ders_w_rank3_j}
    \begin{align}
     J^\mu (x) & := \dv{\mathbb{N}}{\Sigma_\mu} =  \Tr\left[ \hat{\rho}\, \hat{J}^\mu (x) \right] \equiv 
     \left. \pdv{\phi^\mu(x)}{\alpha} \right|_{\mathfrak{b}_\rho, \Omega_{\sigma \tau}}
     \;, \\
     T^{\mu \nu} (x) & := \dv{\mathbb{P}^\nu}{\Sigma_\mu} =  \Tr\left[ \hat{\rho}\, \hat{T}^{\mu \nu} (x) \right] \equiv -  \left. \pdv{\phi^\mu(x)}{\mathfrak{b}_\nu} \right|_{\alpha, \Omega_{\sigma \tau}}
     \;, \\
     J^{\mu, \nu \lambda} (x) & :=   \dv{ \mathbb{J}^{\nu \lambda}}{\Sigma_\mu} =  \Tr\left[\hat{\rho}\, \hat{J}^{\mu, \nu \lambda} (x) \right] \equiv  \left. \pdv{\phi^\mu(x)}{\Omega_{\nu \lambda}}\right|_{\alpha, \mathfrak{b}_\rho}\;,
    \end{align}
\end{subequations}
where the second equalities follow from combining the definitions~\eqref{eq:evals_tot_con_chg_op} with Eqs.~\eqref{eq:total_conserved_operators}, while the right-hand sides follow from the last equalities in Eq.~\eqref{eq:evals_tot_con_chg_op}.\footnote{Note that Eqs.~\eqref{eq:srf_ders_w_rank3_j} are equalities only up to a contribution from the surface of the $\dd \Sigma_\mu$ integral, see Ref.~\cite{shokri_in_prep} for a detailed discussion}.
Due to Eq.~\eqref{eq:thermal_vel}, we have at any given spacetime point $x^\mu$
\begin{equation}
\left. \pdv{\alpha}\right|_{\mathfrak{b}_\rho,\Omega_{\sigma\tau}}
= \left. \pdv{\alpha}\right|_{\beta_\rho,\Omega_{\sigma\tau}} \;, \qquad \left. \pdv{\mathfrak b_\nu}\right|_{\alpha,\Omega_{\sigma\tau}} = \left. \pdv{\beta_\nu}\right|_{\alpha,\Omega_{\sigma\tau}}
\;.
\end{equation}
Furthermore, using Jacobi determinants one proves that
\begin{equation}
 \left. \pdv{\Omega_{\nu\lambda}}\right|_{\alpha, \mathfrak{b}_\rho}  = \left. \pdv{\Omega_{\nu\lambda}}\right|_{\alpha, \beta_\rho}  -  x^\nu\left. \pdv{\mathfrak{b}_\lambda}\right|_{\alpha, \Omega_{\sigma\rho}}+ x^\lambda\left. \pdv{\mathfrak{b}_\nu}\right|_{\alpha, \Omega_{\sigma\rho}}\;.
\end{equation}
Using Eq.~\eqref{eq:tot_ang_mom}, we then arrive at
\begin{subequations} \label{eq:currents_as_ders}
    \begin{align}
     J^\mu (x) & =
     \left. \pdv{\phi^\mu(x)}{\alpha} \right|_{\beta_\rho, \Omega_{\sigma \tau}} 
     \;, \\
     T^{\mu \nu} (x) & =  -  \left. \pdv{\phi^\mu(x)}{\beta_\nu(x)} \right|_{\alpha, \Omega_{\sigma \tau}} 
     \;, \label{eq:Tmunu_as_der}\\
     \hbar \, S^{\mu, \nu \lambda} (x) & =  \left. \pdv{\phi^\mu(x)}{\Omega_{\nu \lambda}}\right|_{\alpha, \beta_\rho}  \;.\label{eq:spin-current_tensor_general}
    \end{align}
\end{subequations}
From Eqs.~\eqref{eq:srf_ders_w_rank3_j} or \eqref{eq:currents_as_ders}, respectively, we derive the covariant form of the Gibbs-Duhem relation
\begin{equation}
    \label{eq:vec_gibbs_duhem}
    \dd{\phi}^\mu
    =J^\mu \dd{\alpha}
    -T^{\mu\nu}\dd{\mathfrak{b}_\nu}
    +\frac{1}{2}J^{\mu,\nu\lambda}\dd{\Omega}_{\nu\lambda}    
    =J^\mu \dd{\alpha}
    -T^{\mu\nu}\dd{\beta_\nu}
    +\frac{\hbar}{2}S^{\mu,\nu\lambda}\dd{\Omega}_{\nu\lambda}
    \;.
\end{equation}

We define the entropy current $S^\mu$ through the relation
\begin{equation}
\label{eq:def_entropy_current}
    \int\dd{\Sigma}_\mu S^\mu = S = -\Tr(\hat{\rho}\ln\hat{\rho})\;,
\end{equation}
where $S$ is the von Neumann entropy.
Using Eqs.~\eqref{eq:statistical_operator} or \eqref{eq:statistical_operator_2}, respectively, as well as Eq.~\eqref{eq:th_dyn_potential_vector} and the definitions~\eqref{eq:srf_ders_w_rank3_j} of the expectation values of particle-number current, energy-momentum tensor, and total angular-momentum tensor we obtain the covariant Euler relation~\cite{Becattini:2023ouz}
\begin{equation}
\label{eq:vec_euler}
  \phi^\mu - S^\mu -\alpha J^\mu + b_\nu T^{\mu \nu} - \frac{1}{2}\, \Omega_{\nu \lambda} J^{\mu,\nu \lambda} =  \phi^\mu - S^\mu -\alpha J^\mu + \beta_\nu T^{\mu \nu} - \frac{\hbar}{2}\, \Omega_{\nu \lambda} S^{\mu,\nu \lambda} =0\;,
\end{equation}
where the expectation value of the spin-current tensor $S^{\mu, \nu \lambda} \equiv \Tr(\hat{\rho}\, \hat{S}^{\mu, \nu \lambda} )$ is identical to Eq.~\eqref{eq:spin-current_tensor_general}.
In global thermodynamic equilibrium the entropy current is conserved, $\partial_\mu S^\mu =0$, on account of the conservation laws~\eqref{eq:conservation_laws}, Eq.~\eqref{eq:phi_hypsur_ind}, as well as the conditions \eqref{eq:gteq_conditions}.

Combining Eqs.~\eqref{eq:vec_gibbs_duhem} and \eqref{eq:vec_euler}, we obtain the covariant form of the first law of thermodynamics,
\begin{equation}\label{eq:vec_first_law}
    \dd S^\mu
    =-\alpha\, \dd{J}^\mu +
    \mathfrak{b}_\nu\, \dd{T}^{\mu\nu}
    -\frac{1}{2}\Omega_{\nu \lambda}\,\dd{J}^{\mu,\nu \lambda}
    = -\alpha\, \dd{J}^\mu +
    \beta_\nu\, \dd{T}^{\mu\nu}
    -\frac{\hbar}{2}\Omega_{\nu \lambda}\,\dd{S}^{\mu,\nu \lambda}    \;.
\end{equation}
Since the current $\phi^\mu$ is timelike, it has a nonvanishing component along the four-velocity.
In an isotropic global-equilibrium state, no contribution orthogonal to $u^\mu$ can exist, because it would single out a particular spatial direction.
On the other hand, a global-equilibrium state with nonzero thermal vorticity is anisotropic, since acceleration and vorticity break rotational symmetry by selecting particular spatial directions.
In this case, the current $\phi^\mu$ can be decomposed as
\begin{equation}\label{eq:phi_gen_tetrad_decomp}
    \phi^\mu = \phi_u u^\mu + \phi_\xi \xi^\mu + \phi_\omega \omega^\mu + \phi_\times \ell^\mu\;.
\end{equation}
In App.~\ref{app:th_dyn_curr} we show that
$\phi_\xi = \phi_\omega =0$, while 
\begin{equation} \label{eq:phi_comps}
    \phi_u \equiv P\, \beta\;, \quad
    \phi_\times = -\ell_\mu \phi^\mu  \equiv P_\times \beta\;. 
\end{equation}
Here,
\begin{equation} \label{eq:def_thdyn_press}
    P:= T u_\mu \phi^\mu
\end{equation}  
is identified with the \textit{thermodynamic pressure}~\cite{Becattini:2023ouz}.
This identification is motivated by the fact that, in isotropic global equilibrium, $\beta_\mu \equiv \beta u^\mu$ is independent of spacetime, such that
\begin{equation}
    \ln \mcZ = \int \dd\Sigma_\mu \phi^\mu = \beta P
    \int \dd\Sigma_\mu u^\mu
    \equiv \beta P V\;.
\end{equation}
Here, the spatial volume of the system is identified with $V \equiv \int \dd\Sigma_\mu u^\mu \equiv \int \dd[3]{\vb{x}}$, where we have chosen the rest frame of the system, $u^\mu = (1, \vb{0})$, to evaluate the integral over the hypersurface.
In this way, the standard relation between grand partition function and thermodynamic pressure is recovered.

Projecting Eq.~\eqref{eq:vec_euler} onto $u_\mu$, we obtain the standard (\textit{longitudinal}) Euler relation of thermodynamics, augmented by a spin contribution,
\begin{equation}
    \label{eq:ortho_euler} 
     \varepsilon + P - T s - \mu \, n -\frac{\hbar}{2}\, \omega_{\nu \lambda}  S^{\nu \lambda} =0\;,
\end{equation}
where $\mu := T \alpha$ is the \textit{chemical potential} and
\begin{equation}
\label{eq:ortho_thdyn_densities}
       \varepsilon  := u_\mu T^{\mu \nu}  u_\nu  \;, \qquad 
        s:= u_\mu S^\mu \;, \qquad 
        n  := u_\mu J^\mu  \;, \qquad S^{\nu \lambda}  := u_\mu \, S^{\mu, \nu \lambda}  \;, 
\end{equation}
are the \textit{energy density}, \textit{entropy density}, \textit{(net) particle-number density}, and the \textit{spin tensor}, respectively.

On the other hand, contracting Eq.~\eqref{eq:vec_euler} with $\ell_\mu$ results in the \textit{transverse} Euler relation previously found in Ref.\ \cite{Florkowski:2024bfw},
\begin{equation}\label{eq:trans_euler}
    q_\times +P_\times -  T s_\times - \mu n_\times - \frac{\hbar}{2}\, \omega_{\nu \lambda}S^{\nu\lambda}_\times  =0\;,
\end{equation}
where
\begin{equation}
\label{eq:trans_thdyn_densities}
     q_\times := -\ell_\mu u_\nu
     T^{\mu\nu}\;, \qquad s_\times := -\ell_\mu S^\mu\;,\qquad n_\times := -\ell_\mu J^\mu\;,\qquad S^{\nu \lambda}_\times := -\ell_\mu S^{\mu,\nu \lambda}\;.
\end{equation}

Projecting the covariant Gibbs-Duhem relation \eqref{eq:vec_gibbs_duhem} onto $u_\mu$ we arrive at
\begin{equation}
\label{eq:ortho_gibbs_duhem}
    \dd P = s \,\dd T + n \,\dd \mu + \frac{\hbar}{2} \,S^{\nu \lambda} \,\dd \omega_{\nu \lambda} +
    \left( P_\times \ell^\mu - u_\nu T^{\nu \lambda} \Delta_\lambda^{\hspace*{0.1cm} \mu} \right) \dd u_\mu\;,
\end{equation}
where we used $u_\mu \dd \ell^\mu \equiv - \ell^\mu \dd u_\mu$, as well as Eqs.~\eqref{eq:decomp_red_spin_potential}, \eqref{eq:phi_comps}, \eqref{eq:ortho_euler}, and \eqref{eq:ortho_thdyn_densities}.
In isotropic global thermodynamic equilibrium, where $u^\mu = const.$ and $\omega_{\nu \lambda} \equiv 0$, the last two terms in Eq.~\eqref{eq:ortho_gibbs_duhem} vanish and one obtains the standard Gibbs-Duhem relation.
In global thermodynamic equilibrium with rotation and acceleration, however, \textit{both} terms are present.
This is actually not surprising considering the fact that the covariant Gibbs-Duhem relation~\eqref{eq:vec_gibbs_duhem} involves 11 independent variables (i.v.s): $\alpha$ (1 i.v.), $\beta_\nu$ (4 i.v.s), and $\Omega_{\nu \lambda}$ (6 i.v.s).
No independent variable can be lost simply by projecting onto $u_\mu$.
Therefore, the longitudinal Gibbs-Duhem relation~\eqref{eq:ortho_gibbs_duhem} also features 11 i.v.s: $T$ (1 i.v.), $\mu$ (1 i.v.), $\omega_{\nu \lambda}$ (6 i.v.s), and $u_\mu$ (3 i.v.s). 
Note that the coefficient of $\dd u_\mu$ is the momentum density introduced in Ref.~\cite{Gallegos:2022jow}.

Equation~\eqref{eq:ortho_gibbs_duhem} has the consequence that the standard way of computing entropy density, particle-number density, and spin density from the thermodynamic pressure by partial differentiation has to be modified.
Obviously, in addition to the standard variables one also has to keep the fluid four-velocity constant,
\begin{subequations} \label{eq:ortho_thdyn_densities_as_ders}
    \begin{align}
     s  & \equiv  \left. \pdv{P}{T} \at[\right]{\mu,\omega_{\sigma \tau},u_\rho}
     \;, \label{eq:s_as_der_p} \\
     n  & \equiv  \left. \pdv{ P}{\mu} \at[\right]{T,\omega_{\sigma \tau},u_\rho}
     \;, \label{eq:n_as_der_p}\\
     \hbar\, S^{\nu \lambda}  & \equiv \left. \pdv{P}{\omega_{\nu \lambda}} \at[\right]{T,\mu,u_\rho}  \;. 
     \label{eq:ortho_spin_as_der_p}    
    \end{align}
\end{subequations}
Note that keeping $T$ and $u_\rho$ constant is equivalent to keeping $\beta_\rho$ constant, and thus also keeping the thermal vorticity $\varpi_{\sigma \tau}$ constant.
This fact will become important in the explicit calculations of thermodynamic quantities performed below.

We now contract the covariant Gibbs-Duhem relation \eqref{eq:vec_gibbs_duhem} with $\ell_\mu$ and obtain with Eqs.~\eqref{eq:decomp_red_spin_potential},  \eqref{eq:phi_comps}, \eqref{eq:trans_euler}, and \eqref{eq:trans_thdyn_densities} the \textit{transverse} Gibbs-Duhem relation
\begin{equation}
\label{eq:trans_gibbs_duhem}
    \dd P_\times = s_\times \dd T + n_\times \dd \mu + \frac{\hbar}{2} S_\times^{\nu \lambda} \dd \omega_{\nu \lambda} +
    \left( P \ell^\mu + \ell_\nu T^{\nu \lambda} \Delta_\lambda^{\hspace*{0.1cm} \mu}\right) \dd u_\mu\;.
\end{equation}
In isotropic global thermodynamic equilibrium, this is a trivial identity (all quantities with index ``$\times$'' are zero and $\dd u^\mu \equiv 0$).
The relations analogous to Eqs.~\eqref{eq:ortho_thdyn_densities_as_ders} read
\begin{equation}\label{eq:trans_thdyn_densities_as_ders}
     s_\times   \equiv  \left. \pdv{P_\times}{T} \at[\right]{\mu,\omega_{\sigma \tau},u_\rho}
     \;, \qquad
     n_\times   \equiv  \left. \pdv{P_\times}{\mu} \at[\right]{T,\omega_{\sigma \tau},u_\rho}
     \;, \qquad
     \hbar\, S_\times^{\nu \lambda}   \equiv \left. \pdv{P_\times}{\omega_{\nu \lambda}} \at[\right]{T,\mu,u_\rho}  \;. 
\end{equation}

Finally, also the first law of thermodynamics is modified in the presence of rotation and acceleration. 
Contracting the covariant first law of thermodynamics~\eqref{eq:vec_first_law} with $u_\mu$ and using Eqs.~\eqref{eq:vec_euler} and \eqref{eq:phi_comps}, we obtain the standard (\textit{longitudinal}) first law of thermodynamics,
\begin{equation}\label{eq:ortho_first_law}
\dd\varepsilon
=T\,\dd s + \mu\, \dd n + \frac{\hbar}{2}\,\omega_{\nu \lambda}\,\dd S^{\nu\lambda} -
\left(P_\times \ell^\mu-u_\nu T^{\nu\lambda}\Delta_\lambda^{\hspace*{0.1cm} \mu}\right)\dd u_\mu\;.
\end{equation}
The last two terms are novel for systems in global thermodynamic equilibrium with rotation and acceleration.

On the other hand, projecting Eq.~\eqref{eq:vec_first_law} onto $\ell_\mu$ and using Eqs.~\eqref{eq:vec_euler} and \eqref{eq:phi_comps}, we obtain the \textit{transverse} first law
\begin{equation}\label{eq:trans_first_law}
  \dd q_\times
=T\,\dd s_\times + \mu\, \dd n_\times + \frac{\hbar}{2}\,\omega_{\nu \lambda}\,\dd S^{\nu \lambda}_{\times} -
\left(P \ell^\mu +\ell_\nu T^{\nu\lambda}\Delta_\lambda^{\hspace*{0.1cm} \mu}\right)\dd u_\mu\;.  
\end{equation}

\subsection{Integral identities and Maxwell relations}

In this paper, we focus on the thermodynamics of an ideal gas of Boltzmann particles, for which $\lim_{T \rightarrow 0} \ln \mcZ =0$.
In this case, it was shown in Ref.~\cite{Becattini:2023ouz} that the thermodynamic-potential four-current defined in Eq.~\eqref{eq:th_dyn_potential_vector} can be computed as
\begin{align} 
\label{eq:phi_as_int_lambda}
    \phi^\mu &\equiv \int_1^\infty 
    \dd \lambda \, \left[ T^{\mu\nu} (\lambda) \beta_\nu -\alpha J^{\mu}(\lambda)- \frac{\hbar}{2} \Omega_{\nu\lambda} S^{\mu,\nu\lambda }(\lambda)\right]\;.
\end{align}
The $\lambda$-dependent energy-momentum tensor, (net) particle-number current, and spin-density current are computed as expectation values of the corresponding operators in an ensemble given by the $\lambda$-dependent grand-canonical density matrix
\begin{align}
       \hat{\rho} (\lambda) & = \frac{1}{\mcZ(\lambda)} \, \exp \left[ -\lambda \int_\Sigma \dd\Sigma_\mu  \left( \hat{T}^{\mu \nu} \beta_\nu - \alpha \,\hat{J}^\mu - \frac{\hbar}{2} \,\Omega_{\nu \lambda} \,\hat{S}^{\mu, \nu \lambda} \right) \right] \;,
\label{eq:statistical_operator_lambda}
\end{align}
with 
\begin{equation}
  \mcZ(\lambda) \equiv \Tr \, \exp \left[ -\lambda \int_\Sigma \mathrm{d}\Sigma_\mu  \left( \hat{T}^{\mu \nu} \beta_\nu - \alpha \,\hat{J}^\mu - \frac{\hbar}{2} \,\Omega_{\nu \lambda} \,\hat{S}^{\mu, \nu \lambda} \right) \right]\;,
\end{equation} 
i.e.,
\begin{subequations} \label{eq:evals_lambda}
    \begin{align}
        J^\mu(\lambda) & := \Tr \left[ \hat{\rho}(\lambda) \,\hat{J}^\mu\right]\;, \\
        T^{\mu\nu}(\lambda) & := \Tr \left[ \hat{\rho}(\lambda) \,\hat{T}^{\mu\nu}\right]\;, \\  
        S^{\mu,\nu \lambda}(\lambda) & := \Tr \left[ \hat{\rho}(\lambda) \,\hat{S}^{\mu,\nu\lambda}\right]\;.
    \end{align}
\end{subequations}
The particular dependence in which $\lambda$ appears in Eq.~\eqref{eq:statistical_operator_lambda} suggests rescaling the inverse temperature as $\beta \rightarrow \beta' := \lambda \beta$, while keeping $\mu$, $\omega_{\nu \lambda}$, and $u_\nu$ constant, i.e.,
$\alpha = \beta \mu \rightarrow \beta' \mu$, $\Omega_{\nu \lambda} = \beta \omega_{\nu \lambda} \rightarrow \beta' \omega_{\nu \lambda}$, and $\beta_\nu = \beta u_\nu \rightarrow \beta' u_\nu$.
In effect, the expectation values \eqref{eq:evals_lambda} are then computed in an ensemble with the same $\mu$, $\omega_{\nu \lambda}$, and $u_\nu$, but at a temperature $T' \equiv T/\lambda$ instead of $T$.
Consequently, one may substitute the integration variable $\lambda$ in Eq.~\eqref{eq:phi_as_int_lambda} by $\beta'$, or $T'=1/\beta'$.
\begin{align} 
\label{eq:phi_as_int_temp}
    \phi^\mu &\equiv \int_0^T \frac{\dd T'}{T^{\prime 2}}\, \left( T^{\mu\nu} u_\nu -\mu J^{\mu}- \frac{\hbar}{2} \omega_{\nu\lambda} S^{\mu,\nu\lambda }\right)_{\mu,\omega_{\sigma \tau},u_\rho}\;.
\end{align}
Here, the integrand, and in particular $T^{\mu\nu}$, $J^\mu$, and $S^{\mu,\nu \lambda}$, are considered to be functions of $T'$ at constant $\mu$, $\omega_{\sigma\tau}$, and $u_{\rho}$.

The thermodynamic pressure then follows from Eq.~\eqref{eq:def_thdyn_press} by contracting Eq.~\eqref{eq:phi_as_int_temp} with $u_\mu$ and multiplying by $T$,
\begin{equation} 
P  =   T \int_0^T \frac{\dd T'}{T^{\prime\,2}} \, \left( \varepsilon - \mu n - \frac{\hbar}{2} \omega_{\nu\lambda} S^{\nu\lambda} \right)_{\mu, \omega_{\sigma \tau},u_\rho} \label{eq:p_as_int_temp}\;,
\end{equation}
where we have used Eq.~\eqref{eq:ortho_thdyn_densities}.
If the integrand contains no explicit dependence on $T$, we immediately derive 
\begin{equation}
\left. \frac{\partial P}{\partial T} \right|_{\mu, \omega_{\sigma \tau},u_\rho} =
\frac{1}{T}\left( \varepsilon +P - \mu n - \frac{\hbar}{2} \omega_{\nu\lambda} S^{\nu\lambda} \right)\;.
\label{eq:dPdT_Euler}
\end{equation}
Therefore, the thermodynamic identity~\eqref{eq:s_as_der_p} implies the validity of the Euler relation~\eqref{eq:ortho_euler}, and vice versa.

The transverse pressure $P_\times$ is found analogously to the thermodynamic pressure from contracting Eq.~\eqref{eq:phi_as_int_temp} with $\ell_\mu$,
\begin{equation} 
P_\times  =   T \int_0^T \frac{\dd T'}{T^{\prime\,2}} \, \left( q_\times - \mu n_\times - \frac{\hbar}{2} \omega_{\nu\lambda} S_\times^{\nu\lambda} \right)_{\mu, \omega_{\sigma \tau},u_\rho} \label{eq:tans_p_as_int_temp}\;.
\end{equation}

Equations~\eqref{eq:p_as_int_temp} and \eqref{eq:tans_p_as_int_temp} can also be directly derived from Eq.~\eqref{eq:s_as_der_p} and the first equation~\eqref{eq:trans_thdyn_densities_as_ders}, respectively, using the Euler relations~\eqref{eq:ortho_euler} and \eqref{eq:trans_euler}.
Namely,
\begin{subequations}
\begin{align}
    T^2 \left. \frac{\partial (P/T)}{\partial T} \right|_{\mu, \omega_{\sigma \tau},u_\rho}& = - P + T
    \left. \frac{\partial P}{\partial T} \right|_{\mu, \omega_{\sigma \tau},u_\rho} = \varepsilon- \mu n- \frac{\hbar}{2} \omega_{\nu\lambda} S^{\nu\lambda} \;, \\
    T^2 \left. \frac{\partial (P_\times/T)}{\partial T} \right|_{\mu, \omega_{\sigma \tau},u_\rho} & = - P_\times + T
    \left. \frac{\partial P_\times}{\partial T} \right|_{\mu, \omega_{\sigma \tau},u_\rho} = q_\times - \mu n_\times- \frac{\hbar}{2} \omega_{\nu\lambda} S_\times^{\nu\lambda} \;.
\end{align}
\end{subequations}
Dividing both sides by $T^2$ and integrating over temperature immediately yields Eqs.~\eqref{eq:p_as_int_temp} and \eqref{eq:tans_p_as_int_temp}.

Following this strategy, an alternative formula for computing the thermodynamic pressure can be derived. 
Contracting both sides of Eq.~\eqref{eq:Tmunu_as_der} with $u_\mu u_\nu$, we obtain the energy density as
\begin{equation} \label{eq:energydens_as_der_of_phimu}
    \varepsilon = - u_\mu u_\nu \left. \frac{\partial \phi^\mu}{\partial \beta_\nu} \right|_{\alpha, \Omega_{\sigma \tau}}\;.
\end{equation}
Using Eq.~\eqref{eq:phi_gen_tetrad_decomp} with Eq.~\eqref{eq:phi_comps} we obtain for the partial derivative of the thermodynamic-potential current
\begin{equation} \label{eq:dphimu_dbetanu}
    \left. \frac{\partial \phi^\mu}{\partial \beta_\nu} \right|_{\alpha, \Omega_{\sigma \tau}} = \beta^\mu \left. \frac{\partial P}{\partial \beta_\nu} \right|_{\alpha, \Omega_{\sigma \tau}} +
    P  \frac{\partial \beta^\mu}{\partial \beta_\nu} +
    \tilde{L}^\mu \left. \frac{\partial (P_\times/\bar{L}) }{\partial \beta_\nu} \right|_{\alpha, \Omega_{\sigma \tau}}+
    \frac{P_\times}{\bar{L}}\, \left. \frac{\partial \tilde{L}^\mu}{\partial \beta_\nu} \right|_{\alpha, \Omega_{\sigma \tau}}\;,
\end{equation}
where we have defined the auxiliary quantities
\begin{equation}
    \tilde{L}^\mu := \beta^3 L^\mu \equiv \epsilon^{\mu \nu \alpha \beta} \beta_\nu (\beta \varkappa_\alpha) (\beta \omega_\beta)\;, \qquad \bar{L}
    := \beta^2 L \equiv \sqrt{\beta^2 \varkappa^2 \beta^2 \omega^2 - [(\beta \varkappa) \cdot (\beta \omega)]^2}\;.
\end{equation}
Since $\beta \varkappa_\mu$ and $\beta \omega_\mu$ are constant if $\Omega_{\sigma \tau}$ is constant, the last partial derivative is simply $\beta^2 \epsilon^{\mu \nu \alpha \beta} \varkappa_\alpha \omega_\beta$.
Therefore, contracting with $u_\mu u_\nu$ to obtain the energy density~\eqref{eq:energydens_as_der_of_phimu}, the last two terms vanish identically.
For the partial derivative in the first term we write

\begin{equation}
    \left. \frac{\partial P}{\partial \beta_\nu} \right|_{\alpha, \Omega_{\sigma \tau}} =
    \frac{\partial \beta}{\partial \beta_\nu} \left. \frac{\partial P}{\partial \beta} \right|_{\alpha, \Omega_{\sigma \tau},u_\rho} +
    \frac{1}{\beta} \left. \frac{\partial P}{\partial u_\nu} \right|_{\beta,\alpha, \Omega_{\sigma \tau}}\;.
\end{equation}
With $\partial \beta/\partial \beta_\nu = u^\nu$ and using Eq.~\eqref{eq:ortho_gibbs_duhem} to compute the second term we obtain

\begin{equation}
    \left. \frac{\partial P}{\partial \beta_\nu} \right|_{\alpha, \Omega_{\sigma \tau}} =
    u^\nu \left. \frac{\partial P}{\partial \beta} \right|_{\alpha, \Omega_{\sigma \tau},u_\rho} +
    \frac{1}{\beta} \left. \frac{\partial P}{\partial u_\nu} \right|_{T,\mu,\omega_{\sigma \tau}}
    =
    u^\nu \left. \frac{\partial P}{\partial \beta} \right|_{\alpha, \Omega_{\sigma \tau},u_\rho} +
    T \left( P_\times \ell^\nu - u_\alpha T^{\alpha \beta} \Delta_{\beta}^{\hspace*{0.1cm} \nu} \right)\;.
\end{equation}
After contraction with $u_\mu u_\nu$, the second term vanishes and we obtain for the energy density~\eqref{eq:energydens_as_der_of_phimu}

\begin{equation}
    \varepsilon = - \beta \left. \frac{\partial P}{\partial \beta} \right|_{\alpha, \Omega_{\sigma \tau},u_\rho} -P \equiv - \left. \frac{\partial (\beta P)}{\partial \beta} \right|_{\alpha, \Omega_{\sigma \tau},u_\rho}\;.
\end{equation}
Integrating both sides over $\beta$ we finally obtain

\begin{equation} \label{eq:thdynpress_as_int_over_enerdens}
    P = \frac{1}{\beta} \int_\beta^\infty \dd \beta'  \left. \varepsilon \, \right|_{\alpha, \Omega_{\sigma \tau},u_\rho} \;,
\end{equation}
where the integral is performed at constant $\alpha, \Omega_{\sigma \tau}$, and $u_\rho$ and we assumed that the pressure vanishes at $\beta \rightarrow \infty$.
Equation~\eqref{eq:thdynpress_as_int_over_enerdens} for the thermodynamic pressure, also suggested in Ref.~\cite{Ambrus:2025dca}, is considerably simpler than Eq.~\eqref{eq:p_as_int_temp}.

Similarly, contracting Eq.~\eqref{eq:Tmunu_as_der} with $-\ell_\mu u_\nu$ we derive an alternative formula for the transverse pressure,

\begin{equation}
    P_\times = \frac{1}{\beta} \int_\beta^\infty \dd \beta'  \left. q_\times \, \right|_{\alpha, \Omega_{\sigma \tau},u_\rho} \;.
\end{equation}

%\begin{figure}[htbp!]
%     \includegraphics[width=10cm]{int_Tmuomega.pdf}
%\caption{\label{fig:int_Tmuomega} Integration path in $(T,\mu,\omega_{\mu \nu})$ space for the $T'$ integral in Eq.~\eqref{eq:p_as_int_temp} (green) and for the $\beta'$ integral in Eq.~\eqref{eq:thdynpress_as_int_over_enerdens} (red). For the sake of illustration, only one of the six independent components of $\omega_{\mu \nu}$ is shown. We also do not show the axes corresponding to the three independent components of $u_\rho$, as this quantity is kept constant for both integrals.}
%\end{figure}

\begin{figure}[htbp!]
     \includegraphics[width=12cm]{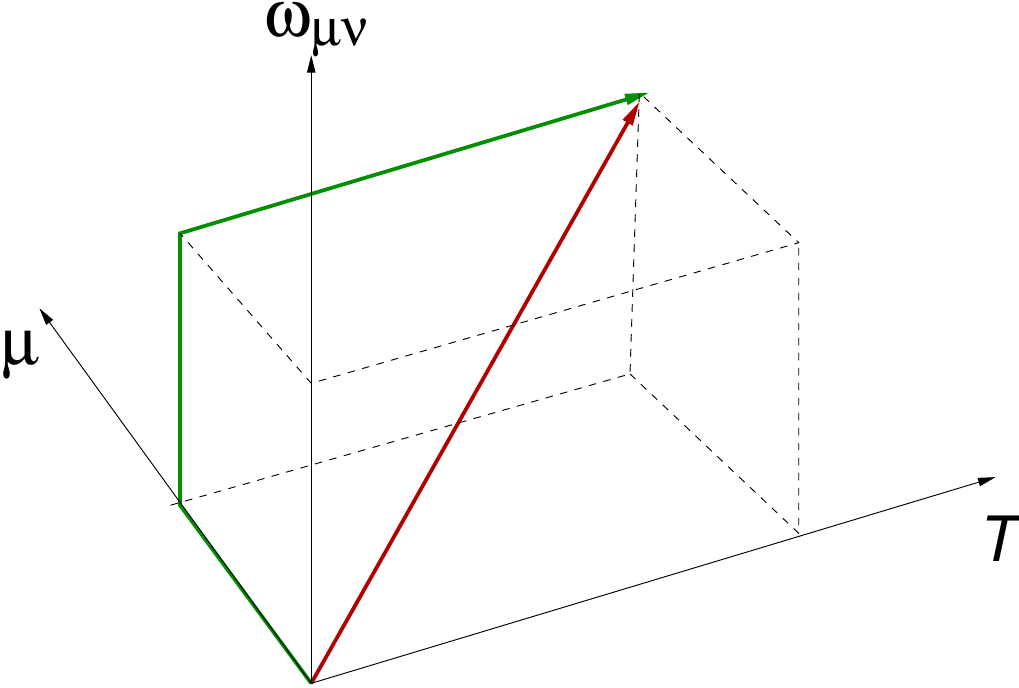}
\caption{\label{fig:int_Tmuomega} Integration path in $(T,\mu,\omega_{\mu \nu})$ space for the $T'$ integral in Eq.~\eqref{eq:p_as_int_temp} (green) and for the $\beta'$ integral in Eq.~\eqref{eq:thdynpress_as_int_over_enerdens} (red). For the sake of illustration, only one of the six independent components of $\omega_{\mu \nu}$ is shown. We also do not show the axes corresponding to the three independent components of $u_\rho$, as this quantity is kept constant for both integrals.}
\end{figure}

Computing the thermodynamic pressure using Eq.~\eqref{eq:p_as_int_temp} or \eqref{eq:thdynpress_as_int_over_enerdens} yields identical results, provided the thermodynamic Maxwell relations hold, i.e.,\footnote{We only quote the relations for $u_\rho = const.$, as this variable is kept constant for both the integral in Eq.~\eqref{eq:p_as_int_temp} and in Eq.~\eqref{eq:thdynpress_as_int_over_enerdens}.}
%\textcolor{blue}{I have added here the full set of variables that were held constant when performing the derivatives.}\\

\begin{subequations} \label{eq:Maxwell}
\begin{align}
      \left. \frac{\partial n}{\partial T} \right|_{\omega_{\sigma\tau}} & = \left. \frac{\partial s}{\partial \mu}\right|_{\omega_{\sigma\tau}}\;, \label{eq:Maxwell_1}\\[.5em]
     \left. \frac{\partial s}{\partial \omega_{\mu \nu}} \right|_{\mu} & = \hbar \left. \frac{\partial S^{\mu \nu}}{\partial T}\right|_{\mu}\;, \label{eq:Maxwell_2}\\[.5em]
      \left. \frac{\partial n}{\partial \omega_{\mu \nu}} \right|_{T} & = \hbar \left. \frac{\partial S^{\mu \nu}}{\partial \mu}\right|_{T}\;. \label{eq:Maxwell_3}
\end{align}
\end{subequations}
\newline
\noindent
In this case, the pressure is independent of the path in $(T,\mu, \omega_{\mu \nu})$ space along which one computes the integrals.
Equation~\eqref{eq:thdynpress_as_int_over_enerdens} corresponds to the red path in Fig.~\ref{fig:int_Tmuomega}, because $\alpha \equiv \mu/T$ and $\Omega_{\sigma \tau} \equiv \omega_{\sigma \tau}/T$ are to be kept constant when integrating over $\beta' \equiv 1/T'$.
On the other hand, Eq.~\eqref{eq:p_as_int_temp} corresponds to the last part of the green path in Fig.~\ref{fig:int_Tmuomega}, because here $\mu$ and $\omega_{\mu \nu}$ are kept constant.
Since the pressure for a Boltzmann gas always vanishes at $T=0$, irrespective of the values of $\mu$ and $\omega_{\mu \nu}$, there is no contribution from the first two parts of the green path in the $(\mu, \omega_{\mu \nu})$-plane (i.e., the part along the $\mu$-axis and the one perpendicular to it parallel to the $\omega_{\mu \nu}$-axis).
\newline

The strategy for the remainder of this paper is then the following. 
We first derive particle-number current, energy-momentum tensor, and spin-current tensor in various pseudo-gauges from the general quantum field-theoretical expressions in terms of components of the Wigner function.
We then check the validity of the Maxwell relations~\eqref{eq:Maxwell}.
In case they are fulfilled, we compute the thermodynamic pressure, which yields identical results when using Eq.~\eqref{eq:p_as_int_temp} or Eq.~\eqref{eq:thdynpress_as_int_over_enerdens}.
On the other hand, if they are violated, one obtains different results when computing the thermodynamic pressure from Eq.~\eqref{eq:p_as_int_temp} or from Eq.~\eqref{eq:thdynpress_as_int_over_enerdens}.
In case the Maxwell relations hold, we confirm the validity of the thermodynamic relations~\eqref{eq:ortho_thdyn_densities_as_ders}.
On the other hand, if they are violated, these thermodynamic relations no longer hold.

%
%**********************************************************************************************************************************************************************************************************************************************************************************************************************************************************************************
%
%********************************************************************************************************************************************************************************************************************************************************************************************************************
%

\section{Wigner function and extended phase space}
\label{sec:Wigner}

The central quantity for the computation of the (net) particle-number current, the energy-momentum tensor, and the spin-current tensor of a given system is the Wigner function.
For spin-1/2 particles, it is defined as the Fourier transform of the normal-ordered two-point correlation function with respect to the relative coordinate~\cite{Speranza:2020ilk,Weickgenannt:2022jes,Wagner:2022amr},\footnote{Note that we use the convention of Ref.~\cite{Wagner:2022amr}, i.e., we define the spacetime measure as $\di^4x$ and the momentum-space measure as $\di^4 k/(2 \pi \hbar)^4$.
Instead, Refs.~\cite{Speranza:2020ilk,Weickgenannt:2022jes} follow Ref.~\cite{degroot1980relativistic}, which defines the spacetime measure as $\di^4x /(2 \pi \hbar)^4$ and the momentum-space measure as $\di^4 k$. 
The former convention has the advantage that Pauli-blocking and Bose-enhancement factors assume the simple form $1 \mp f$ instead of $1\mp (2 \pi \hbar)^3 f$. 
(Here, $f$ is the single-particle distribution function.)}
\begin{align}
W_{\alpha\beta}(x,k) & :=  \int \di^4 y \,e^{-i k\cdot y/\hbar} \,\Tr\left[ \hat{\rho}\, : \hat{\overline{\psi}}_{\beta}\left(x+\frac{y}{2}\right)\hat{\psi}_{\alpha}\left(x-\frac{y}{2}\right):
\right]\; .
\label{eq:wigner-def}
\end{align}
The Wigner function obeys a Dyson-Schwinger equation~\cite{Sheng:2021kfc},
\begin{equation} \label{eq:EoM_Wigner_coll}
    \left[ \gamma^\mu \left(k_\mu + \frac{i \hbar}{2} \partial_\mu^x \right) - m \right] W(x,k) = I_{\text{coll}}\;,
\end{equation}
where $I_{\text{coll}}$ is the collision term. 
We assume that the system is in global equilibrium. 
While this state is reached through interactions, the global-equilibrium distribution function is equivalent to that of an effectively non-interacting system.
Therefore, we may set the collision term to zero in the following\footnote{This collision term is given in full generality by the right-hand side Eq.~(18) of Ref.~\cite{Sheng:2021kfc}, which, in the quasi-particle approximation, simplifies to Eq.~(24) of that reference. Since we neglect interactions in global equilibrium, even the right-hand side of Eq.~(18) vanishes, so there is in principle no need to make the quasi-particle approximation.}.

We act on Eq.~\eqref{eq:EoM_Wigner_coll} from the left with the operator $\gamma^\mu (k_\mu +  \frac{i}{2} \hbar \partial_\mu^x) +m$.
Then we take the Hermitian conjugate of the resulting equation and multiply it from the left and right with $\gamma^0$.
Using $W \equiv \gamma^0 W^\dagger \gamma^0$, one can take linear combinations of the result with the original equation to obtain the mass-shell constraint
\begin{equation} \label{eq:massshell}
    \left(k^2 - m^2 - \frac{\hbar^2}{4} \, \Box_x\right) W(x,k) =0\;,
\end{equation}
and the transport equation
\begin{equation} \label{eq:transport}
    k \cdot \partial_x\, W(x,k) = 0\;.
\end{equation}
Equation \eqref{eq:massshell} tells us that the classical mass-shell of non-interacting particles, $k^2 = m^2$, receives corrections at order $\mathcal{O}(\hbar^2)$.
Since in this work we are interested in corrections to thermodynamic quantities at this order in $\hbar$, we need to account for this modification throughout the following considerations.
We note that Eqs.~\eqref{eq:massshell} and \eqref{eq:transport} also hold for particles with spin different from 1/2. 
The mass-shell modification at order $\mathcal{O}(\hbar^2)$ is thus generic to all particles, independent of their spin.

As a $4\times 4$ Dirac matrix, the Wigner function can be decomposed into components of the Clifford algebra,
\begin{equation}
W =\frac{1}{4}\left(\mathcal{F}+i\gamma^{5}\mathcal{P}
+ \gamma^\mu\mathcal{V}_{\mu}
+\gamma^{5}\gamma^\mu \mathcal{A}_{\mu}
+\frac{1}{2}\sigma^{\mu\nu}\mathcal{S}_{\mu\nu}\right)\;,
\label{eq:wigner-decomp}
\end{equation}
For vanishing collision term, the Clifford components of the Wigner function then obey~\cite{Sheng:2021kfc,Wagner:2024fhf}
\begin{subequations} \label{eq:EoM_Wigner}
\begin{align}
k^\mu \mathcal{V}_\mu - m \mathcal{F} & =  0\;,
\label{eq:scalar_real} \\
m \mathcal{P} + \frac{\hbar}{2}\, \partial^x_\mu \mathcal{A}^\mu & =   0\;,
\label{eq:pseudoscalar_real}  \\
k_\mu \mathcal{F} - m \mathcal{V}_\mu + \frac{\hbar}{2} \, \partial_x^\nu \mathcal{S}_{\mu \nu} & =  0\;,
\label{eq:vector_real}  \\
\frac{1}{2} \, \epsilon_{\mu \nu \rho\sigma} k^\nu \mathcal{S}^{\rho \sigma} + m \mathcal{A}_\mu
- \frac{\hbar}{2} \, \partial_\mu^x \mathcal{P} & =  0\;,
\label{eq:axialvector_real}  \\
\epsilon_{\mu \nu \rho\sigma} k^\rho \mathcal{A}^\sigma + m \mathcal{S}_{\mu \nu} - \frac{\hbar}{2} \left( \partial_{\mu} \mathcal{V}_{\nu}- \partial_{\nu} \mathcal{V}_{\mu}\right) & =  0\;, 
\label{eq:tensor_real}\\
\frac{\hbar}{2}\, \partial_\mu^x \mathcal{V}^\mu & = 0\;,
\label{eq:scalar_im} \\
k^\mu \mathcal{A}_\mu & =  0\;,
\label{eq:pseudoscalar_im} \\
k^\nu \mathcal{S}_{\nu \mu} + \frac{\hbar}{2} \, \partial_\mu^x \mathcal{F} & =  0\;, \label{eq:vector_im} \\
k_\mu \mathcal{P} + \frac{\hbar}{4} \, \epsilon_{\mu \nu \rho\sigma} \partial_x^\nu \mathcal{S}^{\rho\sigma} & =  0\;,
\label{eq:axialvector_im} \\
k_{\mu} \mathcal{V}_{\nu} - k_{\nu} \mathcal{V}_{\mu} + \frac{\hbar}{2}\, \epsilon_{\mu \nu \rho\sigma} \partial_x^\rho \mathcal{A}^\sigma & =  0\;.
\label{eq:tensor_im}
\end{align}
\end{subequations}

From Eq.~\eqref{eq:pseudoscalar_real} one immediately sees that $\mathcal{P}$ is not an independent degree of freedom and can be expressed via the four-divergence of $\mathcal{A}^\mu$,
\begin{equation} \label{eq:P}
    \mathcal{P} = - \frac{\hbar}{2m}\, \partial_x \cdot \mathcal{A}\;.
\end{equation}

Furthermore, as an antisymmetric rank-2 tensor, $\mathcal{S}^{\mu \nu}$ has six independent components, which can be assembled into an ``electric-like'' and a ``magnetic-like'' component in the rest frame of a particle,
\begin{subequations}
\begin{align}
    \mathcal{S}^{\mu \nu} & \equiv \frac{1}{\sqrt{k^2}} \left( k^\mu \mathcal{E}^\nu - k^\nu \mathcal{E}^\mu + \epsilon^{\mu \nu \alpha \beta} k_\alpha \mathcal{B}_\beta \right)\;, \\
    \mathcal{E}^{\mu} &:= - \frac{1}{\sqrt{k^2}} \,\mathcal{S}^{\mu \nu} k_\nu\;, \quad \mathcal{B}^\mu := \frac{1}{2\sqrt{k^2}}\, \epsilon^{\mu \nu \alpha \beta}k_\nu \mathcal{S}_{\alpha \beta}\;. \label{eq:def_E_B}
\end{align}
\end{subequations}
Both components are orthogonal to $k_\mu$,
\begin{equation}
    k \cdot \mathcal{E} = 0\;, \quad k \cdot \mathcal{B} = 0\;,
\end{equation}
such that $\mathcal{E}^\mu$ and $\mathcal{B}^\mu$ have three independent components each.
From Eq.~\eqref{eq:vector_im} one then immediately observes that 
\begin{equation}
    \mathcal{E}_\mu = - \frac{\hbar}{2\sqrt{k^2}}\, \partial_\mu^x \mathcal{F}\;,
\end{equation}
while from Eq.~\eqref{eq:axialvector_real} one deduces that
\begin{equation}
    \mathcal{B}_\mu = - \frac{m}{\sqrt{k^2}}\, \mathcal{A}_\mu + \frac{\hbar}{2\sqrt{k^2}}\, \partial_\mu^x \mathcal{P} = -\frac{m}{\sqrt{k^2}}\left( g_{\mu \nu} + \frac{\hbar^2}{4m^2}\, \partial_\mu^x \, \partial_\nu^x \right) \mathcal{A}^\nu\;,
\end{equation}
where we have used Eq.~\eqref{eq:P}.
Thus, $\mathcal{S}^{\mu \nu}$ can be expressed solely in terms of $\mathcal{F}$ and $\mathcal{A}^\mu$.

Finally, decomposing $\mathcal{V}^\mu$ into components longitudinal and transverse to $k^\mu$,
\begin{equation}
    \mathcal{V}^\mu := \frac{k^\mu}{\sqrt{k^2}}\, \mathcal{V}_\parallel + \mathcal{V}^\mu_{\perp}\;,
\end{equation}
Eq.~\eqref{eq:scalar_real} tells us that 
$\mathcal{V}_\parallel \equiv m\mathcal{F}/\sqrt{k^2}$, while from the projection of Eq.~\eqref{eq:tensor_im} onto $k_\mu$ one deduces that
\begin{equation}\label{eq:mcvperp}
    \mathcal{V}^\mu_\perp = \frac{\hbar}{2k^2}\, \epsilon^{\mu \nu \alpha \beta} k_\nu \partial^x_\alpha \mathcal{A}_\beta\;,
\end{equation}
i.e., $\mathcal{V}^\mu$ can be also expressed in terms of $\mathcal{F}$ and $\mathcal{A}^\mu$.
In summary, the four independent components of the Wigner function are $\mathcal{F}$ and, because of Eq.~\eqref{eq:pseudoscalar_im}, the three components of $\mathcal{A}^\mu$ which are transverse to $k^\mu$.

It is convenient to assign a spin four-vector $\mathfrak{s}^\mu$ to the particles and extend momentum space by spin degrees of freedom~\cite{Weickgenannt:2020aaf}.
For on-shell spin-1/2 particles, this is done as follows:
\begin{equation}
    \di \Gamma := \di K\, \di S(K)\;, \quad \di K := \frac{\di^4 k}{(2 \pi \hbar)^4}\;, \quad \di S(K) :=\sqrt{\frac{k^2}{3\pi^2}}\,\di^4 \mathfrak{s} \, \delta(\mathfrak{s} \cdot \mathfrak{s}+3)\,\delta(k \cdot \mathfrak{s})\;,
\end{equation}
where
\begin{equation} \label{eq:spin-space_int}
    \int \di S(K) = 2\;,\quad \int \di S(K)\, \mathfrak{s}^\mu = 0\;, \quad \int \di S(K)\, \mathfrak{s}^\mu \mathfrak{s}^\nu = - 2 \left( g^{\mu \nu} - \frac{k^\mu k^\nu}{k^2} \right)\;.
\end{equation}
For non-interacting particles, one then defines a \emph{distribution} (in the mathematical sense) in extended phase space via
\begin{equation} \label{eq:dist_phase_space}
    \mathfrak{f}(x,k,\mathfrak{s}) : = \frac{1}{2} \left[ \mathcal{F} (x,k)- \mathfrak{s}^\mu \mathcal{A}_\mu (x,k)\right] \;,
\end{equation}
such that
\begin{subequations} \label{eq:F_and_A}
\begin{align} \label{eq:fF}
    \int \di S(K)\, \mathfrak{f}(x,k,\mathfrak{s}) = \mathcal{F}(x,k)\;, \\ \int \di S(K)\, \mathfrak{s}^\mu\, \mathfrak{f} (x,k,\mathfrak{s}) = \mathcal{A}^\mu(x,k)\;.
    \label{eq:fA}
\end{align}
\end{subequations}
This distribution encompasses the full information of the Wigner function, through its independent components $\mathcal{F}$ and $\mathcal{A}^\mu$.

For on-shell particles, we now define the single-particle distribution \emph{functions} (in the mathematical sense) for particles (+) and antiparticles ($-$) $f^\pm(x,k,\mathfrak{s})$ via\footnote{Here, we neglect the contribution from the vacuum and choose the energy and the momentum of the antiparticles to have the same sign as that of particles.}
\begin{equation} \label{eq:frak_f}
    \mathfrak{f}(x,k,\mathfrak{s}) \equiv 4\pi \hbar\, m \, \theta(k_0)\, \delta\left(k^ 2 -m^2 - F(k)\right)\, \left[f^+(x,k,\mathfrak{s})+ f^-(x,k,\mathfrak{s}) \right]\;.
\end{equation}
The components $\mathcal{F}$ and $\mathcal{A}^\mu$ of the Wigner function obey the same mass-shell constraint~\eqref{eq:massshell} as the latter.
Therefore, acting with the mass-shell operator on Eqs.~\eqref{eq:F_and_A}, we conclude that the function $F(k)$ obeys the equation
\begin{equation} \label{eq:wave_eq_F}
    \frac{\hbar^2}{4} \, \Box_x \, f^{\pm}(x,k,\mathfrak{s}) = F(k) \, f^{\pm}(x,k,\mathfrak{s})\;.
\end{equation}

In thermodynamic equilibrium and for Boltzmann statistics, the single-particle distribution functions for particles and antiparticles read 
\begin{align}\label{eq:sing_part_dist}
f_{\rm eq}^{\pm}(x,k,\mathfrak{s})&=  \exp\left(-\beta_\mu k^\mu \pm \alpha+\frac{\hbar}{4}\Omega_{\mu\nu}\Sigma^{\mu\nu}_{k\mathfrak{s}}\right)\;.
\end{align}
Here, the dipole-moment tensor is defined as 
\begin{equation} \label{eq:dipolemoment}
\Sigma^{\mu\nu}_{k\mathfrak{s}}=-\frac{1}{m}\epsilon^{\mu\nu\alpha\beta}k_{\alpha}\mathfrak{s}_{\beta}
\end{equation} 
and satisfies $k_{\mu}\Sigma^{\mu\nu}_{k\mathfrak{s}}=0$. 
Note that Eq.~\eqref{eq:sing_part_dist} was proven to be correct to order $\order{\hbar}$.
In the following, we assume that it remains correct to order $\order{\hbar^2}$, i.e., we may use the Taylor expansion of the exponential function to extract the contribution of order $\order{\hbar^2}$~\cite{Florkowski:2024bfw}.

In global equilibrium, the conditions~\eqref{eq:gteq_conditions} hold.
Therefore, inserting Eq.~\eqref{eq:sing_part_dist} into Eq.~\eqref{eq:wave_eq_F}, we derive
\begin{equation} \label{eq:F_eq}
    F_{\mathrm{eq}}(k) \equiv - \frac{\hbar^2}{4}\, \varpi_{\mu}^{\hspace*{0.15cm} \lambda} \varpi_{\lambda \nu} k^\mu k^\nu \;.
\end{equation}
Thus, by construction, the global-equilibrium phase-space distribution $\mathfrak{f}_{\mathrm{eq}}(x,k,\mathfrak{s})$ fulfills the mass-shell constraint equation \eqref{eq:massshell}.
It also fulfills the transport equation \eqref{eq:transport}, because
\begin{equation}
    k^\mu \partial_\mu^x f_{\mathrm{eq}}^\pm (x,k, \mathfrak{s}) = - k^\mu \varpi_{\mu \nu} k^\nu \, f_{\mathrm{eq}}^\pm (x,k, \mathfrak{s}) = 0\;.
\end{equation}

With Eq.~\eqref{eq:F_eq}, the mass-shell constraint in momentum space reads
\begin{align}
    m^2 & = k^\mu \left(g_{\mu \nu} + \frac{\hbar^2}{4} \, \varpi_{\mu}^{\hspace*{0.1cm} \lambda} \varpi_{\lambda \nu} \right) k^\nu\;, \label{eq:massshell_mod}
\end{align}
We note that changing the momentum variable according to
\begin{align}
 k^{\prime \mu} &:= \left( g^{\mu}_{\hspace*{0.1cm} \nu} + \frac{\hbar^2}{8}\, \varpi^{\mu \lambda} \varpi_{\lambda \nu} \right) k^\nu \;, \quad 
    k^{\prime}_{\mu} := k_\nu \left( g_{\hspace*{0.1cm} \mu}^{\nu} + \frac{\hbar^2}{8}\, \varpi^{\nu \lambda} \varpi_{\lambda\mu} \right) \;,
    \label{eq:transform_mom}
\end{align}
allows one to recover the standard mass-shell constraint for the momentum $k^{\prime \mu}$ (up to terms of order $\mathcal{O}(\hbar^4)$),
\begin{align}\label{eq:standmassshell}
    k^{\prime 2} & \equiv k^{\prime}_\mu k^{\prime \mu} = m^2\;, 
\end{align}
Since we compute up to order $\mathcal{O}(\hbar^2)$, we will require the inversion of Eq.~\eqref{eq:transform_mom} up to the same order,
\begin{equation}\label{eq:k=kprime}
    k^\mu = \left( g^{\mu}_{\hspace*{0.1cm} \nu} - \frac{\hbar^2}{8}\, \varpi^{\mu \lambda} \varpi_{\lambda \nu} 
\right) k^{\prime \nu} + \mathcal{O}(\hbar^4)\;.
\end{equation}
The corresponding change in the momentum-space measure is
\begin{align} \label{eq:mod_mom_space_measure}
    \di K \equiv \frac{\di^4 k}{(2 \pi \hbar)^4} =
    \frac{\di^4 k'}{(2 \pi \hbar)^4}\, \mathrm{det}\left( g^{\mu}_{\hspace*{0.1cm} \nu} - \frac{\hbar^2}{8}\, \varpi^{\mu \lambda} \varpi_{\lambda \nu} \right)
    \equiv \di K'\,  \left(1 + \frac{\hbar^2}{8} \varpi^{\mu \nu}\varpi_{\mu \nu} \right) + \mathcal{O}(\hbar^4)\;.
\end{align}
In the explicit calculations of thermodynamic quantities in Secs.~\ref{sec:em_s_kin_theory}, 
\ref{Sec-can-psuedo}, and \ref{Sec:GLWPG}, this change of variables will only be employed after the integration over spin space has been performed.
Note that, since we account for the modification of the mass-shell constraint at order $\mathcal{O}(\hbar^2)$, our results for thermodynamic quantities differ from those of Ref.~\cite{Florkowski:2024bfw}.

\section{Pseudo-gauge dependence of particle-number current, energy-momentum tensor, and spin-current tensor}
\label{sec:pseudogauge}

The pseudo-gauge dependence of the energy-momentum and the spin-current tensor has been extensively discussed in Refs.~\cite{Speranza:2020ilk,Weickgenannt:2022jes}.
There, these tensors were expressed as integrals of the Clifford components of the Wigner function over momentum space, or as integrals of the phase-space distribution~\eqref{eq:dist_phase_space} over phase space. 
Here we quote the results insofar as they are relevant for this paper.

Since we compute consistently to order $\mathcal{O}(\hbar^2)$, we require the energy-momentum tensor up to that order, while the spin-current tensor, being multiplied by an overall factor $\hbar$, is needed only up to order $\mathcal{O}(\hbar)$.
Extending previous discussions, we also study the pseudo-gauge dependence of the (net) particle-number current.

\subsection{Canonical pseudo-gauge}
\label{sec:canonical_PG}
In the canonical pseudo-gauge, the (net) particle-number current, the energy-momentum tensor, and the spin-current tensor read
\begin{subequations}
\label{eq:currents_as_ders_can_5}
    \begin{align}
        J_C^\mu &  = \int \di K\, \frac{\partial \mathcal{V}^\mu}{\partial \alpha}\;, \label{eq:part_curr_can} \\
        T^{\mu \nu}_C & = \int \di K\, k^\nu \mathcal{V}^\mu \;,  \label{eq:enmom_can} \\
        S^{\lambda, \mu \nu}_C & = - \frac{1}{2} \epsilon^{\lambda \mu \nu \alpha} \int \di K\, \mathcal{A}_\alpha\;.
    \end{align}
\end{subequations}
Note that the canonical net particle-number current is usually defined as $J_C^\mu = \int \di K\, \mathcal{V}^\mu$. 
Since we choose the energy and momentum of antiparticles to have the same sign as that of particles, cf.~Eq.~\eqref{eq:frak_f}, we need to invoke a change of sign for the antiparticle contribution in the net particle-number current as compared to the energy-momentum tensor. In Boltzmann approximation, this is most conveniently affected by taking a derivative with respect to the reduced chemical potential, which differs in sign for antiparticles as compared to particles.

In order to express the currents~\eqref{eq:currents_as_ders_can_5} in terms of the phase-space distribution $\mathfrak{f}(x,k,\mathfrak{s})$, we need to replace $\mathcal{V}^\mu$ by the independent components $\mathcal{F}$ and $\mathcal{A}^\mu$.
On account of Eqs.~\eqref{eq:vector_real}, \eqref{eq:tensor_real}, and \eqref{eq:scalar_im} the vector component of the Wigner function can be expressed as
\begin{equation} \label{eq:Vmu}
    \mathcal{V}^\mu = \frac{k^\mu}{m} \left( 1- \frac{\hbar^2}{4m^2}\, \Box_x \right) \mathcal{F}  - \frac{\hbar}{2m^2}\, \epsilon^{\mu \nu \alpha \beta} \partial_\nu^x k_\alpha \mathcal{A}_\beta + \mathcal{O}(\hbar^3)\;.
\end{equation}
Inserting this into Eqs.~\eqref{eq:part_curr_can} and \eqref{eq:enmom_can} and using Eqs.~\eqref{eq:F_and_A}, and  \eqref{eq:dipolemoment}, the (net) particle-number current and energy-momentum tensor can be expressed in terms of the phase-space distribution,
\begin{subequations} \label{eq:JTS_C}
\begin{align}
J_C^\mu & = \frac{1}{m} \int \di \Gamma \left[ k^\mu  \left( 1 - \frac{\hbar^2}{4m^2}\, \Box_x \right) +
    \frac{\hbar}{2} \, \Sigma_{k\mathfrak{s}}^{\mu \lambda}  \partial_\lambda^x + \mathcal{O}(\hbar^3)\right] \frac{\partial \, \mathfrak{f}(x,k,\mathfrak{s})}{\partial \alpha}  \;,\label{eq:particle-number_current_can} \\
    T_C^{\mu \nu} & = \frac{1}{m} \int \di \Gamma\, k^\nu  \left[ k^\mu  \left( 1 - \frac{\hbar^2}{4m^2}\, \Box_x \right) +
    \frac{\hbar}{2} \, \Sigma_{k\mathfrak{s}}^{\mu \lambda}  \partial_\lambda^x + \mathcal{O}(\hbar^3)\right] \mathfrak{f}(x,k,\mathfrak{s})  \;.\label{eq:energy-mom_tensor_can}
\end{align}
Similarly, using Eqs.~\eqref{eq:fA} and \eqref{eq:dipolemoment}, the spin-current tensor can be written as
\begin{equation}\label{eq:spin-current_tensor_can}
    S_C^{\lambda, \mu \nu} = - \frac{1}{2} \, \epsilon^{\lambda \mu \nu \alpha}
    \int \di \Gamma\, \mathfrak{s}_\alpha \, \mathfrak{f}(x,k,\mathfrak{s})\;.
\end{equation}
\end{subequations}
Note that, using the Schouten identity, this can be shown to be identical to Eq.~(25b) of Ref.~\cite{Weickgenannt:2022jes}.

%We remark that, when inserting the equilibrium distribution function \eqref{eq:sing_part_dist} into Eq.~\eqref{eq:frak_f} and the latter into Eqs.~\eqref{eq:JTS_C}, the spacetime derivatives will give rise to factors of thermal vorticity.
%However, when scaling the density operator with a factor $\lambda$ as in Eq.~(10) of Ref.~\cite{Becattini:2023ouz}, only 

\subsection{GLW pseudo-gauge}
\label{subSECGLWW}
The de Groot--van Leeuwen--van Weert (GLW) pseudo-gauge can be obtained from the canonical one by choosing
\begin{subequations} \label{eq:pg_C-GLW}
\begin{align}
B^{\lambda \mu}_{C \rightarrow \text{GLW}} & =\frac{\hbar}{4m^{3}}\int\di K\,\left(k^{\mu}\partial_x^{\lambda}-k^{\lambda}\partial_x^{\mu}\right)\frac{\partial\mathcal{F}}{\partial\alpha}-\frac{1}{2m^{2}}\int \di K\epsilon^{\lambda\mu\rho\sigma}k_{\rho}\frac{\partial\mathcal{A}_{\sigma}}{\partial\alpha} \;,\label{BCGLWW} \\
    \Phi_{C \rightarrow \text{GLW}}^{\lambda, \mu \nu} & = \frac{1}{2m} \int \di K\, \left( k^{\mu} \mathcal{S}^{\nu \lambda} - k^{\nu} \mathcal{S}^{\mu \lambda} \right)\;, \\
    Z^{\mu \nu, \lambda \rho}_{C \rightarrow \text{GLW}} & =0\;.   
\end{align}
\end{subequations}
Note that the (net) particle-number current is usually not discussed in the context of pseudo-gauge transformations.
In fact, the pseudo-gauge transformation of the (net) particle-number current is independent of that of the energy-momentum tensor and the spin-current tensor.
We exploit this freedom to choose a pseudo-gauge transformation that makes the (net) particle-number current in GLW pseudo-gauge identical to that in kinetic-theory pseudo-gauge, see Sec.~\ref{sec:KT_PG}.
The explicit derivation of $B^{\lambda \mu}_{C \rightarrow \text{GLW}}$ is given in App.~\ref{AppedndixBlambdamu}.

Using the equation of motion \eqref{eq:transport} for $\mathcal{F}$ as well as Eq.~\eqref{eq:Vmu}, we obtain for the particle-number current
\begin{align}
J^\mu_{\text{GLW}} & = \frac{1}{m} \int \di K\, k^\mu \frac{\partial \mathcal{F}}{\partial \alpha}
= \frac{1}{m} \int \di \Gamma\, k^\mu \frac{\partial \, \mathfrak{f}(x,k, \mathfrak{s})}{\partial \alpha}\;, \label{eq:part_curr_GLW}
\end{align}
where we have used Eq.~\eqref{eq:fF} in the last step.

For the energy-momentum tensor we obtain
\begin{align}
        T^{\mu \nu}_{\text{GLW}} & = \int \di K\, k^\nu \left( \mathcal{V}^\mu + \frac{\hbar}{2m}\, \partial_\lambda^x \mathcal{S}^{\lambda\mu }\right) = \frac{1}{m} \int \di \Gamma\, k^\mu k^\nu\, \mathfrak{f}(x,k,\mathfrak{s})\;,  \label{eq:enmom_GLW}
\end{align}
where we have used Eqs.~\eqref{eq:vector_real} and \eqref{eq:fF} in the last step.

Finally, the pseudo-gauge transformation \eqref{eq:pg_C-GLW} results in the spin tensor
\begin{equation}
        S^{\lambda, \mu \nu}_{\text{GLW}} = - \frac{1}{2} \int \di K\,\left[ \epsilon^{\lambda \mu \nu \alpha}  \mathcal{A}_\alpha + \frac{1}{m} \left( k^\mu \mathcal{S}^{\nu\lambda} - k^\nu \mathcal{S}^{\mu \lambda} \right) \right]\;.
\end{equation}
Using Eqs.~\eqref{eq:vector_real}, \eqref{eq:tensor_real}, \eqref{eq:pseudoscalar_im}, the Schouten identity, as well as Eqs.~\eqref{eq:F_and_A}, \eqref{eq:dipolemoment}, and the mass-shell constraint~\eqref{eq:massshell_mod}, one derives
\begin{equation} \label{eq:spintensor_GLW}
    S^{\lambda, \mu \nu}_{\text{GLW}} = 
    \frac{1}{2m} \int \di \Gamma\, k^\lambda \left[ \Sigma_{k\mathfrak{s}}^{\mu \nu}- \frac{\hbar}{2m^2} \left( k^\mu \partial_x^\nu - k^\nu \partial_x^\mu \right) + \mathcal{O}(\hbar^2) \right] \mathfrak{f}(x,k,\mathfrak{s}) \;.
\end{equation}

\subsection{Kinetic-theory pseudo-gauge}
\label{sec:KT_PG}
While the (net) particle-number current, Eq.~\eqref{eq:part_curr_GLW}, and the energy-momentum tensor in GLW pseudo-gauge, Eq.~\eqref{eq:enmom_GLW}, assume the form well known in kinetic theory, the spin-current tensor~\eqref{eq:spintensor_GLW} has an $\mathcal{O}(\hbar)$ contribution which is absent in kinetic theory.
It was first noted in Ref.~\cite{Wagner:2024fry} (see footnote 1 in that reference) that this can be transformed away by another pseudo-gauge transformation,
\begin{equation}
    \Phi^{\lambda, \mu \nu}_{\text{GLW} \rightarrow \text{kin}} \equiv 0\;\label{PhiGLW-KIN}, \quad 
    Z^{\mu \nu, \lambda \rho}_{\text{GLW}\rightarrow \text{kin}} \equiv \frac{1}{4 m^3} \int \di K\, \left(  k^\lambda g^{\rho \nu} k^\mu - k^\lambda g^{\rho \mu} k^\nu + k^\rho g^{\lambda \mu} k^\nu
     - k^\rho g^{\lambda \nu} k^\mu \right) \mathcal{F}\;.
\end{equation}
When inserted into Eq.~\eqref{eq:pgt_s_current_eval} the first two terms cancel the second and third terms in Eq.~\eqref{eq:spintensor_GLW}, while the last two terms vanish on account of the equation of motion~\eqref{eq:transport} for $\mathcal{F}$.
Thus, the (net) particle-number current, the energy-momentum tensor, and the spin-current tensor in the pseudo-gauge corresponding to kinetic theory assume the form 
\begin{subequations} \label{eq:currents_as_ders_kintheory_PG}
    \begin{align}
    J^\mu & = \frac{1}{m} \int \di \Gamma\, k^\mu \, \frac{\partial\, \mathfrak{f}(x,k,\mathfrak{s})}{\partial \alpha}\;, \label{eq:part_curr_kin_th} \\
            T^{\mu \nu}  & = \frac{1}{m} \int \di \Gamma\, k^\mu k^\nu\, \mathfrak{f}(x,k,\mathfrak{s})\;, \label{eq:en_mom_tensor_kin_th}\\
             S^{\lambda, \mu \nu} & =     \frac{1}{2m} \int \di \Gamma\, k^\lambda  \Sigma_{k\mathfrak{s}}^{\mu \nu}\,\mathfrak{f}(x,k,\mathfrak{s})\;.
             \label{eq:spin-current_tensor}
    \end{align}
\end{subequations}
In this case, we omit the index indicating a particular pseudo-gauge.
The (net) particle-number current and energy-momentum tensor in kinetic-theory pseudo-gauge are identical to those in the GLW pseudo-gauge, $J^\mu \equiv J^\mu_{\text{GLW}}$, $T^{\mu \nu} \equiv T^{\mu \nu}_{\text{GLW}}$, while the spin-current tensor is different.

In the following sections, we explicitly compute the (net) particle-number current, the energy-momentum tensor, and the spin-current tensor, as well as related thermodynamic quantities, in the pseudo-gauges discussed above for a system of massive spin-1/2 particles following Boltzmann statistics.
We first discuss the kinetic-theory pseudo-gauge, as it is the physically most intuitive one, followed by the canonical pseudo-gauge, and we conclude with the GLW pseudo-gauge.

\section{Kinetic-theory pseudo-gauge}
\label{sec:em_s_kin_theory}
In this section, we compute  the (net) particle-number current, the energy-momentum tensor, and the spin-current tensor, as well as related thermodynamic quantities, in the kinetic-theory pseudo-gauge.
We work in global thermodynamic equilibrium, where the distribution function \eqref{eq:sing_part_dist} is an exact solution of the Boltzmann equation with vanishing collision term.
However, as we will see, we need to be careful in distinguishing the spin potential $\omega_{\mu \nu}$, or reduced spin potential $\Omega_{\mu \nu}$, which is an independent thermodynamic variable, from the thermal vorticity $\varpi_{\mu \nu}$.
Albeit $\Omega_{\mu \nu}$ assumes the value $\varpi_{\mu \nu}$ in global thermodynamic equilibrium, the latter is not an independent thermodynamic variable, but instead given by the antisymmetric combination of partial derivatives of $\beta_\mu$, $\varpi_{\mu \nu} = - \frac{1}{2} \left( \partial_\mu \beta_\nu - \partial_\nu \beta_\mu \right)$.

\subsection{Particle-number current}
\label{Sec3a}
In kinetic theory, the (net) particle-number current is given by Eq.~\eqref{eq:part_curr_kin_th}.
We insert the phase-space distribution  \eqref{eq:frak_f} with the equilibrium single-particle distribution function \eqref{eq:sing_part_dist} and expand the latter to second order in $\hbar$.
We then perform the integration over spin space using Eq.~\eqref{eq:spin-space_int} and obtain
\begin{align}
    J^\mu & = 2 \sinh \alpha \int \mathrm{d}K \, 4 \pi \hbar \, \theta(k_0) \, \delta (k^{\prime 2} - m^2)\, k^\mu\, e^{- \beta \cdot k}
\left[ 2 + \frac{\hbar^2}{8 m^2}\, \Omega_{\lambda \rho} \Omega^{\lambda}_{\hspace*{0.1cm} \sigma} \left( g^{\rho\sigma} k^2 - 2 k^\rho k^\sigma \right) \right]\;.
\end{align}
Note that the mass-shell condition is written in terms of the momentum $k^{\prime \mu}$, Eq.~\eqref{eq:transform_mom}.
Since we compute to order $\mathcal{O}(\hbar^2)$, we need to account for this mass-shell modification in the first term in square brackets.
On the other hand, the second term is already of order $\mathcal{O}(\hbar^2)$, such that we can safely replace $k^{\prime \mu} \rightarrow k^\mu$ in the mass-shell condition,
\begin{equation}
    J^\mu \equiv 4 \sinh \alpha \left( \mathcal{J}_1^\mu +  \mathcal{J}_2^\mu \right)\;,
\end{equation}
with
\begin{subequations}
    \begin{align}
        \mathcal{J}_1^\mu & := \int \di K\,
        4 \pi \hbar\, \theta(k_0)\, \delta (k^{\prime 2} - m^2)\, k^\mu \, e^{-\beta \cdot k}\;, \\
        \mathcal{J}_2^{\mu} & :=
        \frac{\hbar^2}{16 m^2} \, \Omega_{\lambda \rho} \Omega^{\lambda}_{\hspace*{0.1cm} \sigma} \int \di K \,
        4 \pi \hbar\, \theta(k_0)\, \delta (k^2 - m^2)\, k^\mu \left( g^{\rho\sigma} k^2 - 2 k^\rho k^\sigma \right)  e^{-\beta \cdot k}\;.
    \end{align}
\end{subequations}
The second term is readily evaluated using the decomposition~\eqref{eq:decomp_red_spin_potential} of the reduced spin potential and the tensor decompositions \eqref{eq:tensor_decomp}, together with the thermodynamic integrals \eqref{eq:Inq},
\begin{equation}
    \mathcal{J}_2^\mu =  \frac{\hbar^2 \beta^2}{8 m^2} \left\{ \left[ 2 I_{31} \,\varkappa^2 +  (I_{30}-I_{31}) \,\omega^2\right] u^\mu -
    2 I_{31}\, L^\mu \right\}\;,
\end{equation}
where $L^\mu$ is defined in Eq.~\eqref{eq:def_cap_l_vec}.
The first term is evaluated by substituting the integration variable $k^{\prime \mu}$, Eq.~\eqref{eq:transform_mom}, for $k^\mu$.
With Eqs.~\eqref{eq:decomp_thermal_vort}, \eqref{eq:k=kprime}, \eqref{eq:mod_mom_space_measure},  \eqref{eq:tensor_decomp}, and \eqref{eq:Inq}, as well as employing the identities \eqref{eq:useful_identities}, this gives, after some lengthy calculation,
\begin{equation}
    \mathcal{J}_1^\mu = I_{10} \, u^\mu +
    \frac{\hbar^2 \beta^2}{8} \left[\left( m^2 \beta\, I_{00}\, a^2 + 2I_{10} \,b^2 \right)  u^\mu - 2 I_{10} \, K^\mu \right]\;,   
\end{equation}
where $K^\mu$ is defined in Eq.~\eqref{eq:def_cap_k_vector}.
Therefore,
\begin{align}\label{eq:charge_current_kin_th}
J^\mu = n\, u^\mu + n_{\times}^\mu\;,
\end{align}
where 
\begin{align} \label{eq:chargedens}
n =  u_\mu J^\mu =  4 \sinh \alpha \left\{ I_{10}
+ \frac{\hbar^2 \beta^2}{8m^2} \left[
2 I_{31} \,\varkappa^2 +  (I_{30}-I_{31}) \,\omega^2 + m^2 \left( m^2 \beta \, I_{00}\, a^2 + 2I_{10} \,b^2 \right) \right] \right\} %
\end{align}
is the (net) particle-number density, while
\begin{align} \label{eq:chargeHall}
n_{\times}^\mu = - 4 \sinh \alpha  \,\frac{\hbar^2 \beta^2}{4 m^2} \left(  I_{31} \, L^\mu +m^2 I_{10}\, K^\mu \right)\;.
\end{align}
In global thermodynamic equilibrium (GTE),
$a^\mu \equiv \varkappa^\mu$ and $b^\mu \equiv \omega^\mu$, and the identity \eqref{eq:identity_2} yields
\begin{align}
\mathrm{GTE:} \qquad    n & = 4 \sinh \alpha \left\{ I_{10}
+ \frac{\hbar^2 \beta^2}{8m^2} \left[
\left(2 I_{31} + m^4 \beta I_{00} \right) \varkappa^2 +  \left(2I_{31}+ 3m^2 I_{10}\right) \omega^2  \right] \right\}\;, \\
n_\times^\mu & = - 4 \sinh \alpha  \,\frac{\hbar^2 \beta^2}{4 m^2} \left(I_{31} + m^2 I_{10} \right) L^\mu\;.
\end{align}
In the fluid rest frame,
\begin{align}
\mathbf{n}_{\times} =  4 \sinh \alpha \,   \frac{\hbar^2 \beta^2}{4 m^2} \left(I_{31} + m^2 I_{10} \right)  \boldsymbol{\varkappa} \times \boldsymbol{\omega}\;.
\end{align}
This structure resembles a Hall-type current generated by the interplay between acceleration and rotation. In particular, the current vanishes whenever either $\varkappa^\mu$ or $\omega^\mu$ is zero, indicating that it originates purely from their mutual coupling. 

\subsection{Energy-momentum tensor}
The energy-momentum tensor in kinetic-theory pseudo-gauge is defined in Eq.~\eqref{eq:en_mom_tensor_kin_th}.
Inserting the phase-space distribution function \eqref{eq:frak_f} and performing the integration over spin space using Eqs.~\eqref{eq:spin-space_int}, we obtain
\begin{align}
    T^{\mu \nu} & = 2 \cosh \alpha \int \mathrm{d}K \, 4 \pi \hbar \, \theta(k_0) \, \delta (k^{\prime 2} - m^2)\, k^\mu k^\nu\, e^{- \beta \cdot k}
\left[ 2 + \frac{\hbar^2}{8 m^2}\, \Omega_{\lambda \rho} \Omega^{\lambda}_{\hspace*{0.1cm} \sigma} \left( g^{\rho\sigma} k^2 - 2 k^\rho k^\sigma \right) \right]\;.
\end{align}
As in the calculation of the particle-number current, we need to account for the $\mathcal{O}(\hbar^2)$ mass-shell modification in the first term in square brackets, while the second term is already of order $\mathcal{O}(\hbar^2)$, such that we can safely replace $k^{\prime \mu} \rightarrow k^\mu$ in the argument of the delta function,
\begin{equation}
    T^{\mu \nu} \equiv 4 \cosh \alpha \left( \mathcal{T}_1^{\mu \nu} +  \mathcal{T}_2^{\mu \nu} \right)\;,
\end{equation}
with
\begin{subequations}
    \begin{align}
        \mathcal{T}_1^{\mu \nu} & := \int \di K\,
        4 \pi \hbar\, \theta(k_0)\, \delta (k^{\prime 2} - m^2)\, k^\mu k^\nu\, e^{-\beta \cdot k}\;, \\
        \mathcal{T}_2^{\mu \nu} & :=
        \frac{\hbar^2}{16 m^2} \, \Omega_{\lambda \rho} \Omega^{\lambda}_{\hspace*{0.1cm} \sigma} \int \di K \,
        4 \pi \hbar\, \theta(k_0)\, \delta (k^2 - m^2)\, k^\mu k^\nu \left( g^{\rho\sigma} k^2 - 2 k^\rho k^\sigma \right)  e^{-\beta \cdot k}\;.
    \end{align}
\end{subequations}
The second term is evaluated using the decompositions~\eqref{eq:decomp_red_spin_potential} and \eqref{eq:tensor_decomp}, as well as the definition~\eqref{eq:def_cap_l_vec},
\begin{align}
    \mathcal{T}_2^{\mu \nu} & =  \frac{\hbar^2 \beta^2}{8 m^2} \left\{ \left[ 2 I_{41} \,\varkappa^2 +  (I_{40}-I_{41}) \,\omega^2\right] u^\mu u^\nu 
    - \left[ 4 I_{42} \,\varkappa^2 +  (I_{41}-I_{42}) \,\omega^2\right]\Delta^{\mu \nu} \right. \nonumber \\
    & \hspace*{1.2cm} - \left. 
    2 I_{41} \left( u^\mu L^\nu + u^\nu L^\mu \right) - 2 I_{42} \left( \varkappa^\mu \varkappa^\nu + \omega^\mu \omega^\nu \right) \right\}\;.
\end{align}
The first term is again evaluated by substituting the integration variable $k^{\prime \mu}$, Eq.~\eqref{eq:transform_mom}, for $k^\mu$.
With Eqs.~\eqref{eq:decomp_thermal_vort}, \eqref{eq:def_cap_k_vector}, \eqref{eq:k=kprime}, \eqref{eq:mod_mom_space_measure}, \eqref{eq:tensor_decomp}, and \eqref{eq:Inq} this gives after some tedious calculation
\begin{align}
    \mathcal{T}_1^{\mu \nu} & = I_{20} \, u^\mu u^\nu  - I_{21} \Delta^{\mu \nu}
    \nonumber \\ & - \frac{\hbar^2 \beta^2}{8} \left\{ \left[ \left(4I_{20} - \beta \, I_{30}\right) a^2 - 2I_{20} \,b^2 \right]  u^\mu u^\nu+ \left[(I_{20}- I_{21} \right) a^2 + 4 I_{21} \, b^2] \Delta^{\mu \nu} \right.
    \nonumber \\
    & \hspace*{1.2cm} + \left. 2 I_{20} \left(u^\mu K^\nu + u^\nu K^\mu \right) + 2 I_{21} \left( a^\mu a^\nu + b^\mu b^\nu\right) \right\}\;,
\end{align}
such that 
\begin{align} \label{eq:en_mom_tensor_kin_th_2}
T^{\mu \nu} & = \varepsilon \, u^\mu u^\nu - P\, \Delta^{\mu \nu} + q_\times^\mu u^\nu + q_\times^\nu u^\mu \nonumber \\
& +\Pi_\varpi \left[ \Delta^{\mu \nu} (a^2 + b^2)
+ a^\mu a^\nu + b^\mu b^\nu \right]  +\Pi_\Omega \left[ \Delta^{\mu \nu} (\varkappa^2 + \omega^2)
+ \varkappa^\mu \varkappa^\nu + \omega^\mu \omega^\nu \right] \;.
\end{align}
Here, 
\begin{align} 
\varepsilon &= u_\mu u_\nu T^{\mu \nu}   = 4 \cosh \alpha \left( I_{20} + \frac{\hbar^2 \beta^2}{8 m^2}\left\{  2 I_{41} \,\varkappa^2 +   (I_{40}-I_{41}) \omega^2 - m^2 \left[ m^2\left( I_{00} - \beta \, I_{10} \right) a^2 - 2I_{20}\, b^2 \right]\right\} \right) \label{eq:energydens_kin_th}
\end{align}
is the energy density.
Furthermore, $P$ is the thermodynamic pressure, which will be explicitly computed in Sec.~\ref{sec3b}, with the result
\begin{align} \label{eq:thdyn_press}
P := 4 \cosh \alpha \left\{ I_{21} + \frac{\hbar^2 \beta^2}{8m^2}\left[ 2 I_{42} \, \varkappa^2 + \left( I_{41} - 3 I_{42} \right) \omega^2 + m^2\left( m^2 \, I_{00}\, a^2 + 2 I_{21}\, b^2 \right) \right] \right\} \;.
\end{align}
Note that, by applying Eqs.~\eqref{eq:useful_identities} to Eq.~\eqref{eq:chargedens}, we readily obtain
\begin{equation} \label{eq:idgas_id}
    n = \beta\, P\, \tanh \alpha\;,
\end{equation}
which is the usual relation between pressure and particle-number density for an ideal gas of Boltzmann particles.
We also defined a ``Hall-like'' energy current
\begin{align}\label{eq:qtimes}
q_{\times}^\mu &:= \Delta^{\mu \alpha} T_{\alpha \beta} u^\beta = - 4 \cosh \alpha \, \frac{\hbar^2 \beta^2}{4 m^2}\left(  I_{41}\, L^\mu + m^2 I_{20}\, K^\mu \right) \;,
\end{align}
as well as
\begin{equation}\label{eq:Pi1_Pi2}
\Pi_\varpi := -4 \cosh \alpha \, \frac{\hbar^2 \beta^2}{4}\, I_{21}\;, \quad \Pi_\Omega := -4 \cosh \alpha \,\frac{\hbar^2 \beta^2}{4m^2}\, I_{42}\;.
\end{equation}
Due to the terms in the last line of Eq.~\eqref{eq:en_mom_tensor_kin_th_2}, the pressure tensor becomes anisotropic at order $\mathcal{O}(\hbar^2)$.
The isotropic pressure is defined as $P_{\text{iso}} := - \frac{1}{3} \Delta_{\mu \nu} T^{\mu \nu}$ and is therefore computed as
\begin{align}\label{eq:isotropic_pressure}
P_{\text{iso}} & = 4\cosh\alpha \left( I_{21}+\frac{\hbar^2\beta^2}{24m^2} \left\{ 10 I_{42}\,\varkappa^{2}+(3I_{41}-5I_{42})\,\omega^{2}
+ m^2 \left[ \left(3I_{20} - 5 I_{21}\right) a^2 + 10 I_{21} \, b^2 \right] \right\} \right)\;.
\end{align}
Note that 
\begin{equation} \label{eq:P_iso_P}
    P_{\text{iso}} \equiv P- \frac{2}{3} \left[ \Pi_\varpi \left( a^2 + b^2 \right) + \Pi_\Omega \left( \varkappa^2 + \omega^2 \right) \right]\;.
\end{equation}
The pressure in the direction of $\omega^\mu$ is computed as
\begin{align}
P_{\omega} := \frac{\omega_\mu \omega_\nu}{\omega^2}\, T^{\mu \nu}  = P - \Pi \, \xi^2\;,
\end{align}
where $\Pi := \Pi_\varpi + \Pi_\Omega$.
The pressure in the direction of $\ell^\mu$ is
\begin{align}
P_{\ell}  &:= \ell_\mu \ell_\nu T^{\mu \nu}  = P - \Pi \, (\varkappa^2 + \omega^2)\;. 
\end{align}
Finally, the pressure in the direction of $\xi^\mu$ (cf.~Eq.~\eqref{eq:def_xi}) is computed as
\begin{align}
P_{\xi}  &:= \frac{\xi_\mu \xi_\nu}{\xi^2}\, T^{\mu \nu}  = P - \Pi \, (\varkappa^2 + \omega^2 - \xi^2)\;.   
\end{align}
Obviously,
\begin{align}
P_{\omega} + P_{\ell} + P_{\xi}= 3 P- 2 \Pi\,(\varkappa^2 + \omega^2) = 3 P_{\rm iso}\;,
\end{align}
in accordance with Eq.~\eqref{eq:P_iso_P}.

\subsection{Spin-current tensor}
In this subsection, we compute the spin-current tensor
\eqref{eq:spin-current_tensor}.
This quantity is multiplied by a factor of $\hbar$ and thus needs to be computed only up to order $\mathcal{O}(\hbar)$.
Therefore, the $\mathcal{O}(\hbar^2)$ mass-shell correction can be neglected.
We obtain
\begin{align}\label{eq:spin-current_tensor_2}
S^{\lambda,\mu \nu}& = 4 \cosh \alpha\, \frac{\hbar \beta}{4 m^2}  \left\{ u^\lambda \left[ (I_{30}- I_{31}) \epsilon^{\mu \nu \alpha \beta} u_\alpha \omega_\beta  -  2 I_{31}( u^\mu \varkappa^\nu - u^\nu \varkappa^\mu) \right] \right. \nonumber\\
&  \hspace*{2.5cm} + \left.  I_{31} (\Delta^{\lambda \mu} \varkappa^\nu - \Delta^{\lambda \nu} \varkappa^\mu  - u^\mu \epsilon^{\nu \lambda \alpha \beta} u_\alpha \omega_\beta  + u^\nu \epsilon^{\mu \lambda \alpha \beta} u_\alpha \omega_\beta) \right\}\;.
\end{align}
From this, we compute the spin tensor as
\begin{align} 
S^{\mu \nu} & := u_\lambda S^{\lambda, \mu \nu} = 4 \cosh \alpha\, \frac{\hbar \beta}{4 m^2}  \left[ (I_{30}- I_{31}) \epsilon^{\mu \nu \alpha \beta} u_\alpha \omega_\beta - 2 I_{31}( u^\mu \varkappa^\nu - u^\nu \varkappa^\mu) \right]\;. \label{eq:spintensor}
\end{align}

%
%%%%%%%%%%%%%%%%%%%%%%%%%%%%%%%%%%%%%%%%%%%%%%%%%%%%%%%%%%%%%%%%%%%%%%%%%%%%%%%%%%

\subsection{Entropy density}
\label{sec:entropydens_KT}
Given the expression~\eqref{eq:thdyn_press} for the thermodynamic pressure, the entropy density can be computed via the Euler relation~\eqref{eq:ortho_euler}.
In this way, using Eqs.~\eqref{eq:chargedens}, \eqref{eq:energydens_kin_th}, and \eqref{eq:spintensor}, we obtain
\begin{align}\label{Eq:kineticentropydens}
s&  = 4\beta\cosh\alpha\left\{I_{20}+ I_{21}+\frac{\hbar^{2}\beta^{2}}{8m^{2}}\left[2(I_{41}-I_{42})\varkappa^{2}+(I_{40}-2I_{41} + 3 I_{42})\omega^{2}+m^{2}\beta(m^{2}I_{10}a^{2}+2 I_{31}b^{2})\right]\right\}\nonumber \\
&-\, 4\mu\beta\sinh\alpha\left\{I_{10}+\frac{\hbar^{2}\beta^{2}}{8m^{2}}\left[2I_{31}\varkappa^{2}+(I_{30}-I_{31})\omega^{2}+m^{2}\beta(m^{2}I_{00}a^{2}+2I_{21}b^{2})\right]\right\}\;.
\end{align}

%%%%%%%%%%%%%%%%%%%%%%%%%%%%%%%%%%%%%%%%%%%%%%%%%%%%%%%%%%%%%%%%%%%%%%%%%%%%%%%%%%%%%%%%%%%%
%

\subsection{Thermodynamic relations}
\label{sec3b}
In this subsection, we confirm that the Maxwell relations~\eqref{eq:Maxwell} are fulfilled and compute the thermodynamic pressure and thermodynamic relations in the kinetic-theory pseudo-gauge.
We then show that the thermodynamic relations~\eqref{eq:ortho_thdyn_densities_as_ders} are fulfilled.

The validity of the Maxwell relations~\eqref{eq:Maxwell} can be straightforwardly checked using Eqs.~\eqref{eq:chargedens}, \eqref{eq:spintensor}, and \eqref{Eq:kineticentropydens}, together with the relations~\eqref{eq:useful_identities}.
Since the calculation is somewhat lengthy, we defer this to App.~\ref{app:Maxwell_KT} for the sake of conciseness.
The conclusion is that the Maxwell relations~\eqref{eq:Maxwell} \textit{are fulfilled} in the kinetic-theory pseudo-gauge.

The validity of the Maxwell relations ensures that the computation of the thermodynamic pressure via an integral relation is independent of the path in $(T,\mu,\omega_{\mu\nu})$ space, e.g., it can be computed either from Eq.~\eqref{eq:p_as_int_temp} or from \eqref{eq:thdynpress_as_int_over_enerdens}.
In the first case, we require the (net) particle-number density, the energy density, and the spin tensor from Eqs.~\eqref{eq:chargedens}, \eqref{eq:energydens_kin_th}, and \eqref{eq:spintensor}, respectively, and integrate at fixed $\mu$ and $\omega_{\sigma \tau}$ (and $u_\rho$).
In the second case, one only needs the energy density~\eqref{eq:energydens_kin_th}  and one integrates at fixed $\alpha$ and $\Omega_{\sigma \tau}$ (and $u_\rho$).
In both cases, one can also exchange the $T'$ (or $\beta'$) integration with the phase-space integration, i.e., one employs the general expressions \eqref{eq:ortho_thdyn_densities}, with the currents from Eqs.~\eqref{eq:currents_as_ders_kintheory_PG}, and first performs the $T'$ (or $\beta'$) integral and then the integral over phase space.
All approaches give identical results, namely Eq.~\eqref{eq:thdyn_press}.
However, when first integrating over phase space and then over $T'$ (or $\beta'$), some care has to be taken to distinguish the reduced spin potential $\Omega^{\mu \nu}$ from the thermal vorticity $\varpi^{\mu \nu}$, as already mentioned above.
Although both are equal in global thermodynamic equilibrium, one always needs to keep $\varpi_{\mu \nu}$ constant.
For the $\beta'$ integral in Eq.~\eqref{eq:thdynpress_as_int_over_enerdens}, this is more natural, as also $\Omega_{\mu \nu}$ is kept constant.
However, for the $T'$ integral in Eq.~\eqref{eq:p_as_int_temp} this is less obvious, as the thermal vorticity, being a function of gradients of $\beta^\mu \equiv u^\mu/T$, in principle involves the temperature.
At this point, we view this as a working prescription to obtain consistent results.
A more general discussion will be presented elsewhere~\cite{shokri_in_prep}.
The details of the calculation of the thermodynamic pressure can be found in App.~\ref{AppTPHIUp}.

From the thermodynamic pressure, one can derive the (net) particle-number density via Eq.~\eqref{eq:n_as_der_p}.
As mentioned in the context of that equation, since $T$ and $u_\rho$ are supposed to be kept constant when performing the partial derivative, also the thermal vorticity has to be kept constant.
With the help of Eqs.~\eqref{eq:useful_identities}, one confirms that the result agrees with Eq.~\eqref{eq:chargedens}, or Eq.~\eqref{eq:idgas_id}, respectively.

Next we consider the thermodynamic relation \eqref{eq:ortho_spin_as_der_p}.
Using Eqs.~\eqref{eq:thdyn_press}, and \eqref{eq:decomp_derivs}, we compute
\begin{align}\label{Eq:kinetictherspin}
   \left. \frac{\partial P}{\partial \omega_{\mu \nu}} \right|_{T,\mu,\varpi_{\sigma \tau}}  & = 4 \cosh \alpha\, \frac{\hbar^2 \beta^2}{4 m^2}  \left[ (I_{41}- 3I_{42}) \epsilon^{\mu \nu \alpha \beta} u_\alpha \omega_\beta - 2 I_{42}( u^\mu \varkappa^\nu - u^\nu \varkappa^\mu) \right]\;.
\end{align}
With Eq.~\eqref{eq:identity_1}, this is seen to be equal to $\hbar S^{\mu \nu}$, proving the validity of the thermodynamic relation \eqref{eq:ortho_spin_as_der_p}.
Since $T$ and $u_\rho$ are supposed to be kept constant when performing the partial derivative with respect to $\omega_{\mu \nu}$, we have again kept the thermal vorticity constant in the above calculation.
For that reason it is also important to distinguish $\Omega^{\mu \nu}$ from $\varpi^{\mu \nu}$ prior to taking derivatives with respect to $\Omega_{\mu \nu}$ (or $\omega_{\mu \nu}$).

Finally, one convinces oneself that the entropy density~\eqref{Eq:kineticentropydens} can also be obtained via Eq.~\eqref{eq:s_as_der_p}.
Again, the thermal vorticity has to be kept constant when taking the partial derivative with respect to $T$.
Taking Eq.~\eqref{eq:s_as_der_p} at face value, one is supposed to keep $\mu$, $\omega_{\sigma \tau}$, and $u_\rho$ constant.
One would therefore be tempted to also take derivatives of the thermal vorticity with respect to $T$, as it is a function of gradients of $\beta^\mu \equiv u^\mu/T$.
As this would lead to a discrepancy with the computation of the entropy density via the Euler relation in Sec.~\ref{sec:entropydens_KT}, this is, however, not correct: the thermal vorticity has to be kept constant both when computing the thermodynamic pressure via its integral representation~\eqref{eq:p_as_int_temp} and when taking derivatives of the thermodynamic pressure. 

%stopped here 13:01, 07-08-2026
\section{Canonical pseudo-gauge}
\label{Sec-can-psuedo}
In this section, we derive the (net) particle-number current, the energy-momentum tensor, and the spin-current tensor, as well as related thermodynamic quantities, in the canonical pseudo-gauge. 
\subsection{Particle-number current}
The particle-number current in the canonical pseudo-gauge is given by Eq.~\eqref{eq:part_curr_can}.
Inserting Eq.~\eqref{eq:Vmu} as well as Eqs.~\eqref{eq:F_and_A} results in
\begin{equation}\label{eq:canonical-J}
    J_C^\mu = J^\mu - \frac{1}{m} \int \mathrm{d}\Gamma \left( k^\mu\, \frac{\hbar^2}{4m^2}\, \Box_x  - \frac{\hbar}{2}\, \Sigma_{k \mathfrak{s}}^{\mu \nu} \partial^x_\nu \right) \frac{\partial \, \mathfrak{f}(x,k,\mathfrak{s})}{\partial \alpha}\;,
\end{equation}
where $J^\mu$ is the (net) particle-number current in the kinetic-theory pseudo-gauge.
We insert the phase-space distribution function \eqref{eq:frak_f} and, as the first term under the integral is already of order $\mathcal{O}(\hbar^2)$, we can neglect corrections of order $\hbar$, while for the second term we need to expand the single-particle distribution functions \eqref{eq:sing_part_dist} to first order in $\hbar$.
Then, performing the spin-space integration with the help of Eqs.~\eqref{eq:spin-space_int}, we arrive at
\begin{equation}
    J_C^\mu = J^\mu + 4 \sinh \alpha \left(\mathcal{J}_3^\mu + \mathcal{J}_4^\mu \right)\;,
\end{equation}
where
\begin{subequations}
    \begin{align}
        \mathcal{J}_3^\mu & : = -
    \frac{\hbar^2}{4m^2} \, \varpi_{\lambda\rho} \varpi^{\lambda}_{\hspace*{0.1cm} \sigma}
    \int \di K\, 4 \pi \hbar\, \theta(k_0) \, \delta(k^2 - m^2)\, k^\mu k^\rho k^\sigma\, e^{-\beta \cdot k}\;, \\
    \mathcal{J}_4^{\mu} & :=  -\frac{\hbar^2}{4m^2} \,   \varpi_{\lambda \rho} \, \Omega^{\lambda}_{\hspace*{0.1cm} \sigma} 
    \int \di K\, 4 \pi \hbar\, \theta(k_0) \, \delta(k^2 - m^2)\, \left( g^{\mu \sigma} m^2 - k^\mu k^\sigma \right) k^\rho \, e^{-\beta \cdot k}\;.
    \end{align}
\end{subequations}
Here, we have made use of the global-equilibrium conditions \eqref{eq:gteq_conditions} (while, as mentioned earlier, still carefully distinguishing between reduced spin potential and thermal vorticity), as well as Eq.~\eqref{eq:wave_eq_F} with Eq.~\eqref{eq:F_eq}.
Since these terms are already of order $\mathcal{O}(\hbar^2)$, the $\mathcal{O}(\hbar^2)$ modification \eqref{eq:massshell_mod} of the mass-shell constraint can be neglected, as it is of higher order.
Using Eqs.~\eqref{eq:tensor_decomp} and \eqref{eq:useful_identities} we arrive at
\begin{subequations}
    \begin{align}
        \mathcal{J}_3^\mu & = \frac{\hbar^2 \beta^2}{4 m^2} \left\{ \left[ \left(I_{30} - I_{31} \right) a^2 + 2 I_{31} \, b^2 \right] u^\mu - 2 I_{31}\, K^\mu \right\} \;, \\
        \mathcal{J}_4^\mu & = \frac{\hbar^2 \beta^2}{4 m^2} \left[2 I_{31} u^\mu \left( a \cdot \varkappa + b \cdot \omega \right) + \left( I_{30} - 2 I_{31} \right) M^\mu + I_{31}\, N^\mu \right]\;.
    \end{align}
\end{subequations}
The canonical particle-number current can thus be written as
\begin{equation}
    J_C^\mu = n_C \, u^\mu + n_{C \times}^\mu\;,
\end{equation}
where
\begin{equation} \label{eq:dens_can}
    n_C : = u_\mu J_C^\mu  = n + 4 \sinh \alpha \, \frac{\hbar^2\beta^2}{4m^2} \left[ \left(I_{30} - I_{31} \right) a^2 + 2I_{31} \left(b^2 + a \cdot \varkappa + b \cdot \omega \right) \right]
\end{equation}
is the (net) particle-number density in the canonical pseudo-gauge and
\begin{equation}
    n_{C\times}^\mu = n_\times^\mu + 4 \sinh \alpha \, \frac{\hbar^2\beta^2}{4m^2} \left[\left( I_{30} - 2 I_{31} \right) M^\mu + I_{31}\, N^\mu - 2I_{31}\, K^\mu \right]\;.
\end{equation}
Equation~\eqref{eq:dens_can} clearly proves the statement made in  the introduction that local thermodynamic quantities are pseudo-gauge dependent.

In global thermodynamic equilibrium (GTE), $a^\mu \equiv \varkappa^\mu$, $b^\mu \equiv \omega^\mu$, and these expressions simplify with the help of Eq.~\eqref{eq:identity_2},
\begin{subequations}
    \begin{align}
        \mathrm{GTE:} \qquad n_C & = n + 4 \sinh \alpha \, \frac{\hbar^2 \beta^2}{4 }\, I_{10}\, \varkappa^2 \;, \\
        \mathrm{GTE:} \qquad n_{C\times}^\mu & = n_\times^\mu + 4 \sinh \alpha \, \frac{\hbar^2 \beta^2}{4 }\, I_{10}\, L^\mu\;.
    \end{align}
\end{subequations}

\subsection{Energy-momentum tensor}
The energy-momentum tensor in the canonical pseudo-gauge is given by Eq.~\eqref{eq:energy-mom_tensor_can}. 
One immediately observes that up to order $\mathcal{O}(\hbar^2)$
\begin{equation}
    T^{\mu \nu}_C = T^{\mu \nu} - \frac{1}{m} \int \di \Gamma \left( k^\mu k^\nu   \frac{\hbar^2}{4m^2}\, \Box_x -
    \frac{\hbar}{2} \, \Sigma_{k\mathfrak{s}}^{\mu \lambda} k^\nu \partial_\lambda^x \right) \mathfrak{f}(x,k,\mathfrak{s})  \;,
\end{equation}
where $T^{\mu \nu}$ is the energy-momentum tensor in the kinetic-theory pseudo-gauge.
We insert the phase-space distribution function \eqref{eq:frak_f} and, for the second term under the integral, expand the single-particle distribution functions \eqref{eq:sing_part_dist} to first order in $\hbar$.
Performing the spin-space integration with the help of Eqs.~\eqref{eq:spin-space_int}, we then arrive at 
\begin{align}
    T^{\mu \nu}_C & = T^{\mu \nu} +
    4 \cosh \alpha \left( \mathcal{T}_3^{\mu \nu} + \mathcal{T}_4^{\mu \nu} \right)\;,
\end{align}
where
\begin{align}
    \mathcal{T}_3^{\mu \nu} & := -
    \frac{\hbar^2}{4m^2} \, \varpi_{\lambda\rho} \varpi^{\lambda}_{\hspace*{0.1cm} \sigma}
    \int \di K\, 4 \pi \hbar\, \theta(k_0) \, \delta(k^2 - m^2)\, k^\mu k^\nu k^\rho k^\sigma\, e^{-\beta \cdot k}\;, \\
    \mathcal{T}_4^{\mu \nu} & :=  -\frac{\hbar^2}{4m^2} \,   \varpi_{\lambda \rho} \, \Omega^{\lambda}_{\hspace*{0.1cm} \sigma} 
    \int \di K\, 4 \pi \hbar\, \theta(k_0) \, \delta(k^2 - m^2)\, \left( g^{\mu \sigma} m^2 - k^\mu k^\sigma \right) k^\nu k^\rho \, e^{-\beta \cdot k}\;.
\end{align}
Here, we have made use of the global-equilibrium conditions \eqref{eq:gteq_conditions} (but keeping the distinction between $\Omega_{\mu \nu}$ and $\varpi_{\mu \nu}$), as well as Eq.~\eqref{eq:wave_eq_F} with Eq.~\eqref{eq:F_eq}.
Since these terms are already of order $\mathcal{O}(\hbar^2)$, the $\mathcal{O}(\hbar^2)$ modification \eqref{eq:massshell_mod} of the mass-shell constraint can be neglected, as it is of higher order.
Using Eqs.~\eqref{eq:tensor_decomp} and \eqref{eq:useful_identities} we arrive at
\begin{subequations}
\begin{align}
    \mathcal{T}_3^{\mu \nu} & =  \frac{\hbar^2 \beta^2}{4 m^2} \left\{ \left[ \left(I_{40}-I_{41} \right)a^2 +
    2 I_{41}\, b^2 \right] u^\mu u^\nu 
    - \left[ \left( I_{41} - I_{42} \right) a^2 + 4 I_{42}\, b^2 \right] \Delta^{\mu \nu} \right. \nonumber \\
    & \hspace*{1.4cm} - \left. 
    2 I_{41}\, \left( u^\mu K^\nu + u^\nu K^\mu \right) - 2 I_{42} \left( a^\mu a^\nu + b^\mu b^\nu \right) \right\}\;, \\
    \mathcal{T}_4^{\mu \nu} & =  \frac{\hbar^2 \beta^2}{4 m^2} \left\{ \left[  2I_{41} \, u^\mu u^\nu -  \left(I_{41} - I_{42} \right) \Delta^{\mu \nu}\right] \left( a \cdot \varkappa + b \cdot \omega \right)
    + I_{41} \left[ u^\mu \left( M^\nu + N^\nu \right) + u^\nu \left(M^\mu + N^\mu \right) \right] \right. \nonumber \\
    & \hspace*{1.4cm} + \left. I_{42} \left( a^\mu \varkappa^\nu + a^\nu \varkappa^\mu + b^\mu \omega^\nu + b^\nu \omega^\mu \right)
    + m^2\, I_{20} \, u^\nu M^\mu
    - m^2\, I_{21}\, \left( u^\mu N^\nu  - a^\nu \varkappa^\mu - b^\mu \omega^\nu \right) \right\}\;.
\end{align}
\end{subequations}
Note that, on account of the last three terms in $\mathcal{T}_4^{\mu \nu}$, the canonical energy-momentum tensor is no longer symmetric.
The canonical energy-momentum tensor thus reads
\begin{align} \label{eq:en_mom_tensor_can}
T_C^{\mu \nu} & = \varepsilon_C \, u^\mu u^\nu - P_C\, \Delta^{\mu \nu} + q_{C\times}^\mu u^\nu + \tilde{q}_{C\times}^\nu u^\mu \nonumber \\
& +\Pi_{C\varpi} \left[ \Delta^{\mu \nu} (a^2 + b^2)
+ a^\mu a^\nu + b^\mu b^\nu \right]  +\Pi_{\Omega} \left[ \Delta^{\mu \nu} (\varkappa^2 + \omega^2)
+ \varkappa^\mu \varkappa^\nu + \omega^\mu \omega^\nu \right] \nonumber \\
& + \Pi_{a\varkappa} \left[ - \Delta^{\mu \nu} \left( a \cdot \varkappa + b\cdot \omega \right) + a^\mu \varkappa^\nu + \omega^\mu b^\nu \right]
+ \Pi_{b \omega}\left[ - \Delta^{\mu \nu} \left( a \cdot \varkappa + b \cdot \omega \right) + \varkappa^\mu a^\nu + b^\mu \omega^\nu \right] \;.
\end{align}
Here, 
\begin{align} 
\varepsilon_C &:= u_\mu u_\nu  T_C^{\mu \nu} = \varepsilon + 4 \cosh \alpha \, \frac{\hbar^2 \beta^2}{4 m^2}\left[ \left( I_{40} - I_{41} \right) a^2 + 2 I_{41} \left( b^2 + a \cdot \varkappa + b \cdot \omega \right) \right]
\label{eq:energydens_can}
\end{align}
is the canonical energy density and
\begin{align} \label{eq:thdyn_press_can}
P_C := P + 4 \cosh \alpha \, \frac{\hbar^2 \beta^2}{4m^2}\left[  \left( I_{41} - 3 I_{42} \right) a^2 + 2 I_{42} \left( b^2 + a \cdot \varkappa + b \cdot \omega \right) \right] 
\end{align}
is the thermodynamic pressure computed from Eq.~\eqref{eq:thdynpress_as_int_over_enerdens}, as will be shown in Sec.~\ref{thermocann}.
As one readily proves using Eqs.~\eqref{eq:useful_identities}, $P_C$ fulfills the ideal-gas equation of state,
\begin{equation}
n_C = \beta P_C \, \tanh \alpha\;,
\end{equation}
with $n_C$ given by Eq.~\eqref{eq:dens_can}.

In Eq.~\eqref{eq:en_mom_tensor_can} we also defined 
\begin{subequations}
\begin{align}\label{eq:qCtimes}
q_{C \times}^\mu &:= \Delta^{\mu \alpha} T_{C,\alpha \beta} u^\beta = q_\times^\mu- 4 \cosh \alpha \, \frac{\hbar^2 \beta^2}{4 m^2}\left[2 I_{41}\, K^\mu - \left(I_{40}-2 I_{41} \right) M^\mu - I_{41} N^\mu \right] \;, \\
\tilde{q}_{C \times}^\mu & := u^{\alpha} T_{C,\alpha \beta} \Delta^{\beta \mu} =  q_\times^\mu - 4 \cosh \alpha \, \frac{\hbar^2 \beta^2}{4 m^2}\left[2 I_{41}\, K^\mu - I_{41}\, M^\mu -5 I_{42} N^\mu \right] \;, \\
\Pi_{C\varpi} & :=  4 \cosh \alpha \, \frac{\hbar^2 \beta^2}{4m^2}\left( I_{41} - 7 I_{42}\right)\;, \\
\Pi_{a\varkappa} & :=  4 \cosh \alpha \, \frac{\hbar^2 \beta^2}{4m^2}\, I_{42}\;, \\
\Pi_{b\omega} & :=  4 \cosh \alpha \, \frac{\hbar^2 \beta^2}{4m^2}\left( I_{41} - 4I_{42}\right)\;. 
\end{align}
\end{subequations}
The canonical isotropic pressure is defined as $P_{C, \text{iso}} := - \frac{1}{3} \Delta_{\mu \nu} T_C^{\mu \nu}$ and is therefore computed as 
\begin{align}\label{eq:isotropic_pressure_can}
P_{C, \text{iso}} & = P_{\text{iso}} + 4\cosh\alpha \, \frac{\hbar^2\beta^2}{12 m^2} \left[ \left( 3 I_{41} -  5I_{42} \right) a^2 + 10 I_{42}\,b^{2}+2 I_{41} \left( a \cdot \varkappa + b \cdot \omega \right) \right]\;.
\end{align}
Equations~\eqref{eq:energydens_can} and \eqref{eq:thdyn_press_can} explicitly corroborate the statement made in the introduction that local thermodynamic quantities are pseudo-gauge dependent.

\subsection{Spin-current tensor}
The spin-current tensor in the canonical pseudo-gauge is given by Eq.~\eqref{eq:spin-current_tensor_can}.
Inserting the phase-space distribution function \eqref{eq:frak_f}, expanding the single-particle distribution functions \eqref{eq:sing_part_dist} to first order in $\hbar$, and performing the spin-space integration with the help of Eqs.~\eqref{eq:spin-space_int} we arrive at
\begin{align}
S_C^{\lambda, \mu \nu} & = 4 \cosh \alpha\, \frac{\hbar \beta}{4}\, I_{10}\left( u^\lambda \omega^{\mu \nu} + u^{\mu} \omega^{\nu \lambda} + u^\nu \omega^{\lambda \mu} \right)\;.
\end{align}
From this, we compute the spin tensor using the decomposition \eqref{eq:decomp_red_spin_potential} as
\begin{equation} \label{eq:spintensor_can}
    S_C^{\mu \nu} = u_\lambda\, S_C^{\lambda, \mu \nu} = 4 \cosh \alpha\, \frac{\hbar \beta}{4}\, I_{10}\, \epsilon^{\mu \nu \alpha \beta} u_\alpha \omega_\beta\;.
\end{equation}

\subsection{Entropy density}
\label{sec:entropydens_C}
Using the Euler relation~\eqref{eq:ortho_euler} with Eqs.~\eqref{eq:dens_can}, \eqref{eq:energydens_can}, \eqref{eq:spintensor_can}, and the thermodynamic pressure~\eqref{eq:thdyn_press_can}, we derive the entropy density as
\begin{align}
   s_C & = \frac{1}{T}\left( \varepsilon_C+P_C - \mu n_C - \frac{\hbar}{2}\, \omega_{\nu \lambda} S_C^{\nu \lambda}\right)  \nonumber \\
   & =  s -4\mu\sinh\alpha\frac{\hbar^{2} \beta^3}{4m^{2}}\left[(I_{30}-I_{31})a^{2}+2I_{31} \left(b^{2}+ a \cdot\varkappa+  b \cdot \omega\right)\right]\nonumber\\
&+4\cosh\alpha\frac{\hbar^{2}\beta^3}{4m^{2}}\left[(I_{40}-3I_{42})a^{2}+2(I_{41}+I_{42}) \left(b^{2} + a \cdot \varkappa + b \cdot \omega\right)
- 2 I_{42}( \varkappa \cdot \varkappa + \omega \cdot \omega)\right]    \;,
\label{eq:entropydens_can_0}
\end{align} 
where $s$ is the entropy density~\eqref{Eq:kineticentropydens} in kinetic-theory pseudo-gauge.

\subsection{Thermodynamic relations}
\label{thermocann}
In this subsection, we check the validity of the Maxwell relations~\eqref{eq:Maxwell} in the canonical pseudo-gauge. 
We then derive the thermodynamic pressure from Eqs.~\eqref{eq:p_as_int_temp} and \eqref{eq:thdynpress_as_int_over_enerdens}.
Finally, we discuss the validity of the thermodynamic relations~\eqref{eq:ortho_thdyn_densities_as_ders}.

The explicit calculation of the Maxwell relations is deferred to App.~\ref{app:maxwell_canonical}.
There, we prove that they are in general \textit{not fulfilled} unless one imposes the global thermodynamic equilibrium conditions $a^\mu = \varkappa^\mu$ and $b^\mu = \omega^\mu$, but only in \textit{carefully selected} terms of the thermodynamic quantities.

Computing the canonical thermodynamic pressure via Eq.~\eqref{eq:thdynpress_as_int_over_enerdens} with the energy density \eqref{eq:energydens_can} immediately gives Eq.~\eqref{eq:thdyn_press_can}, on account of the identity~\eqref{eq:identity_3a}.
Now let us compute the thermodynamic pressure via Eq.~\eqref{eq:p_as_int_temp}.
Substituting the canonical (net) particle-number density~\eqref{eq:dens_can}, the canonical energy density~\eqref{eq:energydens_can}, and the canonical spin tensor~\eqref{eq:spintensor_can} into the integrand of Eq.~\eqref{eq:p_as_int_temp}, we obtain
\begin{align}
   \lefteqn{\varepsilon_C - \mu n_C - \frac{\hbar}{2}\, \omega_{\nu \lambda} S_C^{\nu \lambda}  
   = \varepsilon - \mu \, n - \frac{\hbar}{2}\, \omega_{\nu \lambda} S^{\nu \lambda} }\nonumber \\
  & + 4 \frac{\hbar^2}{4 m^2} \left\{ \left[ \cosh (\beta \mu) (I_{40} - I_{41}) - \mu \sinh (\beta \mu) (I_{30} - I_{31}) \right] \beta^2 a^2 + 2 \left[ \cosh(\beta \mu) I_{41} - \mu \sinh (\beta \mu)  I_{31} \right] \beta^2 b^2 \right\} \nonumber \\
  & 
  + 4 \frac{\hbar^2}{4 m^2} \, 2 \left\{ \left[\cosh(\beta \mu) \beta \, I_{41} - \mu\, \sinh(\beta \mu) \beta\,  I_{31} \right] \left(\beta a \cdot \varkappa  + \beta b \cdot \omega \right)  - \cosh (\beta \mu) \, I_{31} \left( \beta\varkappa \cdot \varkappa +  \beta\omega \cdot \omega\right) \right\}\;.
  \label{eq:intermediate}
\end{align}
When inserting this expression into Eq.~\eqref{eq:p_as_int_temp}, the first line immediately integrates to give the thermodynamic pressure~\eqref{eq:thdyn_press} in kinetic-theory pseudo-gauge.
Since the spin potential and the thermal vorticity have to be kept constant when performing the $T'$ integral, the terms $\beta^2 a^2$, $\beta^2 b^2$ in the second line, as well as $\beta a \cdot \varkappa$, $\beta b \cdot \omega$, $\varkappa \cdot \varkappa$ and $\omega \cdot \omega$ in the third line are kept constant when integrating over $T'$ (note that in all other terms, $\beta = 1/T$ is replaced by $\beta'= 1/T'$ when doing the integration).
Using Eq.~\eqref{eq:betaprime_int} one observes that the second line integrates to a closed form.
However, the last line does not, unless one replaces $\beta \varkappa_\mu$ and $\beta \omega_\mu$ by $\beta a_\mu$ and $\beta b_\mu$ in the last term in the third line (so that they can be kept constant when integrating over $T'$).
With this replacement, one obtains the pressure $P_C$, cf.\ Eq.~\eqref{eq:thdyn_press_can}.
While this replacement can be motivated by the conditions of global thermodynamic equilibrium, it still remains an \textit{ad hoc} prescription (for instance, one may rightfully ask why this replacement should not also be made in other terms involving $\beta \varkappa_\mu$ and $\beta \omega_\mu$).
Without this replacement, the thermodynamic pressure computed from Eq.~\eqref{eq:p_as_int_temp}  does not equal $P_C$.
Since the Maxwell relations~\eqref{eq:Maxwell} are not fulfilled (unless similar replacements are enacted, cf.~App.~\ref{app:maxwell_canonical}), it is not surprising that the two ways to compute the thermodynamic pressure do not agree.
A more detailed calculation of the thermodynamic pressure is deferred to App.~\ref{AppTPHIUp}.

Using Eq.~\eqref{eq:thdyn_press_can} and the identity~\eqref{eq:identity_1} one immediately proves that Eq.~\eqref{eq:n_as_der_p} holds in canonical pseudo-gauge.
We then check the validity of Eq.~\eqref{eq:ortho_spin_as_der_p}.
Using Eqs.~\eqref{eq:thdyn_press}, \eqref{eq:thdyn_press_can}, and \eqref{eq:decomp_derivs} we have
\begin{align}
  \left. \frac{\partial P_C}{\partial \omega_{\nu \lambda}} \right|_{T,\mu,\varpi_{\sigma \tau}} 
    & = 4 \cosh \alpha\, \frac{\hbar^2 \beta^2}{4 m^2}\left\{ 2I_{42} \left[ \left(a^\lambda - \varkappa^\lambda \right)u^\nu -\left(a^\nu - \varkappa^\nu \right)u^\lambda\right] 
    - \epsilon^{\nu \lambda \alpha \beta} u_\alpha \left[ 2I_{42}\, b_\beta - (I_{41}-3 I_{42}) \omega_\beta\right] \right\}\;.
\end{align}
This is not yet identical to Eq.~\eqref{eq:spintensor_can}, but employing $a^\mu \equiv \varkappa^\mu$, $b^\mu \equiv \omega^\mu$, valid in global thermodynamic equilibrium, as well as the identities \eqref{eq:useful_identities}, this agrees with the expression~\eqref{eq:spintensor_can} for the canonical spin tensor.

Finally, we check the validity of Eq.~\eqref{eq:s_as_der_p}.
Using Eqs.~\eqref{eq:useful_identities} we obtain
\begin{align}
s_C  = \left. \frac{\partial P_C}{\partial T} \right|_{\mu, \omega_{\sigma \tau}, \varpi_{\sigma \tau}} & = s-4\mu\sinh\alpha\frac{\hbar^{2} \beta^3}{4m^{2}}\left[(I_{30}-I_{31})a^{2}+2I_{31} \left( b^{2}+ a \cdot\varkappa+  b \cdot \omega\right)\right]\nonumber\\
& \quad \;\;  +4\cosh\alpha\frac{\hbar^{2} \beta^3}{4m^{2}}\left[(I_{40}-3I_{42})a^{2}+2(I_{41}+I_{42})b^{2}+2 I_{41}( a \cdot \varkappa + b \cdot \omega)\right]\;,
\label{eq:entropy_density_can}
\end{align}
This is not identical to Eq.~\eqref{eq:entropydens_can_0}.
However, in global thermodynamic equilibrium, we can replace one factor of $\varkappa_\mu$ and $\omega_\mu$ in the last term in the second line of Eq.~\eqref{eq:entropydens_can_0} with $a_\mu$ and $b_\mu$, respectively, and thus obtain agreement with Eq.~\eqref{eq:entropy_density_can}.
Thus, at least in global thermodynamic equilibrium, the thermodynamic identity~\eqref{eq:s_as_der_p} holds.

\section{GLW pseudo-gauge}
\label{Sec:GLWPG}
In this section, we derive the (net) particle-number current, the energy-momentum tensor, and the spin-current tensor, as well as related thermodynamic quantities, in the GLW pseudo-gauge.

\subsection{Particle-number current}
In the GLW pseudo-gauge, the (net) particle-number current takes the same form as in the kinetic-theory pseudo-gauge (see the discussion in Secs.~\ref{subSECGLWW}--\ref{sec:KT_PG}), 
\begin{align}
J^{\mu}_{\rm GLW}=J^{\mu}\,.
\end{align}
This corresponds to the choice $B^{\lambda\mu}_{\rm GLW\rightarrow kin}=0$ in Eq.~\eqref{eq:op_current_psgt}. 
Consequently, the (net) particle-number density and the transverse current density are given by
\begin{align}
n_{\rm GLW}=n\;,
\end{align}
and 
\begin{align}
n_{\rm GLW,\times}^{\mu}=n^{\mu}_{\times}\;,
\end{align}
where $n$ and $n^{\mu}_{\times}$ are given by Eqs.~\eqref{eq:chargedens} and~\eqref{eq:chargeHall}, respectively.

\subsection{Energy-momentum tensor}
As for the (net) particle-number current, the GLW energy-momentum tensor is identical to that in kinetic-theory pseudo-gauge (see the discussion in Secs.~\ref{subSECGLWW}--\ref{sec:KT_PG}),
\begin{align}
T^{\mu\nu}_{\rm GLW}=T^{\mu\nu}\;.
\end{align}
This can also be inferred from Eq.~\eqref{eq:op_emt_psgt}, since $\Phi_{\rm GLW\rightarrow kin}=0$. 
As a result, the GLW energy density is equal to that in the kinetic-theory pseudo-gauge (see also Eq.~\eqref{eq:pgt_energy_density}), 
\begin{align}\label{EQ:GLWEDEN}
\varepsilon_{\rm GLW}=\varepsilon\;,
\end{align}
where $\varepsilon$ is defined in Eq.~\eqref{eq:energydens_kin_th}. 
Similarly, the isotropic pressure is defined by the relation $P_{\rm iso}=-\tfrac{1}{3}\Delta_{\mu\nu}T^{\mu\nu}$, and thus, since the energy-momentum tensor is the same in GLW pseudo-gauge as in kinetic-theory pseudo-gauge, 
\begin{align}
P_{\rm GLW,iso}=P_{\rm iso}\;,
\end{align}
with $P_{\rm iso}$ given by Eq.~\eqref{eq:isotropic_pressure}.

Since the GLW energy density is identical to that in the kinetic-theory pseudo-gauge, the GLW thermodynamic pressure, computed via Eq.~\eqref{eq:thdynpress_as_int_over_enerdens} is also the same as that in the kinetic-theory pseudo-gauge, cf.~Eq.~\eqref{eq:thdyn_press},
\begin{align} \label{eq:thdyn_press_GLW}
P_{\rm GLW}=P\;.
\end{align}
Consequently, $P_{\rm GLW}$ also fulfills the ideal-gas equation of state
\begin{equation}
n_{\rm GLW} = \beta P_{\rm GLW} \, \tanh \alpha\;.
\end{equation}

\subsection{Spin-current tensor}
The spin-current tensor in the GLW pseudo-gauge is given by Eq.~\eqref{eq:spintensor_GLW}, i.e., it differs from the spin-current tensor in the kinetic-theory pseudo-gauge by an additive contribution,
\begin{equation}
    S_{\mathrm{GLW}}^{\lambda, \mu \nu} = S^{\lambda, \mu \nu} - \frac{\hbar}{4 m^3} \int \mathrm{d}\Gamma\, k^\lambda\left(k^\mu \partial^\nu_x - k^\nu \partial^\mu_x \right) \mathfrak{f}(x,k,\mathfrak{s})\;.
\end{equation}
We insert the phase-space distribution \eqref{eq:frak_f} with the equilibrium single-particle distribution function \eqref{eq:sing_part_dist}.
We only need the latter to zeroth order in $\hbar$, as its contribution to $\hbar S_{\mathrm{GLW}}^{\mu \nu}$ is already of order $\mathcal{O}(\hbar^2)$.
We then perform the integration over spin space using Eq.~\eqref{eq:spin-space_int} together with Eq.~\eqref{eq:tensor_decomp}, to obtain
\begin{align}
    S_{\mathrm{GLW}}^{\lambda, \mu \nu} & = S^{\lambda, \mu \nu} + 4 \cosh \alpha \, \frac{\hbar \beta}{4 m^2}\left\{ u^\lambda \left[ \left(I_{30}-I_{31} \right) \left(u^\mu a^\nu - u^\nu a^\mu\right) - 2I_{31}\, \epsilon^{\mu \nu\alpha \beta} u_\alpha b_\beta \right] \right. \nonumber \\
    & \hspace*{3.5cm} \left.
    + I_{31} \left[ a^\mu \Delta^{\nu \lambda} - a^\nu \Delta^{\mu \lambda} - \left(u^\mu \epsilon^{\nu \lambda \alpha \beta} - u^\nu \epsilon^{\mu \lambda \alpha \beta}\right) u_\alpha b_\beta \right] \right\}\;.
\end{align}
From this the spin tensor follows as
\begin{equation}\label{Eq:spindensitytenGLW}
    S_{\mathrm{GLW}}^{\mu \nu}  = S^{\mu \nu} + 4 \cosh \alpha \, \frac{\hbar \beta}{4 m^2}\left[ \left(I_{30}-I_{31} \right) \left(u^\mu a^\nu - u^\nu a^\mu\right) - 2I_{31}\, \epsilon^{\mu \nu\alpha \beta} u_\alpha b_\beta \right]\;.
\end{equation}

\subsection{Entropy density}
The entropy density in the GLW pseudo-gauge follows from the Euler relation~\eqref{eq:ortho_euler}. 
Since the energy density, thermodynamic pressure, and (net) particle-number density coincide with their kinetic-theory counterparts, we may use the Euler relation for the latter to write
\begin{align}
s_{\rm GLW}
&=
s
-
\beta\frac{\hbar}{2}
\omega_{\mu\nu}
\left(
S_{\rm GLW}^{\mu\nu}-S^{\mu\nu}
\right)\;.
\label{eq:sGLW_from_kin}
\end{align}
Using Eq.~\eqref{Eq:spindensitytenGLW}, one finds
\begin{align}
\frac{\hbar}{2}
\omega_{\mu\nu}
\left(
S_{\rm GLW}^{\mu\nu}-S^{\mu\nu}
\right)
&=
\frac{\hbar^2\beta}{m^2}\cosh\alpha
\left[
(I_{30}-I_{31})\,a \cdot \varkappa
+
2I_{31}\,b \cdot \omega
\right]\;.
\label{eq:spincontract_GLW}
\end{align}
%
%where we have used
%
%\begin{align}
%\frac{1}{2}
%\omega_{\mu\nu}
%\left(
%u^\mu a^\nu-u^\nu a^\mu
%\right)
%&=
%\varkappa\cdot a,
%\\
%\frac{1}{2}
%\omega_{\mu\nu}
%\epsilon^{\mu\nu\alpha\beta}
%u_\alpha b_\beta
%&=
%-\omega\cdot b.
%\end{align}
%
Substituting Eq.~\eqref{eq:spincontract_GLW} into Eq.~\eqref{eq:sGLW_from_kin}, we obtain
\begin{align}
s_{\rm GLW}=s-\frac{\hbar^2\beta}{m^2}
\cosh\alpha
\left[
(I_{30}-I_{31})\,
\beta a \cdot \varkappa
+
2I_{31}\,
\beta b \cdot \omega
\right]\;.
\label{eq:entropydens_GLW}
\end{align}
\subsection{Thermodynamic relations}
\label{thermoglw}
%
%\msc{Most of the following results are quite obvious. We have $\varepsilon_{\rm kinetic} = \varepsilon_{\rm GLW}$, and $n_{\rm kinetic} = n_{\rm GLW}$, while $S^{\lambda\mu\nu}$ changes. Therefore, the result of the integration of $\varepsilon$ should preserve the derivative of $P$ with respect to $\mu$. 
%Also, it is clear that since the two formulas agree in the kinetic gauge, they cannot agree here.}

As elaborated in detail in App.~\ref{AppTPHIUp}, the Maxwell relations~\eqref{eq:Maxwell} are violated in the GLW pseudo-gauge. 
Therefore, we do not expect that the thermodynamic pressure evaluated from Eq.~\eqref{eq:p_as_int_temp} agrees with the result obtained from Eq.~\eqref{eq:thdynpress_as_int_over_enerdens}.
Indeed, the pressure computed via Eq.~\eqref{eq:thdynpress_as_int_over_enerdens} gives the result~\eqref{eq:thdyn_press_GLW}, while the pressure computed via Eq.~\eqref{eq:p_as_int_temp} features an extra term, cf.~Eq.~\eqref{Eq:GLWthermopressure}, which originates from the difference between the GLW spin tensor and the kinetic-theory one, Eq.~\eqref{Eq:spindensitytenGLW}. 
Furthermore, as one can readily convince oneself, the thermodynamic relations~\eqref{eq:s_as_der_p} and \eqref{eq:ortho_spin_as_der_p} are violated in the GLW pseudo-gauge, even when using the expression~\eqref{eq:thdyn_press_GLW} for the pressure.

\section{Conclusions}
\label{conclusion}

In this work, we have studied a system of massive spin-1/2 Boltzmann particles in global thermodynamic equilibrium with nonvanishing acceleration and rotation, consistently including quantum corrections at order $\order{\hbar^{2}}$.
Thermodynamic relations for such a system are conveniently written in terms of a thermodynamic-potential current, which has a component in the direction of the fluid four-velocity, but also one in a direction orthogonal to the latter (which is also orthogonal to the directions of acceleration and rotation).
Standard thermodynamic relations are obtained by projecting the relations for the thermodynamic-potential current onto the fluid four-velocity, but  there also exists a set of such thermodynamic relations from projecting onto the orthogonal direction. 
As a further consequence, we found that the fluid four-velocity appears as an independent thermodynamic variable in the thermodynamic relations.

From investigating the Wigner function at order $\mathcal{O}(\hbar^2)$, we derived a modification of the standard mass-shell condition $k^2 = m^2$, which, in global thermodynamic equilibrium, is quadratic in the thermal vorticity.
We then computed the (net) particle-number current, energy-momentum tensor, spin-current tensor, and associated thermodynamic quantities at order $\order{\hbar^2}$ for various choices of pseudo-gauge. 
Besides the canonical and GLW pseudo-gauges we also studied a pseudo-gauge where the above quantities assume the form well known from kinetic theory and which we therefore termed ``kinetic-theory pseudo-gauge''.
Our explicit calculations prove that, while global quantities and the grand partition function are independent of the choice of pseudo-gauge in global thermodynamic equilibrium, local quantities are pseudo-gauge dependent.

For each pseudo-gauge, we checked whether the thermodynamic Maxwell relations are fulfilled. 
The validity of the Maxwell relations implies that, when computing the thermodynamic pressure via a line integral in the space of independent thermodynamic variables, the result does not depend on the chosen path.
This also ensures thermodynamic consistency, i.e., that thermodynamic quantities can be computed as partial derivatives of the pressure with respect to the conjugate thermodynamic variables.
We found that the kinetic-theory pseudo-gauge is in this sense thermodynamically consistent.
By contrast, the canonical and GLW pseudo-gauges in general violate the Maxwell relations, leading to  inconsistencies of some thermodynamic relations.
This suggests that the kinetic-theory pseudo-gauge is the most natural choice for applications such as spin hydrodynamics.

Our results demonstrate that pseudo-gauge transformations, despite preserving the global conserved charges, in general do not guarantee thermodynamic consistency. 
However, subsequent analysis indicates that the latter can be restored by properly accounting for the surface-term ambiguity in the identification of the thermodynamic currents~\eqref{eq:currents_as_ders} as derivatives of the thermodynamic-potential currents.
This will be reported in more detail elsewhere~\cite{shokri_in_prep}. 

Another consequence of our results is that, when studying spin hydrodynamics at order $\mathcal{O}(\hbar^2)$, one needs to account for the anisotropy of the equilibrium reference state introduced by nonvanishing acceleration and rotation.
As a consequence, the energy-momentum tensor derived in various pseudo-gauges fulfills $T^{\mu \nu} u_\nu = \varepsilon u^\mu + q_\times^\mu$, i.e., the energy density and fluid four-velocity do not fulfill the standard eigenvalue equation $T^{\mu \nu} u_\nu = \varepsilon u^\mu$.
The standard eigenvalue equation cannot be recovered by redefining the fluid four-velocity as in the Landau frame, as the latter quantity is fixed by the equilibrium conditions.

%*******************************************************************************************************************************************************************************************************************************************************************************************************************************************************************************************************************************************************************************************************************************************************************************************************************************************
%
\section*{Acknowledgments}
This work was supported in part by the Deutsche Forschungsgemeinschaft (DFG, German Research Foundation) through the Collaborative Research Center CRC-TR 211 ``Strong-interaction matter under extreme conditions'' - project number 315477589 - TRR 211.
The authors gratefully acknowledge fruitful discussions with A.~Dash, V.~Ambrus, F.~Becattini, and X.-L.~Sheng for a critical reading of the manuscript and useful remarks.
A.D.~acknowledges the financial support provided by the Polish National Agency for Academic Exchange (NAWA) within the Bekker programme (Nr decyzji BPN/BEK/2024/1/00364/DEC/1, Numer umowy~BPN/BEK/2024/1/00364/U/00001). 
A.D.~would like to thank the Institute for Theoretical Physics of Goethe University for the warm hospitality during the completion of this work.
D.H.R.~thanks K.J.~Eskola for discussions and the warm hospitality at Jyv\"askyl\"a University, where part of this work was done. 

%******************** ***********************************************************************************************************************************************************************************************************************************************************************************************************************************************************************************************************************************************************************************************************************************************************************************************************************
%
\appendix

\section{Pseudo-gauge dependence of thermodynamic currents and pseudo-gauge independence of the partition function}
\label{sec:pseudogauge_invariance}
In this appendix, we demonstrate in full generality that all thermodynamic quantities change under a pseudo-gauge transformation~\eqref{eq:pseudogauge_ops}.
On the level of expectation values, the pseudo-gauge transformation~\eqref{eq:pseudogauge_ops} reads
\begin{subequations} \label{eq:pseudogauge_evals}
\begin{align}
J^{\prime\mu} & = J^{\mu} + \hbar\, \partial_{\lambda} B^{\lambda\mu}\;, \label{eq:pgt_n_current_eval}\\
T'^{\mu\nu}& ={T}^{\mu\nu}+\frac{\hbar}{2}\partial_{\lambda}\left({\Phi}^{\lambda,\mu\nu}-{\Phi}^{\mu,\lambda\nu}-{\Phi}^{\nu,\lambda\mu}\right)\;,\label{eq:pgt_emt_eval}\\
S'^{\mu, \nu\lambda}& ={S}^{\mu, \nu\lambda}-{\Phi}^{\mu, \nu\lambda} + \hbar \,\partial_\rho Z^{\nu \lambda, \mu \rho} \;.
\label{eq:pgt_s_current_eval}
\end{align}
\end{subequations}
Under the pseudo-gauge transformation \eqref{eq:pseudogauge_evals}, and assuming that $\alpha$, $\beta_\nu$, and $\Omega_{\nu\lambda}$ do not change under such a transformation, we find that 
\begin{align}
T'^{\mu\nu}\beta_\nu -\alpha J'^{\mu} - \frac{\hbar}{2}\Omega_{\nu\lambda}S'^{\mu,\nu\lambda} &  = T^{\mu\nu} \beta_\nu-\alpha J^{\mu}- \frac{\hbar}{2} \Omega_{\nu\lambda}S^{\mu,\nu\lambda}+ \hbar \,X^{\mu}\;,\label{eq:pgt_all_charges_eval}
\end{align}
where we defined
\begin{align}\label{eq:def_cap_x}
X^{\mu}& :=\partial_{\lambda}A^{\lambda\mu}- \alpha\partial_{\lambda}B^{\lambda\mu}
+ \Phi^{\nu,\lambda \mu} \xi_{\nu\lambda}  + \frac{1}{2} \Phi^{\mu, \nu\lambda} \left( \Omega_{\nu\lambda} - \varpi_{\nu\lambda} \right)- \frac{\hbar}{2}\, \partial_\rho Z^{\nu\lambda ,\mu \rho} \Omega_{\nu \lambda}\;.
\end{align}
Here
\begin{equation}
A^{\lambda\mu}:=\frac{1}{2}\left(\Phi^{\lambda,\mu\nu}-\Phi^{\mu,\lambda \nu}-\Phi^{\nu,\lambda\mu}\right)\beta_{\nu}
\end{equation}
is an antisymmetric tensor~\cite{Becattini:2023ouz}. 
In global thermodynamic equilibrium (GTE), cf.~Eqs.~\eqref{eq:gteq_conditions},
\begin{align}\label{XmuGL}
    \mathrm{GTE:} \quad X^{\mu} \longrightarrow \partial_{\lambda}\left( A^{\lambda\mu}- \alpha B^{\lambda\mu} + \frac{\hbar}{2}\,   Z^{\nu\rho, \lambda \mu} \Omega_{\nu \rho}\right)\;,
\end{align}
which is a four-gradient of an antisymmetric rank-2 tensor, the integral of which vanishes by the relativistic Stokes theorem (provided appropriate boundary conditions are fulfilled).
%Therefore, using Eq.~\eqref{eq:phi_as_int_temp}, the logarithm of the grand partition function,  Eq.~\eqref{eq:th_dyn_potential_vector}, is invariant under a pseudo-gauge transformation in global thermodynamic equilibrium, see Appendix~\ref{app:pgt_invar_of_partfunc}.

From Eq.~\eqref{eq:pgt_all_charges_eval}, we infer the transformation law of the thermodynamic potential current \eqref{eq:phi_as_int_lambda} as
\begin{align}
    \phi^{\prime \mu}& =\phi^{\mu} + \hbar \int_{1}^{\infty} \di \lambda\, X^{\mu}(\lambda;\alpha, \beta_\rho, \Omega_{\sigma \tau})\;.
\label{eq:pgt_phi}
\end{align}
Note that the dependence of $X^\mu$ on $\lambda$ enters as follows: $B^{\lambda \mu}$, $\Phi^{\lambda, \mu \nu}$, and $Z^{\nu \lambda, \mu \rho}$ are functions of components of the Wigner function, see, for instance, Eqs.~\eqref{eq:pg_C-GLW} and \eqref{PhiGLW-KIN}, and the latter depends on the density operator and thus on $\lambda$, cf.~Eq.~\eqref{eq:statistical_operator_lambda}.

Finally, the entropy-density four-current can be computed from the covariant Euler relation~\eqref{eq:vec_euler} as
\begin{align}\label{eq:entropy_current_from_phi}
    S^\mu & = \phi^\mu + T^{\mu\nu} \beta_\nu -\alpha J^{\mu}- \frac{\hbar}{2} \,\Omega_{\nu\lambda} S^{\mu,\nu\lambda}\;.
\end{align}
Then, using the transformation laws \eqref{eq:pgt_all_charges_eval} and \eqref{eq:pgt_phi}, one obtains
\begin{align}\label{eq:pgt_entropy_current}
S'^{\mu}= S^{\mu} + \hbar \, X^{\mu}+ \hbar \int_{1}^{\infty}\di \lambda \; 
X^{\mu}(\lambda;\alpha, \beta_\rho, \Omega_{\sigma \tau})\;.
\end{align}
%However, the total entropy~\eqref{eq:def_entropy_current} is pseudo-gauge invariant in global thermodynamic equilibrium, by the same argument as the grand partition function, see Appendix~\ref{app:pgt_invar_of_partfunc}.

By projecting Eqs.~\eqref{eq:pseudogauge_evals} onto $u^\mu$, we immediately derive the transformation laws for the particle-number density, the energy density, and the spin tensor,
\begin{subequations}
\begin{align}
n' & \equiv u_\mu J^{\prime \mu}  = u_\mu J^\mu + \hbar \, u_\mu\, \partial_\lambda B^{\lambda \mu}  \equiv n + \hbar \, u_\mu\, \partial_\lambda B^{\lambda \mu}\;,\label{eq:pgt_n_density}\\
\varepsilon' & \equiv u_\mu  T^{\prime \mu \nu} u_\nu = u_\mu  T^{\mu \nu} u_\nu + \hbar \, u_\mu u_\nu\, \partial_\lambda \Phi^{\mu, \nu \lambda} \equiv  \varepsilon+ \hbar \, u_\mu u_\nu\, \partial_\lambda \Phi^{\mu, \nu \lambda}\;, \label{eq:pgt_energy_density} \\
S^{\prime \nu\lambda} & \equiv u_\mu S^{\prime \mu, \nu\lambda} = u_\mu S^{\mu, \nu\lambda} - u_\mu \Phi^{\mu, \nu\lambda}  + \hbar\, u_\mu \partial_\rho Z^{\nu\lambda, \mu \rho} \equiv
S^{\nu\lambda }- u_\mu \Phi^{\mu, \nu\lambda} + \hbar\, u_\mu \partial_\rho Z^{\nu\lambda, \mu \rho} \label{eq:pgt_spin_rank2_tensor} \;.
\end{align}
\end{subequations}
These quantities are therefore not invariant under a pseudo-gauge transformation, as explicitly demonstrated by the concrete examples in Secs.~\ref{sec:em_s_kin_theory} -- \ref{Sec:GLWPG}.

From Eqs.~\eqref{eq:def_thdyn_press} and \eqref{eq:pgt_phi}, we furthermore deduce the transformation law for the thermodynamic pressure as
\begin{align}\label{eq:pgt_pressure}
P'& = P+\hbar\, T\int_{1}^{\infty} \di \lambda \; 
u \cdot X (\lambda;\alpha, \beta_\rho, \Omega_{\sigma \tau})\;.
\end{align}
This shows that also the thermodynamic pressure is not invariant under a pseudo-gauge transformation.

Finally, since $s\equiv u_\mu S^\mu$, the pseudo-gauge transformation of the entropy density follows from Eq.~\eqref{eq:pgt_entropy_current} as
\begin{equation} \label{eq:pgt_entropy_density}
    s'  = s + \hbar \, u \cdot X  + \hbar  
    \int_{1}^{\infty}\di \lambda \; 
u \cdot X(\lambda;\alpha, \beta_\rho, \Omega_{\sigma \tau})\;,
\end{equation}
i.e., also the entropy density is not invariant under a pseudo-gauge transformation.

Similarly, by projecting onto $\ell^\mu$, we derive the pseudo-gauge transformations for the following quantities:
\begin{subequations}
\begin{align}
    n_\times' & = n_\times - \hbar \, \ell_\mu\, \partial_\lambda B^{\lambda \mu} \;, \\
    q_\times' & = q_\times - \frac{\hbar}{2}\ell_{\mu}u_{\nu}\partial_{\lambda}\left({\Phi}^{\lambda,\mu\nu}-{\Phi}^{\mu,\lambda\nu}-{\Phi}^{\nu,\lambda\mu}\right) \,  \;, \\
    S_\times^{\prime \lambda \nu} & = S_\times^{\nu\lambda}+\ell_{\mu}\Phi^{\mu,\nu\lambda}-\hbar\ell_{\mu}\partial_{\rho}Z^{\nu\lambda,\mu\rho}\;, \\
    P_\times' & = P_\times -\hbar\, T\int_{1}^{\infty} \di \lambda \; 
\ell \cdot X (\lambda;\alpha, \beta_\rho, \Omega_{\sigma \tau})\;, \\
    s_\times' & = s_\times- \hbar \, \ell \cdot X  - \hbar  
    \int_{1}^{\infty}\dd \lambda \; 
\ell \cdot X(\lambda;\alpha, \beta_\rho, \Omega_{\sigma \tau})\;.
\end{align}
\end{subequations}

We now demonstrate that the logarithm of the grand partition function~\eqref{eq:th_dyn_potential_vector} and the total entropy~\eqref{eq:def_entropy_current}, defined as the integral of the entropy current over a Cauchy spacelike hypersurface, are invariant under a general pseudo-gauge transformation.

By integrating Eq.~\eqref{eq:pgt_phi} over a Cauchy spacelike hypersurface, we obtain 
\begin{align}
\ln \mcZ^{\prime}=\ln \mcZ+\hbar\int_{\Sigma}\di\Sigma_{\mu}
\int_{1}^{\infty}
\di\lambda\, X^{\mu}\,.
\end{align}
Using the global equilibrium form  \eqref{XmuGL} of $X^\mu$, which is a divergence of an antisymmetric rank-2 tensor, together with the relativistic Stokes theorem one finds
\begin{align}\label{eq:secondterm}
\int_{\Sigma}
\di\Sigma_{\mu}
\int_{1}^{\infty}
\di\lambda\,X^{\mu}
=\int_{1}^{\infty}
\di\lambda \int_{\Sigma}
\di\Sigma_{\mu}\,\partial_{\alpha}
\,\left( A^{\alpha\mu}- \alpha B^{\alpha\mu}{+ \frac{\hbar}{2}\,   Z^{\nu\rho, \alpha \mu} \Omega_{\nu \rho}}\right)=0\,.
\end{align}
In the last step, the integral over $\lambda$ was interchanged with the integral over the hypersurface, since $\lambda$ is independent of spacetime point. 
Therefore, 
\begin{align}
\ln \mcZ^{\prime}=\ln \mcZ\,.
\end{align}

Integrating the entropy current~\eqref{eq:pgt_entropy_current} over a spacelike Cauchy hypersurface $\Sigma$, one obtains the corresponding transformation law for the total entropy,
\begin{align}
S'
=
S
+
\hbar
\int_{\Sigma}
\di\Sigma_{\mu}\,X^{\mu}
+
\hbar
\int_{\Sigma}
\di\Sigma_{\mu}
\int_{1}^{\infty}
\di \lambda\,X^{\mu}\,.
\end{align}
Using the same arguments as in Eq.~\eqref{eq:secondterm}, we obtain
\begin{align}
S'=S\,.
\end{align}
\section{Decomposition of the thermodynamic-potential current}
\label{app:th_dyn_curr}
In this appendix, we demonstrate that the hypersurface independence of the thermodynamic-potential current $\phi^\mu$ in global thermodynamic equilibrium, i.e., Eq.~\eqref{eq:phi_hypsur_ind}, implies that $\phi^\mu$ has only components in the directions of $u^\mu$ and $\ell^\mu$.
To this end, we start with the general decomposition \eqref{eq:phi_gen_tetrad_decomp} in terms of the tetrad $v^{a\mu}\in\{u^\mu, \xi^\mu, \omega^\mu, \ell^\mu\}$,
\begin{equation}
    \phi^\mu = \phi_a v^{a\mu}\;,
\end{equation}
where a summation over the tetrad index $a=\{u,\xi,\omega,\times\}$ is implicit.
We postulate that the hypersurface-independence condition \eqref{eq:phi_hypsur_ind} follows solely from the global-equilibrium identities \eqref{eq:gteq_conditions}, without imposing additional differential constraints on the constitutive thermodynamic functions. 
Otherwise, the existence of global thermodynamic equilibrium would depend on the particular microscopic realization of the fluid, implying that only specific microscopic theories would admit equilibrium. 
We refer to this postulate as the \textit{Microscopic Agnosticism of Equilibrium (MAE)}.
\footnote{The MAE is a general postulate and is not limited to the hypersurface-independence condition. 
It formalizes an assumption that is frequently left implicit in hydrodynamic theories, namely that equilibrium conditions should constrain the hydrodynamic fields rather than the microscopic constitutive functions.}

We first note that, since $\phi^\mu$ is a vector and $\omega^\mu$ is a pseudo-vector, parity invariance forbids a contribution proportional to $\omega^\mu$ for a parity-conserving medium, thus $\phi_\omega \equiv 0$.\footnote{The only possibility to have a contribution proportional to $\omega^\mu$ is if its coefficient is a pseudo-scalar (like $\varkappa \cdot \omega$). This, however, is not possible in a parity-conserving medium.}
To investigate the remaining contributions to $\phi^\mu$, subject to the constraint $\partial_\mu \phi^\mu =0$, we first recall that the coefficients $\phi_a$ are functions of $\alpha$, $\beta$, and $\Omega_{\a\b}$.
Therefore, their gradients are given by
\begin{equation} \label{eq:grad_phi_a}
    \partial_\mu \phi_a = \left.\pdv{\phi_a}{\a}\at[\right]{\beta,\Omega_{\a\b}}\partial_\mu \a +\left.\pdv{\phi_a}{\beta}\at[\right]{\a,\Omega_{\a\b}} \partial_\mu \beta    + \left.\pdv{\phi_a}{\Omega_{\a\b}}\at[\right]{\a,\b} \partial_\mu \Omega_{\a\b}\;.
\end{equation}
In global thermodynamic equilibrium, Eqs.~\eqref{eq:gteq_conditions}, 
$\partial_\mu \alpha \equiv \partial_\mu\Omega_{\a\b}\equiv 0$, while
\begin{equation} \label{eq:grad_beta}
    \partial_\mu \beta =
    \frac{1}{\beta}\,\beta^\nu \partial_\mu \beta_\nu=
    - \Omega_{\mu \nu} u^\nu =  \beta \varkappa_\mu \;,
\end{equation}
cf.~Eq.~\eqref{eq:kappa_omega_def}.
Thus, Eq.~\eqref{eq:grad_phi_a} reduces to
\begin{equation}\label{eq:gteq_scalar_grad}
    \partial_\mu \phi_a \equiv \beta \left.\pdv{\phi_a}{\beta}\at[\right]{\a,\Omega_{\a\b}}\varkappa_\mu \;.
\end{equation}

Next, we calculate the divergence of the tetrad vectors in global thermodynamic equilibrium.
First note that the Killing condition \eqref{eq:killing_condition} and Eq.~\eqref{eq:th_vort_def} imply $\partial_\mu \beta_\nu \equiv - \Omega_{\mu \nu}$, so that with the decomposition of the reduced spin potential \eqref{eq:decomp_red_spin_potential} and Eq.~\eqref{eq:grad_beta} we obtain
\begin{equation}\label{eq:gteq-grad-u}
    \partial_\mu u_\nu
\equiv \partial_\mu \left(\frac{1}{\beta} \, \beta_\nu \right) = - \varkappa_\mu u_\nu + \frac{1}{\beta} \partial_\mu \beta _\nu=
-u_\mu\varkappa_\nu-\epsilon_{\mu\nu\alpha\beta}u^\alpha\omega^\beta\;.
\end{equation}
Contracting the Lorentz indices immediately results in the well-known global-equilibrium identity 
\begin{equation}\label{eq:no_expansion_condition}
    \partial_\mu u^\mu = 0\;.
\end{equation}
Using Eqs.\ \eqref{eq:grad_beta} and \eqref{eq:gteq-grad-u} together with $\partial_\mu\Omega_{\a\b} \equiv 0 $ and Eq.~\eqref{eq:kappa_omega_def}, we obtain
\begin{subequations}\label{eq:gteq-grad-kappa-omega}
\begin{align}\label{eq:gteq-grad-kappa}
    \partial_\mu \varkappa_\nu
    &\equiv
    -\varkappa^2 u_\mu u_\nu
    +\omega^2 \Delta_{\mu\nu}
    -\varkappa_\mu\varkappa_\nu
    +\omega_\mu\omega_\nu
    -u_{\mu}L_{\nu}-u_{\nu}L_{\mu}\;,
    \\
    \label{eq:gteq-grad-omega}
    \partial_\mu\omega_\nu
    &\equiv
    -2\varkappa_\mu\omega_\nu
    +(\varkappa\cdot\omega)g_{\mu\nu}\;.
\end{align}
\end{subequations}
Then, we employ Eqs.\ \eqref{eq:gteq-grad-u}, \eqref{eq:gteq-grad-kappa-omega}, and \eqref{eq:def_cap_l_vec} to find (after some lengthy calculation)
\begin{equation}\label{eq:div_ell}
    \partial_\mu \ell^\mu = \frac{1}{L} \left( g^{\mu \nu} + \ell^\mu \ell^\nu \right) \partial_\mu L_\nu = 0\;.
\end{equation}
Finally, Eqs.\ \eqref{eq:gteq-grad-kappa-omega} and \eqref{eq:def_xi} yield (again after some lengthy calculation)
\begin{align}\label{eq:div_xi}
\partial_\mu\xi^\mu \equiv 
2\left[
\omega^2
+
\frac{(\varkappa\cdot\omega)^2}{\omega^2}
\right] = 2\left(\varkappa^2+\omega^2-\xi^2\right)\;.
\end{align}

Using Eqs.\ \eqref{eq:gteq_scalar_grad}, \eqref{eq:no_expansion_condition}, and \eqref{eq:div_ell}, together with $u\cdot \varkappa = \ell\cdot\varkappa = 0$, we find 
\begin{equation}
    \partial_\mu (\phi_u u^\mu) = \partial_\mu (\phi_\times \ell^\mu) = 0\;.
\end{equation}
Note that these relations hold identically only on account of the equilibrium conditions, without any constraints on the thermodynamic coefficients $\phi_u$ and $\phi_\times$.

Let us now consider the remaining contribution, which reads
\begin{equation}\label{eq:phi_xi_eom}
\partial_\mu\left(\phi_\xi\,\xi^\mu\right) = - \beta\pdv{\phi_\xi}{\beta}\at[\Big]{\alpha, \Omega_{\a\b}}\xi^2 + 2 \phi_\xi \left(\varkappa^2+\omega^2-\xi^2\right)\;,
\end{equation}
where we have used Eqs.\ \eqref{eq:def_xi}, \eqref{eq:gteq_scalar_grad}, and \eqref{eq:div_xi}.
Therefore, to satisfy Eq.~\eqref{eq:phi_hypsur_ind}, the $\beta$ dependence of $\phi_\xi$ must obey the following differential equation:
\begin{equation}
    - \beta\pdv{\phi_\xi}{\beta}\at[\Big]{\alpha, \Omega_{\a\b}}\xi^2 + 2 \phi_\xi \left(\varkappa^2+\omega^2-\xi^2\right) = 0\;.
\end{equation}
This equation is precisely an additional differential constraint excluded by the MAE postulate.
We therefore conclude that the only admissible solution is $\phi_\xi\equiv0$.

As a concrete illustration, consider the particular case of rigid cylindrical rotation, where $\varkappa \cdot \omega = 0$ and 
\begin{equation}
    \varkappa^2 = \left(1-\frac{\beta^2}{\beta_0^2}\right) \omega^2\;,
\end{equation}
with $T_0=1/\beta_0$ being the temperature at the origin of the rotation axis \cite{Shokri:2023rpp}.
Then, Eq.~\eqref{eq:phi_xi_eom} becomes
\begin{equation}
    \beta\pdv{\phi_\xi}{\beta}\at[\Big]{\alpha, \Omega_{\a\b}}\left(1-\frac{\beta^2}{\beta_0^2}\right) = 2 \phi_\xi \;.
\end{equation}
The solution of this equation reads
\begin{equation}
    \phi_\xi(\beta) = \frac{\mathcal{C}(\alpha, \Omega_{\a\b})}{\beta^{-2}-\beta_0^{-2}}\;,
\end{equation}
which is regular at $\beta=\beta_0$ only if the integration constant $\mathcal{C}(\alpha, \Omega_{\a\b})$ vanishes and thus $\phi_\xi \equiv 0$.

Therefore, the most general thermodynamic-potential current compatible with the hypersurface-independence condition $\partial_\mu \phi^\mu =0$ and the MAE postulate is
\begin{equation}
   \phi^\mu =\beta \left(P u^\mu + P_\times \ell^\mu\right)\;,
\end{equation}
where we have used the definitions \eqref{eq:phi_comps}.
The condition that $\phi^\mu$ be future-directed timelike then requires \begin{equation}
    P > \abs{P_\times} \geq 0\;.
\end{equation}
Note that our assumptions do not allow us to conclude that $P_\times$ is positive semi-definite.

\section{The tensor $B^{\lambda\mu}_{\rm C\rightarrow GLW}$ in terms of Wigner-function components}
\label{AppedndixBlambdamu}
The goal of this appendix is to derive the expression for the antisymmetric tensor $B^{\lambda\mu}_{\rm C\rightarrow GLW}$~\eqref{BCGLWW}.
This tensor appears in the  transformation of the (net) particle-number current from the canonical to the GLW pseudo-gauge.
Since $J^{\mu}=J^{\mu}_{\rm GLW}$, it follows that $B^{\lambda\mu}_{\rm C\rightarrow GLW}=B^{\lambda\mu}_{\rm C\rightarrow kin}$.
One may therefore equivalently compute $B^{\lambda\mu}_{\rm C\rightarrow kin}$ using the kinetic-theory definition of the (net) particle-number current~\eqref{eq:part_curr_kin_th}.

The pseudo-gauge transformation from the canonical to the kinetic-theory (net) particle-number currents reads, cf.~Eq.~\eqref{eq:pgt_n_current_eval}, 
\begin{align} \label{eq:JmuJmucanB}
J^{\mu}= J^{\mu}_{\rm C} + \hbar\, \partial_{\lambda} B^{\lambda\mu}_{\rm C\rightarrow kin}\;.
\end{align}
Using the identity \eqref{eq:fF}, the kinetic-theory (net) particle-number current defined in Eq.~\eqref{eq:part_curr_kin_th} can be rewritten as
\begin{align}\label{Admsdr}
J^{\mu}=\frac{1}{m}\int \di K\,k^{\mu}\,\frac{\partial\mathcal{F}}{\partial\alpha}~\bigg|_{\beta,\Omega_{\mu\nu}}\,.
\end{align}
On the other hand, the canonical (net) particle-number current \eqref{eq:part_curr_can} involves the vector component $\mathcal{V}^{\mu}$ of the Wigner function. 
Using Eq.~\eqref{eq:Vmu}, the canonical current can be written as
\begin{align} \label{eq:Jmucan}
J^{\mu}_{\rm C} =\frac{1}{m}\int \di K\,k^{\mu}\frac{\partial\mathcal{F}}{\partial\alpha}-\frac{\hbar^{2}}{4m^{3}}\int \di K\,k^{\mu}\Box_{x}\frac{\partial\mathcal{F}}{\partial\alpha}-\frac{\hbar}{2m^{2}}\int \di K\,\epsilon^{\mu\nu\alpha\beta}\partial_{\nu}^{x}k_{\alpha}\frac{\partial\mathcal{A_{\beta}}}{\partial\alpha}\;.
\end{align}
With Eqs.~\eqref{Admsdr} and \eqref{eq:Jmucan}, we obtain from Eq.~\eqref{eq:JmuJmucanB}
\begin{align}\label{asdfvc}
\partial_{\lambda}B^{\lambda\mu}_{\rm C\rightarrow kin}=
\frac{\hbar}{4m^{3}}\int \di K\,k^{\mu}\Box_{x}\frac{\partial\mathcal{F}}{\partial\alpha}+\frac{1}{2m^{2}}\int \di K\,\epsilon^{\mu\nu\alpha\beta}\partial_{\nu}^{x}k_{\alpha}\frac{\partial\mathcal{A_{\beta}}}{\partial\alpha}\,.
\end{align}
It is now possible to infer the form of $B^{\lambda\mu}_{\rm C\rightarrow kin}$, 
\begin{align}\label{Eq;b55}
B^{\lambda\mu}_{\rm C\rightarrow kin}=\frac{\hbar}{4m^{3}}\int\di K\,(k^{\mu}\partial^{\lambda}_{x}-k^{\lambda}\partial^{\mu}_{x})\frac{\partial\mathcal{F}}{\partial\alpha}-\frac{1}{2m^{2}}\int \di K\epsilon^{\lambda\mu\rho\sigma}k_{\rho}\frac{\partial\mathcal{A}_{\sigma}}{\partial\alpha}\,.
\end{align}
This expression is antisymmetric under $\lambda \leftrightarrow \mu$, and its divergence reproduces Eq.~\eqref{asdfvc} upon using the identity
$k\cdot\partial \mathcal{F}=0$, which follows from the equation of motion~\eqref{eq:transport} for the Wigner function. 
Equation~\eqref{Eq;b55} is the form employed in Eq.~\eqref{BCGLWW}.
\section{Tensor decomposition and thermodynamic integrals}
\label{AppendixA}
For the calculations in this paper, the following tensor decompositions are useful:
\begin{subequations} \label{eq:tensor_decomp}
\begin{align}
    \int \frac{\mathrm{d}^3 \mathbf{k}}{(2 \pi \hbar)^3 k_0}\, k^\mu \, e^{-\beta \cdot k} & = I_{10}\, u^\mu\;, \\[1em]
    \int \frac{\mathrm{d}^3 \mathbf{k}}{(2 \pi \hbar)^3 k_0}\, k^\mu k^\nu\, e^{-\beta \cdot k} & = I_{20}\, u^\mu u^\nu - I_{21}\, \Delta^{\mu \nu}\;, \\[1em]
    \int \frac{\mathrm{d}^3 \mathbf{k}}{(2 \pi \hbar)^3 k_0}\, k^\mu k^\nu k^\rho \, e^{-\beta \cdot k} & = I_{30}\, u^\mu u^\nu u^\rho- I_{31}\left( u^\mu \Delta^{\nu \rho} + u^\nu \Delta^{\mu \rho} + u^{\rho} \Delta^{\mu \nu} \right)\;, \\[1em]
    \int \frac{\mathrm{d}^3 \mathbf{k}}{(2 \pi \hbar)^3 k_0}\, k^\mu k^\nu k^\rho k^\sigma \, e^{-\beta \cdot k} & = I_{40}\, u^\mu u^\nu u^\rho u^\sigma \nonumber \\
    & - I_{41}\left( u^\mu u^\nu \Delta^{\rho \sigma} + u^\mu u^\rho \Delta^{\nu \sigma} + u^\mu u^\sigma \Delta^{\nu \rho} + u^\nu u^\rho \Delta^{\mu \sigma} + u^\nu u^\sigma \Delta^{\mu \rho} + u^\rho u^\sigma \Delta^{\mu \nu} \right) \nonumber \\
&    + I_{42} \left( \Delta^{\mu \nu} \Delta^{\rho \sigma} + \Delta^{\mu \rho} \Delta^{\nu \sigma} + \Delta^{\mu \sigma} \Delta^{\nu \rho} \right)\;, \label{eq:4k}
\end{align}
\end{subequations}
where $k_0 = \sqrt{\mathbf{k}^2 + m^2}$ and the thermodynamic integrals are defined as
\begin{align} \label{eq:Inq}
I_{nq}=\frac{1}{(2q+1)!!}\int \frac{\di^3 \mathbf{k}}{(2\pi\hbar)^{3} k_0}\,E_\mathbf{k}^{n-2q}\left(E_\mathbf{k}^2 - m^2\right)^{q}e^{-\beta E_\mathbf{k}}\; ,
\end{align}
with $E_{\mathbf{k}} := u \cdot k$.

The thermodynamic integrals \eqref{eq:Inq} fulfill the following identities:
\begin{subequations} \label{eq:useful_identities}
    \begin{align}
      \label{eq:identity_1}
\beta I_{n+1,q} & = I_{n,q-1} + (n+1-2q) I_{nq}\;, \\[1em]
\label{eq:identity_2}
I_{nq} & = (2q+3) I_{n,q+1} + m^2 I_{n-2,q}\;, \\[1em]
\label{eq:identity_3}
I_{n+1,q} & = - \frac{\di I_{nq}}{\di \beta}\;.
    \end{align}
\end{subequations}
\section{Explicit calculation of the thermodynamic pressure in different pseudo-gauges}
\label{AppTPHIUp}
In this appendix, we provide the details of computing the thermodynamic pressure according to Eqs.~\eqref{eq:p_as_int_temp} and \eqref{eq:thdynpress_as_int_over_enerdens}. 
We perform this derivation separately in the kinetic-theory, canonical, and GLW pseudo-gauges. 
We show that, in the kinetic-theory pseudo-gauge, the resulting expressions agree, whereas in the canonical and GLW pseudo-gauges, they do not.
Here, the integral~\eqref{eq:thdynpress_as_int_over_enerdens} can be evaluated in closed form, whereas the integral~\eqref{eq:p_as_int_temp} does not have a primitive.

For the integral~\eqref{eq:p_as_int_temp}, it is convenient to change the integration variable from $T'$ to \(\beta^{\prime}=1/T^{\prime}\). 
Then, Eq.~\eqref{eq:p_as_int_temp} can be written as
\begin{align}\label{betaTphiuPapp}
P= T \int_{\beta}^{\infty} \di\beta^{\prime} \; \bigg( \varepsilon-\mu n\, - \frac{\hbar}{2} \omega_{\mu\nu} S^{\mu\nu} \bigg)_{\mu,\omega_{\sigma \tau},\varpi_{\sigma \tau}}\,.
\end{align}
As discussed in the main text, this integral is evaluated at constant \(\mu\), as well as at constant spin potential $\omega_{\sigma \tau}$, i.e., constant \(\varkappa\) and \(\omega\), as well as at constant thermal vorticity $\varpi_{\sigma \tau}$, i.e., constant \(\beta a_\mu\) and constant \(\beta b_\mu\).
When evaluating the $\beta'$ integral of Eq.~\eqref{betaTphiuPapp}, the following identity (which is proved using Eq.~\eqref{eq:identity_3}) is useful,
\begin{align} \label{eq:betaprime_int}
   & \int_\beta^\infty \di \beta'\left[ \cosh(\beta' \mu)\, \beta^{\prime \ell} \, I_{n+1,q}(\beta') - \ell\, \cosh(\beta' \mu) \, \beta^{\prime \ell - 1} I_{nq}(\beta') - \mu \, \sinh(\beta' \mu)\, \beta^{\prime \ell}\, I_{nq}(\beta') \right] \nonumber \\
   & = - \int_\beta^\infty \di \beta'\,  \frac{\di}{\di \beta'} \left[ \cosh(\beta' \mu)\, \beta^{\prime \ell}\, I_{nq} (\beta') \right] 
    = \cosh(\beta \mu)\, \beta^\ell\, I_{nq} (\beta)\;,
\end{align}
where the contribution from the upper boundary vanishes, since $I_{nq} (\beta)$ is essentially a combination of modified Bessel functions of the second kind, $K_\nu(\beta m)$, which behave as $\sim e^{-\beta m}/\sqrt{\beta m}$ for $\beta \rightarrow \infty$ (independent of $\nu$).\footnote{The contribution from the upper boundary vanishes as long as $m > \mu$, which is a prerequisite for applying Boltzmann statistics to fermions.}
In performing the $\beta'$ integration we require the identity \eqref{eq:betaprime_int} for $\ell=0$, $\ell = 1$, and $\ell = 2$.

For the integral in Eq.~\eqref{eq:thdynpress_as_int_over_enerdens}, we need the identity
\begin{align} \label{eq:identity_3a}
    \int_\beta^\infty \di \beta' \left[\beta^{\prime \ell} I_{n+1,q}(\beta') - \ell \, \beta^{\prime \ell - 1} I_{nq}(\beta')\right] & = \beta^\ell \, I_{nq}(\beta)\;,
\end{align}
which immediately follows from Eq.~\eqref{eq:betaprime_int} setting $\mu =0$.

\subsection{Kinetic-theory pseudo-gauge}
We first evaluate the pressure integral~\eqref{eq:p_as_int_temp}.
Contracting the kinetic-theory spin tensor~\eqref{eq:spintensor} with the spin potential, we obtain
\begin{align}\label{contractionkineticcbase}
\frac{\hbar}{2}\omega_{\mu\nu}S^{\mu\nu}=2\cosh\alpha\,\frac{\hbar^{2}\beta}{2m^{2}}\,\left[2I_{31}\varkappa^{2}+\left(I_{30}-I_{31}\right)\omega^{2}\right].
\end{align}
Substituting the above quantity, together with the kinetic-theory expressions for the energy density,  Eq.~\eqref{eq:energydens_kin_th}, and the (net) particle-number density, Eq.~\eqref{eq:chargedens}, into Eq.~\eqref{betaTphiuPapp}, we arrive at
\begin{align}\label{Pressurethermoapp}
P&=4T\int_{\beta}^{\infty}\di\beta^{\prime}\,\left[\cosh(\beta' \mu)\,I_{20}-\mu\sinh(\beta' \mu)\,I_{10}\right] \nonumber \\[1em]
& +T\frac{\hbar^{2}}{m^{2}}\varkappa^{2}\int_{\beta}^{\infty}\di\beta^{\prime}\,\left[\cosh(\beta' \mu)\,\beta^{\prime\, 2}I_{41}-\mu\sinh(\beta' \mu)\, \beta^{\prime\, 2}I_{31}-2\cosh(\beta'\mu)\,\beta^{\prime}I_{31}\right]\nonumber\\[1em]
&+T\frac{\hbar^{2}}{2m^{2}}\omega^{2} \int_{\beta}^{\infty}\di\beta^{\prime}\,\left[\cosh(\beta'\mu)\,\beta^{\prime\,2}(I_{40}-I_{41})-\mu \sinh(\beta'\mu)\,\beta^{\prime\, 2}(I_{30}-I_{31})-2\cosh(\beta'\mu)\,\beta^{\prime}(I_{30}-I_{31})\right]\nonumber\\[1em]
&+T\frac{\hbar^{2}m^{2}}{2}\beta^{2}a^{2}\int_{\beta}^{\infty}\di\beta^{\prime}\,\left[\cosh(\beta'\mu)\,\beta'I_{10}-\cosh(\beta'\mu)\,I_{00}-\mu\sinh(\beta'\mu)\; \beta'I_{00}\right] \nonumber \\[1em]
&+T\hbar^{2}\beta^{2}b^{2}\int_{\beta}^{\infty}\di\beta^{\prime}\left[\cosh(\beta'\mu)\,I_{20}-\mu\sinh(\beta'\mu)\,I_{10}\right]\;.
\end{align}

All integrals above are of the type~\eqref{eq:betaprime_int} for different values of $n$, $q$, and $\ell$:
The first and the fifth one correspond to Eq.~\eqref{eq:betaprime_int} for $n=1$, $q=0$, and $\ell=0$, the second integral for $n=3,q=1,$ and $\ell=2$, the third term is the difference between the integral for $n=3$, $q=0$, and $\ell=2$ and that for $n=3$, $q=1$, and $\ell=2$.
Finally, the fourth term corresponds to the integral for $n=0$, $q=0$, and $\ell = 1$.
This ultimately yields the thermodynamic pressure given in Eq.~\eqref{eq:thdyn_press}. 
This result also confirms that $P=T\phi \cdot u$.

We now evaluate the pressure integral~\eqref{eq:thdynpress_as_int_over_enerdens}. 
In kinetic-theory pseudo-gauge, the energy density is given by Eq.~\eqref{eq:energydens_kin_th}. 
Since the integration in Eq.~\eqref{eq:thdynpress_as_int_over_enerdens}
is performed at fixed $\alpha$, $\Omega_{\sigma\tau}$, and $u_\rho$, the combinations $\beta a^\mu$, $\beta b^\mu$, $\beta\varkappa^\mu$, and $\beta\omega^\mu$ are kept constant along the integration path. 
Substituting Eq.~\eqref{eq:energydens_kin_th} into Eq.~\eqref{eq:thdynpress_as_int_over_enerdens}, we find
\begin{align}
P
&=
4 T\, \cosh\alpha
\Bigg\{
\int_{\beta}^{\infty}\di\beta'\,I_{20}(\beta')
+
\frac{\hbar^2}{8m^2}
\Bigg[
2\beta^2\varkappa^2
\int_{\beta}^{\infty}\di\beta'\,I_{41}(\beta')
+
\beta^2\omega^2
\int_{\beta}^{\infty}\di\beta'\,
\left[
I_{40}(\beta')-I_{41}(\beta')
\right]
\nonumber\\[1em]
&\qquad
-
m^4\beta^2a^2
\int_{\beta}^{\infty}\di\beta'\,
\left[
I_{00}(\beta')-\beta'I_{10}(\beta')
\right]+
2m^2\beta^2b^2
\int_{\beta}^{\infty}\di\beta'\,I_{20}(\beta')
\Bigg]
\Bigg\}\;.
\label{eq:pressure_integrals_expanded}
\end{align}
All these integrals can be evaluated using Eq.~\eqref{eq:identity_3a} for different values of $n, q,$ and $\ell$: The first and the last correspond to $n=1, q=0,$ and $\ell=0$, and the second integral to $n=3, q=1,$ and $\ell=0$. 
The third integral is the difference between the integrals for $n=3, q=0,$ $\ell=0$ and $n=3, q=1,$ and $\ell=0$.
Finally, the fourth integral corresponds to $n=0$, $q=0$, and $\ell=1$. 
%All the above integrals can be evaluated using~\
%\textcolor{red}{(DHR: simply use~\eqref{eq:identity_3a} for $\ell =1$!)}
Using these results, we obtain the thermodynamic pressure~\eqref{eq:thdyn_press}.
Therefore, the calculation in the kinetic-theory pseudo-gauge shows that the integral representation of the thermodynamic pressure is independent of the path chosen in the space of independent thermodynamic variables, provided the Maxwell relations are fulfilled.

\subsection{Canonical pseudo-gauge}
We first evaluate the canonical thermodynamic-pressure integral~\eqref{eq:p_as_int_temp}. 
The contraction of the spin potential with the canonical spin density given in Eq.~\eqref{eq:spintensor_can} reads
\begin{align}\label{canonicalcontractionmain}
\frac{\hbar}{2}\omega_{\mu\nu}S_C^{\mu\nu}=2\cosh\alpha\frac{\hbar^{2}\beta}{2m^{2}}(I_{30}-3 I_{31})\omega^{2}\;.
\end{align}
In deriving the above expression, we have used Eq.~\eqref{eq:identity_2}. 
Using Eq.~\eqref{contractionkineticcbase}, one may rewrite Eq.~\eqref{canonicalcontractionmain} as
\begin{align}
\frac{\hbar}{2}\omega_{\mu\nu}S_C^{\mu\nu}=\frac{\hbar}{2}\omega_{\mu\nu}S^{\mu\nu}-2\,\cosh\alpha\,\frac{\hbar^{2}\beta}{m^{2}}\,(\varkappa^{2}+\omega^{2})\,I_{31}\;.
\end{align}
The next step is to substitute the above expression, together with the canonical (net) particle-number density~\eqref{eq:dens_can} and energy density~\eqref{eq:energydens_can}, into Eq.~\eqref{betaTphiuPapp}.
Keeping in mind that $\beta a_\mu$ and $\beta b_\mu$ are to be kept constant for the $\beta'$ integral, this results in
\begin{align}\label{PCApp2}
 P_C&=P+ T\frac{\hbar^{2}}{m^{2}}\beta^{2}a^{2}\int_{\beta}^{\infty}\di\beta^{\prime}\,\left[\cosh(\beta'\mu)(I_{40}-I_{41})-\mu\sinh(\beta'\mu)(I_{30}-I_{31})\right] \nonumber \\[1em]
 &+T\frac{2\hbar^{2}}{m^{2}}\beta^{2}b^{2}\int_{\beta}^{\infty}\di\beta^{\prime}\,\left[\cosh(\beta'\mu) I_{41}-\mu\sinh(\beta'\mu) I_{31}\right]\nonumber\\[1em]
 &+T\frac{2\hbar^{2}}{m^{2}}\left(\beta a \cdot\varkappa + \beta b \cdot \omega \right)\int_{\beta}^{\infty}\di\beta^{\prime}\, \beta^{\prime}\left[\cosh(\beta'\mu)\, I_{41}-\mu\sinh(\beta'\mu)I_{31}\right] +T\frac{2\hbar^{2}}{m^{2}}\int_{\beta}^{\infty}\di\beta^{\prime}\,\beta^{\prime}\, \cosh(\beta'\mu)\,I_{31}(\varkappa^{2}+\omega^{2})\;.
\end{align}
The first two integrals can be evaluated directly using Eq.~\eqref{eq:betaprime_int} for $\ell=0$. 
However, the last two integrals  cannot be treated in the same manner, since the corresponding integrands are missing an additional term proportional to $\cosh(\beta'\mu)$, which is required for the direct application of Eq.~\eqref{eq:betaprime_int} for $\ell=1$. 
As explained in the main text, these missing terms can be generated by replacing one factor of $\beta'\varkappa^\mu$ and one factor of $\beta'\omega^\mu$ under the last integral by $\beta a^\mu$ and $\beta b^\mu$, respectively, and then pulling $\beta a \cdot \varkappa + \beta b \cdot \omega$ out of the integral (as this term is to be kept constant for the $\beta'$ integral).
This is, however, an \textit{ad hoc} prescription.
The result is then the same as when computing the thermodynamic pressure via Eq.~\eqref{eq:thdynpress_as_int_over_enerdens}.

We now compute the canonical thermodynamic-pressure integral~\eqref{eq:thdynpress_as_int_over_enerdens}. 
In the canonical pseudo-gauge, the energy density is given by Eq.~\eqref{eq:energydens_can}. 
Hence we obtain
\begin{align}
P_C
&=
P
+
4T\,\cosh\alpha\,
\frac{\hbar^2\beta^2}{4m^2}
\Bigg\{
a^2
\int_\beta^\infty\di\beta'\,
\left[
I_{40}(\beta')-I_{41}(\beta')
\right]+
2
\left(
b^2+a\cdot\varkappa+b\cdot\omega
\right)
\int_\beta^\infty\di\beta'\,
I_{41}(\beta')
\Bigg\}\;,
\label{eq:PC_integral_reduced}
\end{align}
where the integration is performed at fixed
$\alpha$, $\Omega_{\sigma\tau}$, and $u_\rho$\,. 
To evaluate these integrals, we use the identity~\eqref{eq:identity_3a} for $\ell=0$ and $(n,q)=(3,0)$ and $(3,1)$, respectively. 
We thus recover the canonical thermodynamic pressure~\eqref{eq:thdyn_press_can}.

Hence, in the canonical pseudo-gauge, the thermodynamic pressure obtained from an integral representation depends on the path chosen in the space of independent thermodynamic variables. 
This path dependence is due to the violation of the Maxwell relations, as shown in App.~\ref{app:maxwell_canonical}.

%%%%%%%%%%%%%%%%%%%%%%%%%%%%%%%%%%%%%%%%%%%%%%%%%%%%%%%%%%%%%%%%%%%%%%%%%%%%%%%%%%%%%%%%%%%%%%%%%%%%%%%%%%%%%%%%%%%%%%%%%%%%%%%%%%%%%%%%%%%%%%%%%%%%%%%%%%%%
\subsection{GLW pseudo-gauge}
To evaluate the GLW thermodynamic pressure using Eq.~\eqref{eq:p_as_int_temp}, we first compute the contraction of the spin potential with the GLW spin tensor.
This quantity is related to its kinetic counterpart through
\begin{align}
\frac{\hbar}{2}\omega_{\mu\nu}S^{\mu\nu}_{\rm GLW}=\frac{\hbar}{2}\omega_{\mu\nu}S^{\mu\nu}+\frac{\hbar^{2}\beta}{m^{2}}\cosh\alpha\left[(I_{30}-I_{31})a\cdot\varkappa+2I_{31}b\cdot\omega\right]\;.
\end{align}
Substituting this result together with the GLW energy density and (net) particle-number density into the integrand of Eq.~\eqref{eq:p_as_int_temp}, we obtain
\begin{align}\label{Eq:GLWthermopressure}
P_{\rm GLW}
=P-T\frac{\hbar^{2}}{m^{2}} \left[  \beta a \cdot \varkappa \int_{\beta}^{\infty}\di\beta^{\prime}\,\cosh(\beta' \mu)\,(I_{30}-I_{31})+2\beta b\cdot\omega \int_{\beta}^{\infty}\di\beta^{\prime}\,\cosh(\beta' \mu)\, I_{31} \, \right]\,.
\end{align}
Unlike the corresponding integral encountered in the kinetic-theory pseudo-gauge, the remaining integrals in Eq.~\eqref{Eq:GLWthermopressure} cannot be reduced to a closed form using the identity~\eqref{eq:betaprime_int}.
On the other hand, the GLW thermodynamic pressure computed using the integral representation~\eqref{eq:thdynpress_as_int_over_enerdens} is identical to its kinetic-theory pseudo-gauge counterpart because $\varepsilon_{\rm GLW}=\varepsilon$.
As in the canonical pseudo-gauge, in the GLW pseudo-gauge the thermodynamic pressure obtained from an integral representation depends on the path chosen in the space of independent thermodynamic variables. 
This path dependence originates from the violation of the Maxwell relations, as shown in App.~\ref{app:maxwell_GLW}.

%%%%%%%%%%%%%%%%%%%%%%%%%%%%%%%%%%%%%%%%%%%%%%%%%%%%%%%%%%%%%%%%%%%%%%%%%%%%%%%%%%%%%%%%%%%%%%%%%%%%%%%%%%%%%%%%%%%%%%%%%%%%%%%%%%%%%%%%%%%%%%%%%%%%%%%%%%%%%%%%%%%%%%%%%%%%%%%%%%%%%%%%%%%%%%%%%%%%%%%%%%%%%%%%%%%%%%%%%%%%%%%%%%%%%%%%%%%%%%%%%%%%%%%%%%%%%%%%%%%%
%%%%%%%%%%%%%%%%%%%%%%%%%%%%%%%%%%%%%%%%%%%%%%%%%%%%%%%%%%%%%%%%%%%%%%%%%%%%%%%%%%%%%%%%%%%%%%%%%%%%%%%%%%%%%%%%%%%%%%%%%%%%%%%%%%%%%%%%%%%%%%%%%%%%%%%%%%%%%%%%%%%%%%%%%%%%%%%%%%%%%%%%%%%%%%%%%%%%%%%%%%%%%%%%%%%%%%%%%%%%%%%%%%%%%%%%%%%%%%%%%%%%%%%%%%%%%%%%%%%%%

\section{Maxwell relations for the kinetic-theory pseudo-gauge}
\label{app:Maxwell_KT}
In this appendix, we verify the Maxwell relations in the kinetic-theory pseudo-gauge.
In all calculations, the thermal vorticity is kept constant, i.e.,  $\beta^2 a^2= const.$ and $\beta^2 b^2 = const.$.
The left-hand side of the first Maxwell relation~\eqref{eq:Maxwell_1} is computed using Eqs.~\eqref{eq:chargedens} and \eqref{eq:identity_3} as
\begin{align}
\left.
\frac{\partial n}{\partial T}
\right|_{\omega_{\sigma\tau}}
&=
-4\mu\beta^2\cosh\alpha
\left\{I_{10}+ \frac{\hbar^2\beta^2}{8m^2}
\left[ 2I_{31}\varkappa^2 +
(I_{30}-I_{31})\omega^2  +
m^2 \left( m^2\beta I_{00}a^2 +
2I_{10}b^2 \right) \right] \right\}
\nonumber\\
&\quad +
4\beta^2\sinh\alpha \left(
I_{20} + \frac{\hbar^2}{8m^2}
\left\{ 2\beta \left( \beta I_{41}-2I_{31}
\right)\varkappa^2 + \beta \left[
\beta(I_{40}-I_{41}) -
2(I_{30}-I_{31})
\right]\omega^2 \right. \right. 
\nonumber\\
&\hspace{4.3cm}
- \left. \left.
m^2 \left[ m^2  \left( I_{00}-\beta I_{10} \right)\beta^2a^2
-
2 I_{20}\beta^2b^2 \right] 
\right\} \vphantom{\frac{\hbar^2}{8m^2}}
\right) \;.
\label{eq:dn-dT}
\end{align}%
The right-hand side is computed employing Eq.~\eqref{Eq:kineticentropydens} as
\begin{align}
\left.
\frac{\partial s}{\partial\mu}
\right|_{\omega_{\sigma \tau}} & = 4\beta^2\sinh\alpha\left\{I_{20}+ I_{21}+\frac{\hbar^{2}\beta^{2}}{8m^{2}}\left[2(I_{41}-I_{42})\varkappa^{2}+(I_{40}-2I_{41} + 3 I_{42})\omega^{2}+m^{2}\beta(m^{2}I_{10}a^{2}+2 I_{31}b^{2})\right]\right\}\nonumber \\
&-\, 4\beta\sinh\alpha\left\{I_{10}+\frac{\hbar^{2}\beta^{2}}{8m^{2}}\left[2I_{31}\varkappa^{2}+(I_{30}-I_{31})\omega^{2}+m^{2}\beta(m^{2}I_{00}a^{2}+2I_{21}b^{2})\right]\right\} \nonumber \\
& -\, 4\mu\beta^2\cosh\alpha\left\{I_{10}+\frac{\hbar^{2}\beta^{2}}{8m^{2}}\left[2I_{31}\varkappa^{2}+(I_{30}-I_{31})\omega^{2}+m^{2}\beta(m^{2}I_{00}a^{2}+2I_{21}b^{2})\right]\right\}\;.
\end{align}
With Eq.~\eqref{eq:identity_1} this is seen to be identical to Eq.~\eqref{eq:dn-dT}, so the first Maxwell relation~\eqref{eq:Maxwell_1} holds.

For the left-hand side of the second Maxwell relation~\eqref{eq:Maxwell_2} it is convenient to first decompose the derivative with respect to the spin potential using Eq.~\eqref{eq:kappa_omega_def}:
\begin{equation} \label{eq:decomp_derivs}
    \frac{\partial}{\partial \omega_{\mu \nu} } =
    \frac{\partial \varkappa^\lambda}{\partial \omega_{\mu\nu}}\, \frac{\partial}{\partial \varkappa^\lambda }+ \frac{\partial \omega^\lambda}{\partial \omega_{\mu\nu}}
    \,\frac{\partial}{\partial \omega^\lambda }
    = u^\mu \frac{\partial}{\partial \varkappa_\nu }
    - u^\nu \frac{\partial}{\partial \varkappa_\mu }
    - \epsilon^{\mu \nu \alpha \lambda} u_\alpha \frac{\partial}{\partial \omega^\lambda }\;.
\end{equation}
Then, with Eq.~\eqref{Eq:kineticentropydens} we obtain
\begin{align}
\left.
\frac{\partial s}{\partial\omega_{\mu\nu}}
\right|_{\mu}
&=
\frac{\hbar^2\beta^3}{m^2}
\left\{ \left[ \cosh\alpha
\left( I_{40}-2I_{41}+3I_{42} \right)
- \mu\sinh\alpha \left(I_{30}-I_{31}\right)\right]
\epsilon^{\mu\nu\alpha\beta} u_\alpha\omega_\beta
\right. \nonumber\\
&\qquad\qquad
- \left. 
2 \left[ \cosh\alpha \left(I_{41}-I_{42}\right)
- \mu\sinh\alpha\,I_{31} \right]
\left(u^\mu\varkappa^\nu - u^\nu\varkappa^\mu \right) \right\}\;.
\label{eq:ds-domega}
\end{align}
For the right-hand side of the Maxwell relation~\eqref{eq:Maxwell_2} we obtain with Eq.~\eqref{eq:spintensor}
\begin{align}
\hbar
\left.
\frac{\partial S^{\mu\nu}}{\partial T}
\right|_{\mu} & = \frac{\hbar^2\beta^2}{m^2}
\left( \left\{ \cosh\alpha
\left[ \beta \left(I_{40}-I_{41}\right) - I_{30}- I_{31}  \right]
- \beta \mu\sinh\alpha \left(I_{30}-I_{31}\right)\right\}
\epsilon^{\mu\nu\alpha\beta} u_\alpha\omega_\beta
\right. \nonumber\\
&\qquad\qquad
- \left. 
2 \left[ \cosh\alpha \left(\beta I_{41}-I_{31}\right)
- \beta \mu\sinh\alpha\,I_{31} \right]
\left(u^\mu\varkappa^\nu - u^\nu\varkappa^\mu \right) \right)\;.
\label{eq:dS-dT}
\end{align}
With Eq.~\eqref{eq:identity_1} this is seen to agree with Eq.~\eqref{eq:ds-domega}, so that the second Maxwell relation~\eqref{eq:Maxwell_2} is also fulfilled.

Finally, using the decomposition~\eqref{eq:decomp_derivs} and Eq.~\eqref{eq:chargedens}, the left-hand side of the third Maxwell relation~\eqref{eq:Maxwell_3} reads
\begin{align}
\left.
\frac{\partial n}{\partial \omega_{\mu\nu}}
\right|_{T} =
4\sinh\alpha\frac{\hbar^2\beta^2}{4m^2}
\left[ \left(I_{30}-I_{31}\right)
\epsilon^{\mu\nu\alpha\beta}
u_\alpha\omega_\beta
- 2I_{31} \left( u^\mu\varkappa^\nu
- u^\nu\varkappa^\mu \right) \right]\;.
\label{eq:dn-domega}
\end{align}
Taking the derivative of Eq.~\eqref{eq:spintensor} with respect to $\mu$, we observe that this is equal to the right-hand side of the third Maxwell relation~\eqref{eq:Maxwell_3}, thus this relation holds as well.

\section{Maxwell relations for the canonical pseudo-gauge}
\label{app:maxwell_canonical}

In this appendix, we check the Maxwell relations in the canonical pseudo-gauge.
We show that they are not satisfied unless one imposes the global-equilibrium conditions $a^\mu \equiv \varkappa^\mu$ and $b^\mu \equiv \omega^\mu$ in \textit{selected terms}. 
As in the kinetic-theory pseudo-gauge, the thermal vorticity is kept constant when taking partial derivatives.

We first consider the Maxwell relation~\eqref{eq:Maxwell_1}. 
Using Eq.~\eqref{eq:dens_can} and the identities~\eqref{eq:useful_identities},the left-hand side of this relation reads
\begin{align}
\left.
\frac{\partial n_C}{\partial T}
\right|_{\omega_{\sigma\tau}}
& = 
\left.
\frac{\partial n}{\partial T}
\right|_{\omega_{\sigma \tau}}
+
\frac{\hbar^2 \beta^3}{m^2}
\left\{\sinh\alpha
\left[ \beta (I_{40}-I_{41})a^2
+ 2\beta I_{41}b^2 +
2 (I_{30} +I_{31})
\left( a\cdot\varkappa+b\cdot\omega
\right) \right] \right. \nonumber \\
& \hspace*{2.5cm} - \left.\beta \mu\, \cosh\alpha
\left[(I_{30}-I_{31})a^2
+ 2I_{31} \left(
b^2+a\cdot\varkappa+b\cdot\omega \right)
\right] \right\}\;.
\label{eq:dnC_dT_app}
\end{align}
The right-hand side of Eq.~\eqref{eq:Maxwell_1} is calculated with Eq.~\eqref{eq:entropydens_can_0} and the identities~\eqref{eq:useful_identities} as
\begin{align}
\left.
\frac{\partial s_C}{\partial\mu}
\right|_{\omega_{\sigma\tau}}
& =
\left.
\frac{\partial s}{\partial\mu}
\right|_{\omega_{\sigma\tau}} +
\frac{\hbar^2 \beta^3}{m^2}
\left\{\sinh\alpha
\left[ \beta (I_{40}-I_{41})a^2
+ 2\beta I_{41}b^2 +
2 (I_{30} +2I_{31})
\left( a\cdot\varkappa+b\cdot\omega
\right) 
- 2I_{31} \left( \varkappa \cdot \varkappa + \omega\cdot \omega \right) \right] \right. \nonumber \\
& \hspace*{2.5cm} - \left.\beta \mu\, \cosh\alpha
\left[(I_{30}-I_{31})a^2
+ 2I_{31} \left(
b^2+a\cdot\varkappa+b\cdot\omega \right)
\right] \right\}\;.
\label{eq:dsC_dmu_app}
\end{align}
This is not identical to Eq.~\eqref{eq:dnC_dT_app}, i.e., the  Maxwell relation~\eqref{eq:Maxwell_1} is in general \textit{not fulfilled} unless one employs the global-equilibrium conditions and replaces one factor of $\varkappa^\mu$ by $a^\mu$ and one factor of $\omega^\mu$ by $b^\mu$ \textit{exclusively in the last term} in the first line.

We now consider the Maxwell relation~\eqref{eq:Maxwell_2}. 
Using the entropy density~\eqref{eq:entropy_density_can} and Eq.~\eqref{eq:decomp_derivs}, we obtain for the left-hand side
\begin{align}
\left.
\frac{\partial s_C}{\partial\omega_{\mu\nu}}
\right|_{\mu}
& =
\left.
\frac{\partial s}{\partial\omega_{\mu\nu}}
\right|_{\mu}
+
\frac{2\hbar^2\beta^3}{m^2}
\left\{
\left[
\cosh\alpha\,(I_{41}+I_{42})
-
\mu\sinh\alpha\,I_{31}
\right]
\left(
u^\mu a^\nu-u^\nu a^\mu
-
\epsilon^{\mu\nu\alpha\beta}
u_\alpha b_\beta
\right)
\right. \nonumber\\
&\hspace*{3cm}
- \left.
2\cosh\alpha\,I_{42}
\left[
u^\mu\varkappa^\nu-u^\nu\varkappa^\mu
-
\epsilon^{\mu\nu\alpha\beta}
u_\alpha\omega_\beta
\right]
\right\}\;.
\label{eq:dsC_domega_app}
\end{align}
On the other hand, with Eqs.~\eqref{eq:spintensor} and \eqref{eq:spintensor_can}, together with the identities~\eqref{eq:useful_identities}, the right-hand side reads
\begin{align}
\hbar
\left.
\frac{\partial S_C^{\mu\nu}}{\partial T}
\right|_{\mu}
& =
\hbar
\left.
\frac{\partial S^{\mu\nu}}{\partial T}
\right|_{\mu} +
\frac{2\hbar^2\beta^3}{m^2}
\left[
\cosh\alpha\,(I_{41}-I_{42})
-
\mu\sinh\alpha\,I_{31}
\right]
\left(
u^\mu\varkappa^\nu-u^\nu\varkappa^\mu
-
\epsilon^{\mu\nu\alpha\beta}
u_\alpha\omega_\beta
\right)\;.
\label{eq:dSC_dT_app}
\end{align}
Equations~\eqref{eq:dsC_domega_app} and \eqref{eq:dSC_dT_app} are not identical, unless one employs the global-equilibrium condition $a^\mu = \varkappa^\mu$, $b^\mu = \omega^\mu$ \textit{exclusively in the last term} in the first line of Eq.~\eqref{eq:dsC_domega_app}.

Finally, we consider the Maxwell relation~\eqref{eq:Maxwell_3}. 
Using Eqs.~\eqref{eq:dens_can}, one finds for the left-hand side
\begin{align}
\left.
\frac{\partial n_C}{\partial\omega_{\mu\nu}}
\right|_{T}
= \left. \frac{\partial n}{\partial\omega_{\mu\nu}}
\right|_{T} + 
2 \sinh\alpha \,\frac{\hbar^2\beta^2}{m^2}\,
 I_{31}\left( u^\mu a^\nu-u^\nu a^\mu
- \epsilon^{\mu\nu\alpha\beta} u_\alpha b_\beta
\right)\;.
\label{eq:dnC_domega_app}
\end{align}
On the other hand, with Eqs.~\eqref{eq:spintensor} and \eqref{eq:spintensor_can}, as well as Eq.~\eqref{eq:identity_2} the right-hand side becomes
\begin{equation}
\label{eq:dSC_dmu_app}
\hbar
\left.
\frac{\partial S_C^{\mu\nu}}{\partial\mu}
\right|_{T}
=\left.
\frac{\partial S^{\mu\nu}}{\partial\mu}
\right|_{T} + 
2 \sinh\alpha \, \frac{\hbar^2\beta^2}{m^2}\,
I_{31}
\left(u^\mu \varkappa^\nu-u^\nu \varkappa^\mu
-\epsilon^{\mu\nu\alpha\beta} u_\alpha \omega_\beta
\right)\;.
\end{equation}
Equations~\eqref{eq:dnC_domega_app} and \eqref{eq:dSC_dmu_app} only agree after imposing $a^\mu = \varkappa^\mu$, $b^\mu = \omega^\mu$.
Consequently, in general the Maxwell relations~\eqref{eq:Maxwell} are not fulfilled in the canonical pseudo-gauge.
However, they can be fulfilled if the global-equilibrium conditions $a^\mu = \varkappa^\mu$ and $b^\mu = \omega^\mu$ are imposed in \textit{selected terms}.

\section{Maxwell relations for the GLW pseudo-gauge}
\label{app:maxwell_GLW}
In this appendix, we examine the Maxwell relations in the GLW pseudo-gauge. 
We show that, in general, they are not satisfied.  

We first consider Eq.~\eqref{eq:Maxwell_1}.  
Since $n_{\rm GLW}=n$, using Eq.~\eqref{eq:entropydens_GLW} we obtain
\begin{align}
\left.
\frac{\partial n_{\rm GLW}}{\partial T}
\right|_{\omega_{\sigma\tau}}
& =
\left.
\frac{\partial s_{\rm GLW}}{\partial\mu}
\right|_{\omega_{\sigma\tau}} +
\sinh\alpha \,\frac{\hbar^2\beta^3}{m^2}
\left[
(I_{30}-I_{31})\,\varkappa\cdot a
+
2I_{31}\,\omega\cdot b
\right]\;.
\label{eq:Maxwell_GLW_1_diff}
\end{align}
Here, we used the fact that the corresponding Maxwell relation is satisfied in the kinetic-theory pseudo-gauge. 
The Maxwell relation~\eqref{eq:Maxwell_1} is therefore not satisfied, not even when imposing the global-equilibrium conditions.

Next, we consider the Maxwell relation~\eqref{eq:Maxwell_2}. 
It is advantageous to consider the difference between the entropy densities in the GLW and kinetic-theory pseudo-gauges on the left-hand side,
\begin{align}
&
\left.
\frac{\partial
\left(
s_{\rm GLW}-s
\right)}
{\partial\omega_{\mu\nu}}
\right|_{\mu}=
-
\cosh\alpha\,\frac{\hbar^2\beta^2}{m^2}
\left[
(I_{30}-I_{31})
\left(
u^\mu a^\nu-u^\nu a^\mu
\right)-
2I_{31}
\epsilon^{\mu\nu\alpha\beta}
u_\alpha b_\beta
\right]\; .
\label{eq:dsGLW_V_domega}
\end{align}
On the other hand, the difference between the spin tensors in the GLW and kinetic-theory pseudo-gauges yields for the right-hand side 
\begin{align}
&
\hbar
\left.
\frac{\partial
\left(
S_{\rm GLW}^{\mu\nu}-S^{\mu\nu}
\right)}
{\partial T}
\right|_{\mu}\nonumber\\
&=
\frac{\hbar^2\beta^3}{m^2}
\left\{
\left[
\cosh\alpha\,(I_{40}-I_{41})
-
\mu\sinh\alpha\,(I_{30}-I_{31})
\right]
\left(
u^\mu a^\nu-u^\nu a^\mu
\right)
-
2
\left[
\cosh\alpha\,I_{41}
-
\mu\sinh\alpha\,I_{31}
\right]
\epsilon^{\mu\nu\alpha\beta}
u_\alpha b_\beta
\right\}.
\label{eq:dSGLW_V_dT}
\end{align}
This does not agree with Eq.~\eqref{eq:dsGLW_V_domega}, thus the Maxwell relation~\eqref{eq:Maxwell_2} is not satisfied, even when imposing $a^\mu=\varkappa^\mu$ and $b^\mu=\omega^\mu$.

Finally, we examine the Maxwell relation~\eqref{eq:Maxwell_3}. 
Again, since $n_{\rm GLW}=n$ and the corresponding Maxwell relation in the kinetic-theory pseudo-gauge is satisfied, the difference originates entirely from the additional contribution to the GLW spin tensor.
We find
\begin{align}
\left.
\frac{\partial n_{\rm GLW}}
{\partial\omega_{\mu\nu}}
\right|_{T}
-
\hbar
\left.
\frac{\partial S_{\rm GLW}^{\mu\nu}}
{\partial\mu}
\right|_{T}=
-
\sinh\alpha\, \frac{\hbar^2\beta^2}{m^2}
\left[
(I_{30}-I_{31})
\left(
u^\mu a^\nu-u^\nu a^\mu
\right)
-
2I_{31}
\epsilon^{\mu\nu\alpha\beta}
u_\alpha b_\beta
\right]\; .
\label{eq:Maxwell_GLW_3_diff}
\end{align}
Since this does not vanish, even when imposing the global-equilibrium conditions, the Maxwell relation~\eqref{eq:Maxwell_3} is not satisfied. 
%

%\newpage
%************************************************
\bibliography{ref.bib}{}
\bibliographystyle{utphys}
%************************************************
\end{document}